\documentclass[prd,a4paper,onecolumn,preprintnumbers,nofootinbib]{revtex4}
\usepackage[a4paper, hdivide={1.91cm,,1.165cm}, vdivide={1.83cm,,3.8cm}]{geometry}

\usepackage{color}
\usepackage[hyperfootnotes=false]{hyperref}
\usepackage{ulem}
\usepackage{youngtab}
\usepackage[english]{babel} 
\usepackage{amsmath}
\usepackage{float}
\usepackage{amssymb}
\usepackage{braket}
\usepackage{amsfonts,bbm}
\usepackage{mathrsfs}
\usepackage{dsfont,tensor}
\usepackage{enumerate}
\usepackage{graphicx}
\usepackage{bbm}
\usepackage{array}
\usepackage{booktabs}
\usepackage{units}
\usepackage{textcomp}
\usepackage{mathtools}
\usepackage{cancel}
\usepackage{slashed}
\usepackage{multirow}
\usepackage{slashed}
\usepackage{comment}
\usepackage{relsize}
\allowdisplaybreaks

\newcommand{\dd}{\text{d}}

\newcommand{\mx}{\text{m}_{\chi}}

\newcommand{\vf}{v_{f}}

\newcommand{\pts}[1]{\phantom{.}\hfill(\textit{#1~point}\ifthenelse{\equal{#1}{1}}{}{\textit{s}})}
\newcommand{\Lagr}{\mathcal{L}}

\newcommand{\Msun}{{\ifmmode{{\rm{M_{\odot}}}}\else{${\rm{M_{\odot}}}$}\fi}}

\newcommand{\beq}{\begin{equation}}
\newcommand{\eeq}{\end{equation}}
\newcommand{\bea}{\begin{eqnarray}}
\newcommand{\ena}{\end{eqnarray}}

\newcommand{\lsim}{\mathrel{\mathop{\kern 0pt \rlap
{\raise.2ex\hbox{$<$}}}
\lower.9ex\hbox{\kern-.190em $\sim$}}}
\newcommand{\gsim}{\mathrel{\mathop{\kern 0pt \rlap
{\raise.2ex\hbox{$>$}}}
\lower.9ex\hbox{\kern-.190em $\sim$}}}
\begin{document}
%%%%%%%%%%%%%%%%%%%%%%%%%%%%%%%%%

\preprint{ULB-TH/18-16}

\title{New Analysis of Neutron Star Constraints on Asymmetric Dark Matter}

\author{Raghuveer Garani}
\email{rgaranir@ulb.ac.be}
\affiliation{Service de Physique Th\'eorique, Universit\'e Libre de Bruxelles, Boulevard du Triomphe, CP225, 1050 Brussels, Belgium}

\author{Yoann Genolini}
\email{yoann.genolini@ulb.ac.be}
\affiliation{Service de Physique Th\'eorique, Universit\'e Libre de Bruxelles, Boulevard du Triomphe, CP225, 1050 Brussels, Belgium}

\author{Thomas Hambye}
\email{thambye@ulb.ac.be}
\affiliation{Service de Physique Th\'eorique, Universit\'e Libre de Bruxelles, Boulevard du Triomphe, CP225, 1050 Brussels, Belgium}

%%%%%%%%%%%%%%%%%%%%%%%%%%%%%%%%%%
%%%%%%%%%%%%%%%%%%%%%%%%%%%%%%%%%%

\begin{abstract}
Due to their extreme density and low temperature, neutron stars (NS) are efficient probes to unveil interactions between standard model and dark matter (DM) particles. From elastic scatterings on NS material, DM can get gravitationally trapped by the star. The cooling of DM through further collisions may lead to the formation of a dense core which could collapse into a black hole, thus destroying the whole NS. From the observation of old NS, such a scenario leads to very stringent constraints on the parameter space of asymmetric DM. In this work we reexamine this possibility in detail. This includes: (a) 
a new  detailed determination of  the number of DM particles captured, properly taking into account the fact that neutrons form a highly degenerate Fermi material; (b)
the determination of the time evolution of the DM density and energy profiles inside the NS, which allows us to understand how, as a function of time, DM thermalizes with NS material; 
(c) the determination of the corresponding constraints which hold on the DM-neutron cross section, including for the case where a large fraction of DM particles have not thermalized; (d) the first determination of the stringent constraints which also hold in a similar way on the DM-muon cross section, particularly relevant for leptophilic DM models; and (e) the use of realistic NS equations of state in determining these constraints.

\end{abstract}

%%%%%%%%%%%%%%%%%%%%%%%%%%%%%%%%%
%%%%%%%%%%%%%%%%%%%%%%%%%%%%%%%%%
\maketitle

%%%%%%%%%%%

\section{Introduction}

If Dark Matter (DM) interacts with ordinary matter, it could be trapped in astrophysical objects such as the Sun or compact stars. The most straightforward way this could be realized is through elastic scattering of DM particle with ordinary matter in the stellar medium. Such a scattering can reduce the kinetic energy of DM particle such that its velocity falls below the corresponding escape velocity. Once gravitationally trapped, DM can undergo further scatterings and thermalize with the stellar matter. This possibility is interesting in several ways. For the Sun it is mostly interesting because it can lead to DM signals in the form of high energy neutrino flux from DM pair annihilation. For a Neutron Star (NS), the corresponding signal at Earth is expected to be very weak. However, for Asymmetric Dark Matter (ADM, see e.g. the reviews of~\cite{Davoudiasl:2012uw,Boucenna:2013wba,Petraki:2013wwa,Zurek:2013wia}), this is very interesting because DM accumulating in the center could form a core which could further gravitationally collapse into a black hole.
The requirement that such collapses do not occur gives constraints on the mass and interactions of particle DM.
This has been analyzed in a series of works \cite{Goldman:1989nd,Gould:1989gw,Kouvaris:2007ay,Kouvaris:2010vv,deLavallaz:2010wp,McDermott:2011jp,Kouvaris:2011fi,Guver:2012ba,Bell:2013xk,Bramante:2013hn,Bramante:2013nma,Kouvaris:2013kra,Bramante:2014zca,Bramante:2017xlb,Kouvaris:2018wnh,Gresham:2018rqo}. More generally, DM annihilation (and kinetic heating) can heat up the NS leading to an increase of the surface temperature. Observational constraints on this temperature can lead to an upper bound on the annihilation rate which depends on the number of DM particles accreted~\cite{Kouvaris:2007ay,Baryakhtar:2017dbj,Bramante:2013hn,Bell:2013xk,Bertone:2007ae}. 
Also for a NS, the recent first observation of gravitational waves from binary NS merger~\cite{GBM:2017lvd} lead naturally to the question of whether the presence of DM in neutron stars could possibly affect the spectrum of such gravitational waves. 
All these phenomena crucially depend on the amount of DM that can accumulate in neutron stars. 

In this work, we reconsider in detail DM accretion and thermalization in NS and associated constraints from black hole formation for ADM, incorporating a series of effects which had not been considered (or only partially incorporated) before. This includes:
\begin{itemize}
\item A proper treatment of Pauli blocking for DM accretion rate, i.e. including the fact that neutrons form a highly degenerate Fermi plasma in the NS. This allows us to obtain semi-analytical results for the number of accreted DM particles. Interestingly, our formalism takes into account various finite temperature effects. Two such effects turn out to be particularly important for low DM mass: saturation of the Fermi degeneracy suppression and evaporation of accreted DM (i.e.~DM kicked out of the NS by scattering off neutrons). This is detailed in section~\ref{sec:accretion}.

\item The computation of trajectories (orbits) of DM particles, once they are gravitationally trapped. More precisely, we examine the shrinking of orbits due to further scatterings with NS matter, taking into account the effect of Pauli blocking, and also variations of DM velocity along its orbit. DM trajectories can be divided in 2 periods: a first (short) period  where the average orbit is larger than the NS and, a second one where the orbit is fully enclosed inside the star. This enables us to determine as a function of time, both the DM radial density profile and the DM energy distribution, which to our knowledge were never determined before. From the evolution of DM energy distribution, we can compute as a function of time, the number of accreted DM particles which had (and had not) the time to thermalize with the neutrons (i.e. whose kinetic energy does (does not) reach the temperature of the NS). Thus, we treat DM thermalization as a progressive effect rather than an instantaneous process happening for all the accreted DM particles after a characteristic average "thermalization time". This point is the subject of section~\ref{sec:thermalization}.

\item An update, using the two previous points, of the constraints which hold on the DM mass and DM-neutron elastic cross section, from the requirement that the NS is not destroyed due to black hole formation. In particular, the knowledge of the DM energy distribution as a function of time, allows us to determine whether there is black hole formation even when a large part of accreted DM have not had the time to thermalize. This happens to be crucial for the case where DM is a boson and forms a Bose-Einstein condensate. Section~\ref{sec:blackholeconstraints} is devoted to this point.

\item For the first time, a detailed study of DM interactions, not only with neutrons, but with protons and muons which are ineluctable constituents of NS matter. Using our formalism we are able to compute DM accreted by scattering off degenerate protons and muons, and the corresponding constraints from black hole formation. In particular, for muons this allows us to obtain interesting bounds on the DM-muon elastic scattering cross section, which could be the only relevant ones if DM is for instance ``quarkophobic''. This is discussed at the end of section~\ref{sec:blackholeconstraints}.
\end{itemize}

The results presented and discussed in this paper depend on the macroscopic properties of neutron stars such as mass ($M_\star$), radius ($R_\star$) and temperature ($T_\star$), and also on the microscopic properties of its degenerate medium (chemical potential, etc). These different scales can be linked thanks to the equation of state of nuclear matter which, for the extreme conditions met in a neutron star, relies on extrapolations. Hence in this paper, instead of considering a uniform NS, we will consider a realistic profile which corresponds to the low mass configuration of model BSK20 from Ref.~\cite{Potekhin:2013qqa}. This benchmark model we call ``Model A'', is chosen for being the most conservative one.
We discuss how our results change with respect to the NS profile considered, by comparing them to three other benchmark models: B, C and D in Appendix~\ref{app-profiles}. 

%%%%%%%%%%%%%%%%%%%%%%%%%%%%%%%%%%%%%%%

\section{Gravitational Trapping of DM from Elastic Scattering with Neutron Star Matter}
\label{sec:accretion}

In this work, we are interested in models where DM ($\chi$) is asymmetric with negligible DM-DM self interactions, and with heavy mediators ($m_{med} \gg m_{\chi}$). For bosonic DM, we consider the following vectorial effective interaction lagrangian
\beq
\Lagr_{int}  \supset G \left(\chi^\dagger \overset{\leftrightarrow}{\partial_\mu}  \chi \right)  \bar{f}\gamma^\mu f,
\label{eq:lint-bosons}
\eeq
which results in the following differential cross section in the non relativistic limit (i.e. when both DM and fermion momenta are set to zero) 
\beq
\frac{\dd \sigma_{\chi-f}}{\dd \cos \theta_{cm}} = \frac{G^2}{2 \pi} \frac{m_\chi^2 m_f^2}{\left(m_\chi + m_f\right)^2}.
\label{eq:cross-section-bosonic}
\eeq
For fermionic DM (which we also denote by $\chi$) we consider the following effective lagrangian
\beq
\Lagr_{int}  \supset G \left(\bar{\chi}  \gamma_\mu  \chi \right)  \bar{f} \gamma^\mu f,
\label{eq:lint-fermions}
\eeq
which, in the non relativistic limit, turns out to result in the same differential cross section as above
\beq
\frac{\dd \sigma_{\chi-f}}{\dd \cos \theta_{cm}} = \frac{G^2}{2 \pi} \frac{m_\chi^2 m_f^2}{\left(m_\chi + m_f\right)^2 } .
\label{eq:cross-section-fermionic}
\eeq
 Note that small momentum dependence in the cross section has been ignored.
 For scalar interactions, i.e.~$\Lagr_{int}  \supset G_S \left(\chi^\dagger \chi \right)  \bar{f}f$ and $\Lagr_{int}  \supset G_S \left(\bar{\chi} \chi \right)  \bar{f}f$, the cross sections are 
 $\frac{\dd \sigma_{\chi-f}}{\dd \cos \theta_{cm}} = \frac{G_S^2}{8 \pi} \frac{m_f^2}{\left(m_\chi + m_f\right)^2}$ and 
  $\frac{\dd \sigma_{\chi-f}}{\dd \cos \theta_{cm}} = \frac{G_S^2}{2 \pi} \frac{m_\chi^2 m_f^2}{\left(m_\chi + m_f\right)^2}$, for scalar and fermion DM particles, respectively. Thus for scalar interactions the results can be obtained from the vectorial case by the simple rescaling $G\rightarrow G_S/(2 m_\chi)$ and $G\rightarrow G_S$, respectively. Phenomenology for pseudo-scalar, axial-vector mediators and for light mediators is left for future work. As electrons in NS are relativistic (their chemical potential is $\mathcal{O}(0.1)$ GeV, see Appendix~\ref{app-profiles}), constraints on DM-electron cross section is also left for future work since the scattering formalism presented here is only relevant for non-relativistic degenerate neutron star matter.

%%%%%%%%%%%%%%%%%%%
\subsection{Boltzmann Equation for ADM}

The rate of accumulation of DM particles in a star is governed by the
Boltzmann equation incorporating the capture of DM particles from the halo ($C^{\rm w}_{\star}$), the rate of annihilation ($A_{\star}$)  and the rate of evaporation ($E^{\rm w}_{\star}$). In the following we consider the case of {\it completely anti-symmetric} DM candidate, which does not pair annihilate today. In this case the Boltzmann equation reads

\begin{eqnarray}
\label{eq:diffeq-3}
\frac{\dd \text{N}_{\chi}}{\dd \text{t}} &=& C_{\star}^{\rm w} - E_{\star}^{\rm w}\text{N}_{\chi}\; ,
\end{eqnarray}
with the following solution 
\begin{equation}
\label{eq:evolution-asym-sol}
N_\chi (t) =  C_{\star}^{\rm w} t_{\star} \left (\frac{1 - e^{- E_{\star}^{\rm w} t_\star}}{E_{\star}^{\rm w} t_\star} \right).
\end{equation}
In the limit of evaporation being negligible, the number of accreted particles grows linearly with time, $N_\chi (t) =  C_{\star}^{\rm w} t_{\star}$.

%%%%%%%%%%%%%%%%%%%%%
\subsection{Geometric Upper Bound on DM accretion}

Before considering in detail how DM is accreted, it is useful to compute the amount of DM which can be accreted
if every DM particle passing through the neutron star were trapped by it. This gives an absolute upper bound on the possible amount of accreted DM.  To this end, we define the geometric capture rate to be the flux of DM in the halo that intersects a NS with mass $M_\star$ and radius $R_\star$~\cite{Press:1985ug,Bottino:2002pd},
\begin{eqnarray}
\label{eq:capturegeom}
C_\star^{\rm geom} &=& \pi R_\star^2 \, \left(\frac{\rho_\chi}{m_\chi}\right) \, \int_0^{\infty} \dd u_\chi \, f_{v_\star}(u_\chi) \, \frac{\omega^2(R_\star)}{u_\chi} \,,\nonumber\\
&=& \pi R_\star^2 \, \left(\frac{\rho_\chi}{m_\chi}\right)
\, \langle v \rangle_0 \,  \left(1 + \frac{3}{2} \, \frac{v_{esc}^2(R_\star)}{v_d^2}\right) \xi(v_\star,v_d)\; ,
\end{eqnarray}
where $\omega(r)=\sqrt{v_{esc}^2(r)+u_\chi^2}$ is the speed of the DM particle within the NS, $v_{esc}(r)$ is the escape velocity at a given radius, $u_\chi$ is the DM speed at infinity, $\rho_\chi$ is the DM energy density in the star neighborhood and $\langle v \rangle_0 = \sqrt{8/(3\pi)}v_d \,$ is the resulting average speed in the DM rest frame with DM velocity dispersion $v_d$. The factor $\xi(v_\star,v_d)$ takes into account the suppression due to the motion of the NS with velocity $v_\star$ ($\xi(v_\star = 0,v_d) = 1$). For an isotropic flux of DM, $\xi(v_\star,v_d)$ is
\begin{eqnarray}
\label{eq:xi}
\,\,\,\,\,\xi(v_\star,v_d) \equiv\nonumber \nonumber \frac{v_d^2 \, e^{- \frac{3 \, v_\star^2}{2 \, v_d^2}} + \sqrt{\frac{\pi}{6}} \, \frac{v_d}{v_\star} \, \left(v_d^2 + 3 \, v_{esc}^2(R_\star) + 3 \, v_\star^2\right)
\, \textrm{Erf}\left(\sqrt{\frac{3}{2}} \, \frac{v_\star}{v_d}\right)}{2 \, v_d^2 + 3 \, v_{esc}^2(R_\star)}~.
\end{eqnarray}
For the canonical values of $v_\star$ and $v_d$, $\xi(v_\star = 220 \, \textrm{km/s},v_d = 270 \, \textrm{km/s}) \simeq 0.75$. By taking $v_{esc}(R_\star=11.6 {\rm \;km}) = 0.62\,c$ (Model A, see Appendix~\ref{app-profiles}), the geometric rate can be written as

\begin{equation}
	\label{eq:capturegeom-sim}
	C_\star^{\rm geom} = 5.6 \times 10^{25} \left( \frac{\rho_\chi}{\rm{GeV/cm^3}} \cdot\frac{1 \,\rm{GeV}}{m_\chi}\cdot  \frac{R_\star}{11.6 \;{\rm km} }\cdot\frac{ M_\star}{\rm 1.52\;M_\odot} \right){\rm s}^{-1}\; .
	\end{equation}
The mass and radius dependence is obtained assuming the second term dominates the parenthesis of Eq.~(\ref{eq:capturegeom}), which is always true in practice. Strictly speaking, this bound can only be reached for an infinite cross-section $\sigma_{\chi-f}$. Practically, in the following, as soon as $\sigma_{\chi-f}$ becomes larger than $\sigma_{crit}=\pi R_\star^2/N_b$ (the geometrical cross section per baryon, with number of baryons $N_b$), the capture rate is close to this bound. In numbers, we find that $\sigma_{crit} \approx 2.5\cdot 10^{-45}$ for Model A. From $C_\star^{\rm geom}$, one can estimate the total DM mass accreted by a NS after a time $t_\star$,

\begin{eqnarray}
M_\chi^{\rm tot} & \approx & m_\chi\, C_\star^{\rm geom}\, t_\star, \\
& \approx & 1.1\times 10^{-14} \;{\rm M_\odot.}\left( \frac{ M_\star}{\rm 1.52 \;M_\odot}\right )^2\left( \frac{\rho_{\chi}}{\rm GeV.cm^{-3}} \cdot \frac{11.6 \;{\rm km} }{R_\star} \cdot  \frac{t_\star}{\rm 10\;Gyr} \right )\;.\label{eq:est_tot_M}
\end{eqnarray}

Two important comments should be made here. Firstly, the characteristic scale of $\sigma_{crit}$ shows that NS are able to probe very small elastic scattering cross section $\sigma_{\chi-f}$. 
Secondly, from Eq.~(\ref{eq:est_tot_M}), the typical mass accreted is a small fraction of the NS mass. 
To maximize this fraction, one can consider NS in "extreme" DM densities like $\rho_\chi= 10^6$ ${\rm GeV/ cm^3}$, and use the most optimistic NS star profile (i.e.~benchmark model D in the Appendix~\ref{app-profiles}, which increases the captured DM mass by 40\%). Hence, we see right from the start that from ``standard accretion'' this mass fraction will always remain tiny. Thus, as well known, it is not in this way that NS could accrete enough DM to leave an imprint in the gravitational wave spectrum. Even if it were the case, we find it interesting to note, as a side comment, that no such large imprint is expected since the dark matter cores will stay at the center of the neutrons stars, see Appendix~\ref{app-gw}. Finally note that this absolute upper bound also holds for the case where, on top of the capture induced by DM elastic scattering on ordinary matter, DM is also accreted via DM self-interactions.

%%%%%%%%%%%%%%%%%%%%%%
\subsection{DM Scattering off Fermi Sea of Neutrons}

The neutrons in NS form a highly degenerate Fermi plasma whose degeneracy is parametrized by the value of the corresponding chemical potential $\mu_F$. This chemical potential can largely vary depending on the neutron star model considered. The four benchmark NS models we consider in Appendix~\ref{app-profiles} give an idea of the possible variations, with chemical potentials in the core ranging from 0.2 GeV to 0.8 GeV. The chemical potential also varies as a function of radial distance from the core, as shown in Fig.~\ref{fig:chem-pot}.
The effect of Pauli blocking in DM-neutron scattering can be summarized as follows: when a DM particle scatters off a neutron, the neutron must find an energy level which is not already occupied by an other neutron or be completely ejected from the Fermi sea.
This means that not all neutrons are available to scatter along all the kinematical possibilities which, if there were no degeneracy, would be allowed. This leads to a suppression in the scattering rate. For DM accretion in NS this effect has been either neglected~\cite{Kouvaris:2010vv,Kouvaris:2011fi} or estimated by simple means~\cite{McDermott:2011jp,Bell:2013xk}.
In order to account for this effect, one must compute the differential scattering rate integrated over the incoming Fermi-Dirac neutron distribution $f_p(E_p)$ and over the available holes for the final neutron state, $1 - f_{p^\prime}(E_{p^\prime})$, \cite{Reddy:1997yr}. To this end we write the capture rate as follows~\cite{Gould:1987ju, Garani:2017jcj}

\begin{eqnarray}
\label{eq:capture1}
C_{\star}^{\rm w} = \int_0^{R_\star} 4 \pi r^2 \dd r \int_0^\infty \dd u_\chi \, \left(\frac{\rho_\chi}{\mx}\right) \, \frac{f_{v_\star}(u_\chi)}{u_\chi} \, w(r) \int_0^{v_{esc}(r)} R^- (w \to v) \,  \dd v ~,
\end{eqnarray}
where $f_{v_\star}(u_\chi)$ is the DM halo velocity distribution, and $R^- (w \to v)$ is the differential scattering rate in the NS frame for a DM particle with velocity $w$ to scatter to a smaller velocity $v$ ($w > v$). The formal expression for a degenerate medium is given by,
\beq
\label{eq:cap-diff}
R^-(w \rightarrow v) = \int \zeta_n(r) \rho_u \frac{\dd \sigma}{\dd v} |\boldsymbol{w} -\boldsymbol{u}|f_p(E_p,r) (1- f_{p^\prime}(E_p + q_0,r)) \dd^3 u\;,
\eeq
where $\sigma$ is the usual (free) particle cross section. The notations used are summarized in Tab.~\ref{tab:notation}.
\begin{table}
\begin{center}
\begin{tabular}{|c| c|}
\hline\hline
Quantity & Definition  \\ 
\hline
$u_\chi$  &  DM velocity in the DM halo far from the NS \\
$\boldsymbol{w},k^\mu$  & Incoming DM velocity and 4-momentum $(E_k,\boldsymbol{k})$\\
$\boldsymbol{v}, k'^\mu$  & Outgoing  DM velocity and 4-momentum $(E_k',\boldsymbol{k^\prime})$ \\
$\boldsymbol{u}, p^\mu$  &   Incoming target particle velocity and 4-momentum $(E_p ,\boldsymbol{p})$ \\
$\boldsymbol{u'},p'^\mu$  &  Outgoing target particle velocity and 4-momentum $(E_{p'},\boldsymbol{p^\prime})$\\
$ q^\mu$ & 4-momentum transfer for the process $(q_0,\boldsymbol{q})$\\
$p'^\mu = p^\mu + q^\mu$ & 4-momentum conservation \\
$k'^\mu = k^\mu - q^\mu $ & 4-momentum conservation \\
$m_{r}$ & Reduced mass of DM-target system \\
$\mu_F$ & Chemical potential of the target particle \\
$\mu=m_\chi/m_t$ & Ratio of DM mass by the target mass ($m_t$)\\
$f_i(E_i)$ & Fermi-Dirac distribution, function of $(E_i,\mu_i,T)$\\
\hline
\end{tabular}
  \caption{The relevant variables of the problem are listed in the first column,  and the corresponding definition in the second column. Note that we first focus on neutrons as DM targets. These notations are generic and are also used below when muons and protons targets.}
\label{tab:notation}
\end{center}
\end{table}
The Fermi-Dirac distribution for the neutron in the initial and final state is
\beq
f_p(E_p,r) = \left(e^{(E_p -\mu_F(r))/T(r)} +1\right)^{-1}\;,\label{eq:fp}
\eeq
and
\beq
1-f_{p^\prime}(E_p+q_0,r) = 1-\left(e^{(E_p+q_0 -\mu_F(r))/T(r)} +1\right)^{-1}\; ,\label{eq:fpp}
\eeq
with $q_0 = 1/2 \, m_\chi(w^2 -v^2)$, the energy lost by DM in a single scatter in the NS frame. The Eq.~(\ref{eq:cap-diff}) also involves $\rho_u$, the phase space density of neutrons, $\rho_u=g_s m_n^3/(2\pi)^3$ (with $g_s=2$ for spin 1/2 fermions). As detailed in Appendix~\ref{app:numd-resc}, the factor $\zeta_n(r)$ is introduced to account for the correct number density given by the NS profiles. That way we specify the right number density for a given chemical potential.\\

From Eq.~(\ref{eq:cap-diff}) it is possible to derive analytical results for the $R(w \rightarrow v)$ factors, in the non-relativistic and vanishing temperature limit, relevant when the target scattering particles are neutrons, protons and muons. To this end we further generalize the calculation presented in Ref.~\cite{Bertoni:2013bsa,Reddy:1997yr} by providing analytical expressions for the differential scattering rates above. This derivation is quite technical, and we refer the interested reader to Appendix~\ref{app:diff-rate}, where details of this calculation are extensively discussed.
Finally, to compute the capture rate, the remaining three integrals on $r$, $u_\chi$ and $v$ in Eq.~(\ref{eq:capture1}) have to be computed numerically. 

%%%%%%%%%%%%%%%%%%%%%%%%%%%%%%%%%%%
\subsection{Results}

\begin{figure}[t]
	\begin{center}
	  \includegraphics[width=0.5\linewidth]{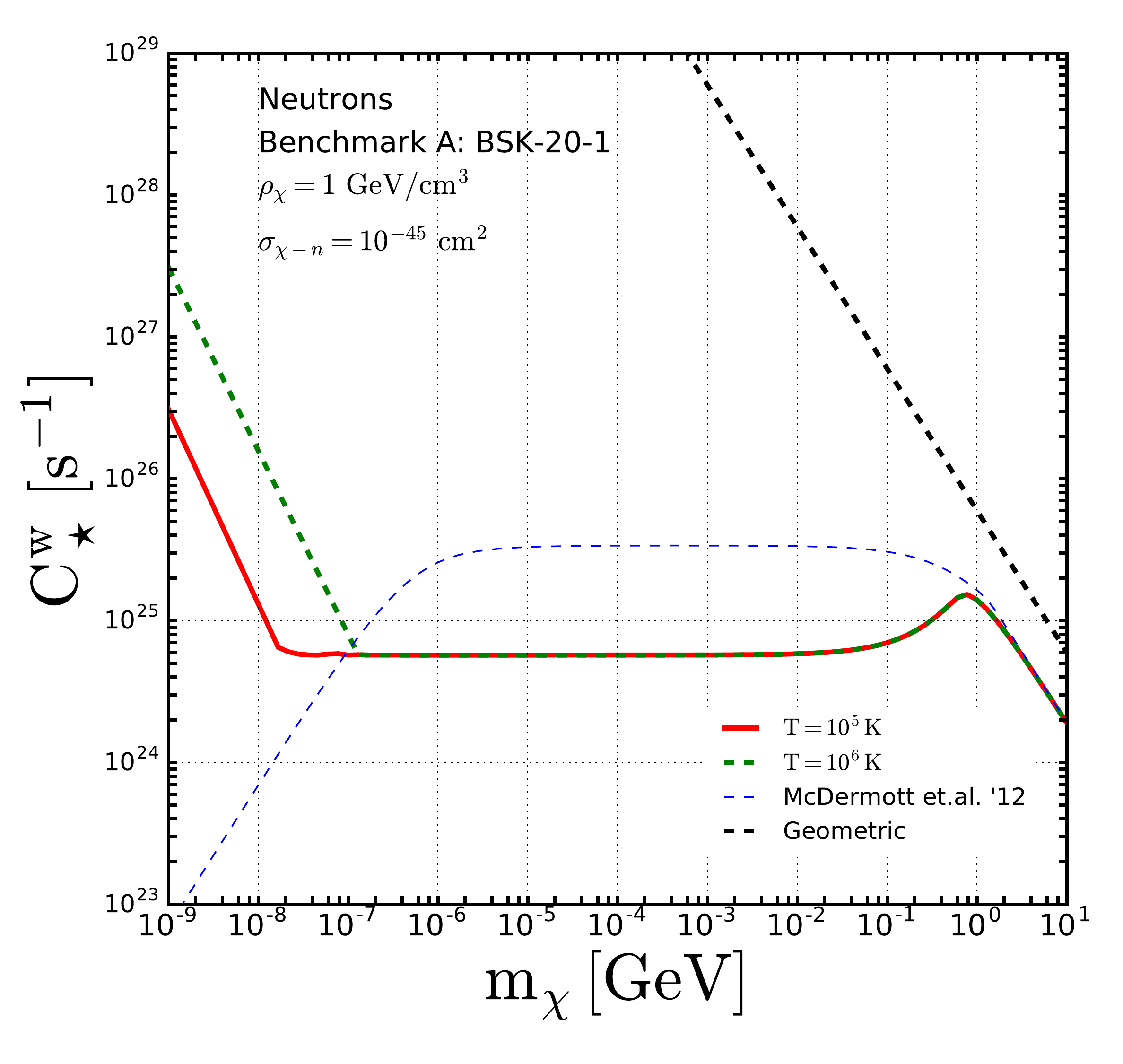}\\
	\end{center}
	\caption{Capture rate as a function of DM mass for benchmark Model A, for DM density $\rho_\chi=1$ GeV/cm$^3$, and $\sigma_{\chi-n}=10^{-45}\rm cm^2$. The dashed-green and solid-red lines results from our accurate treatment of the Fermi sea, for two typical NS temperatures, while the blue line is computed from the heuristic argument used in Ref.~\cite{McDermott:2011jp}. The black dashed line corresponds to the geometric capture rate.}
	\label{fig:cap-bsk20A}
\end{figure}

In Fig.~\ref{fig:cap-bsk20A} we show the numerical results we obtain for the
number of DM particles accreted as a function of DM mass. We use benchmark NS Model A and compute these quantities for two typical values of NS temperature, 10$^5$ K (solid-red lines) and 10$^6$ K (dashed-green lines), respectively.
These results follow from the computation of Eq.~(\ref{eq:cap-diff}).
Also shown in Fig.~\ref{fig:cap-bsk20A}, in black-dashed line, the amount of accreted DM if all DM particles crossing the NS were captured, as described by the geometrical limit Eq.~(\ref{eq:capturegeom}).

We first discuss the mass dependence of the accretion rate. For DM mass $m_\chi \gtrsim$~1~GeV, the mass dependence is the same as if neutrons were described by a Maxwell-Boltzmann distribution, i.e. there is no effect due to Pauli blocking. Strictly this is true when the average energy transfer is larger than $\mu_F$ or in other words, when the energy of DM particle is large enough to kick a neutron out of the Fermi sea, where there are no occupied state. Thus, in this case the DM accretion rate is simply proportional to the cross section and to the DM flux. Since the latter is proportional to $1/m_{\chi}$, the accretion rate goes as $1/m_{\chi}$ too, just as the geometric cross section, Eq.~(\ref{eq:capturegeom}).\footnote{Note that (even if this is not shown in the plots), this behavior is valid for a mass up to $10^6$~GeV, for neutrons/protons. Above this mass, the accretion rate scales as $1/m_{\chi}^2$ for kinematic reasons: it is less likely that DM particles lose enough energy to be captured in a single scatter.} In the example considered in Fig.~\ref{fig:cap-bsk20A}, the accretion rate is $\sim$ 2.5 times smaller than the geometric rate because we took a cross section $\sigma_{\chi-n}=10^{-45}\rm cm^2$ which is 2.5 times smaller than the critical cross section $\sigma_{crit}$. Results in this figure scale as the cross section, except if the accretion rate is larger than the geometric one, in this case the accretion rate saturates to the geometric rate.

For $m_{\chi}\lesssim  1$~GeV the accretion rate no longer scales as $1/m_{\chi}$ but is suppressed because of Pauli blocking. Actually, Pauli blocking effect manifest itself through a complicated interplay of the four energy scales of the problem, namely the temperature ($T_\star$), the neutron mass ($m_n$), the chemical potential of neutron ($\langle \mu_F \rangle$) \footnote{In the following, we are referring to $\langle \mu_F \rangle$, the averaged Fermi chemical potential of neutrons in DM-neutron collision.} and the DM mass ($m_{\chi}$). In Fig.~\ref{fig:cap-bsk20A} one can observe three different regimes, depending on $m_{\chi}$:

\begin{itemize}
\item \underline{Regime I}: for $m_{\chi}$ in the range [0.1 GeV, 1 GeV], whose boundary values correspond to $\langle\mu_F\rangle \simeq 10^{-1}$~GeV (i.e.~$v_f \simeq 0.44$) and the neutron mass, respectively. In this regime, the Fermi suppression is relatively moderate and increases when $m_{\chi}$ decreases. For instance, for $m_{\chi}=100$~MeV, the suppression is of order $100$ with respect to what we would obtain neglecting the Fermi sea suppression. It is of order of a few when compared to the results of~\cite{McDermott:2011jp} (blue line in Fig.~\ref{fig:cap-bsk20A}). This regime does not depend on the NS temperature.

\item \underline{Regime II}: for $m_{\chi}$ in the range [10$^{-8,-7}$ GeV, 0.1 GeV], whose boundary values correspond to ${\cal O}(T)$ and $\langle\mu_F\rangle \simeq 10^{-1}$~GeV. In this regime the suppression is more important and leads to a flat behavior of the capture rate as the DM mass $m_{\chi}$ decreases. Here, the Fermi suppression brings the capture rate orders of magnitude below the geometrical rate and, about an order of magnitude below the result of Ref.~\cite{McDermott:2011jp}. Similar to \underline{regime I}, this regime does not depend on the NS temperature.  

\item \underline{Regime III}: for $m_{\chi}$ below ${\cal O}(T)$. The accretion rate is no longer constant but increases as $1/m_{\chi}$ with decreasing mass. This behavior is different from the $T=0$ K limiting case, which would corresponding to a flat behavior in the capture rate. Note that using approximation of Ref.~\cite{McDermott:2011jp} (blue-dashed line) the accretion rate decreases with mass. Clearly, regime III is due to finite temperature effects.  
\end{itemize}

To understand the behavior of these regimes we will start by discussing the $T=0$ K case which is relevant to describe \underline{regimes I and II},  and subsequently discuss the effects of non-vanishing temperature relevant for \underline{regime III}.

\begin{figure}[htb]
	\begin{center}
	  \includegraphics[width=0.32\linewidth]{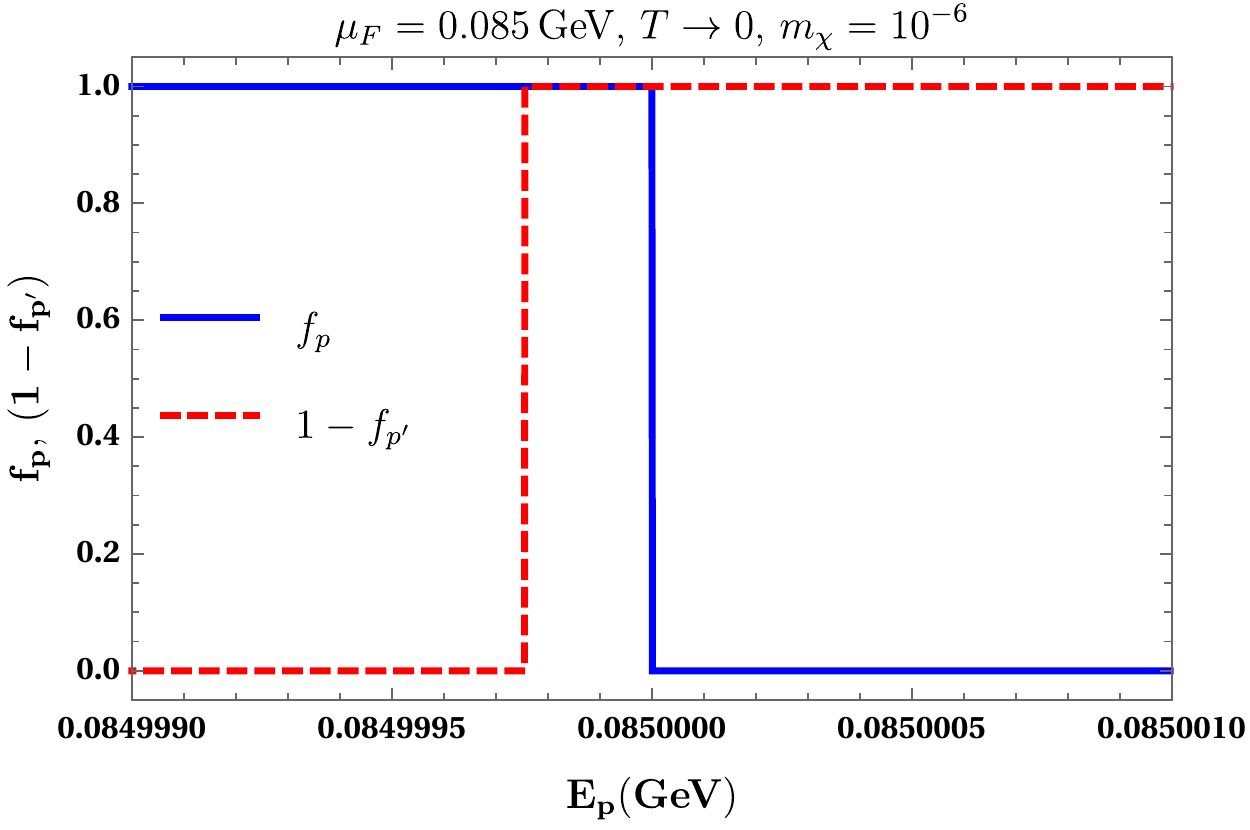}\quad
	   \includegraphics[width=0.32\linewidth]{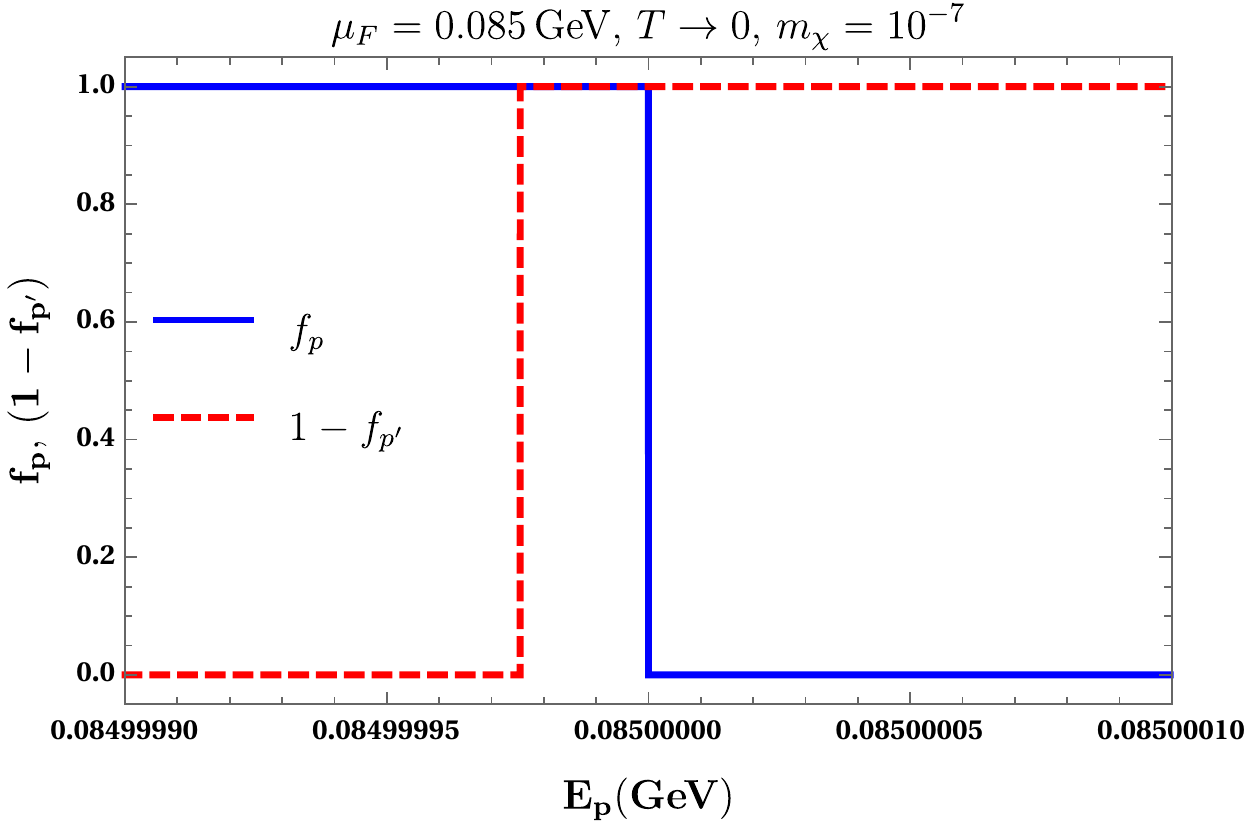}\quad
	   \includegraphics[width=0.32\linewidth]{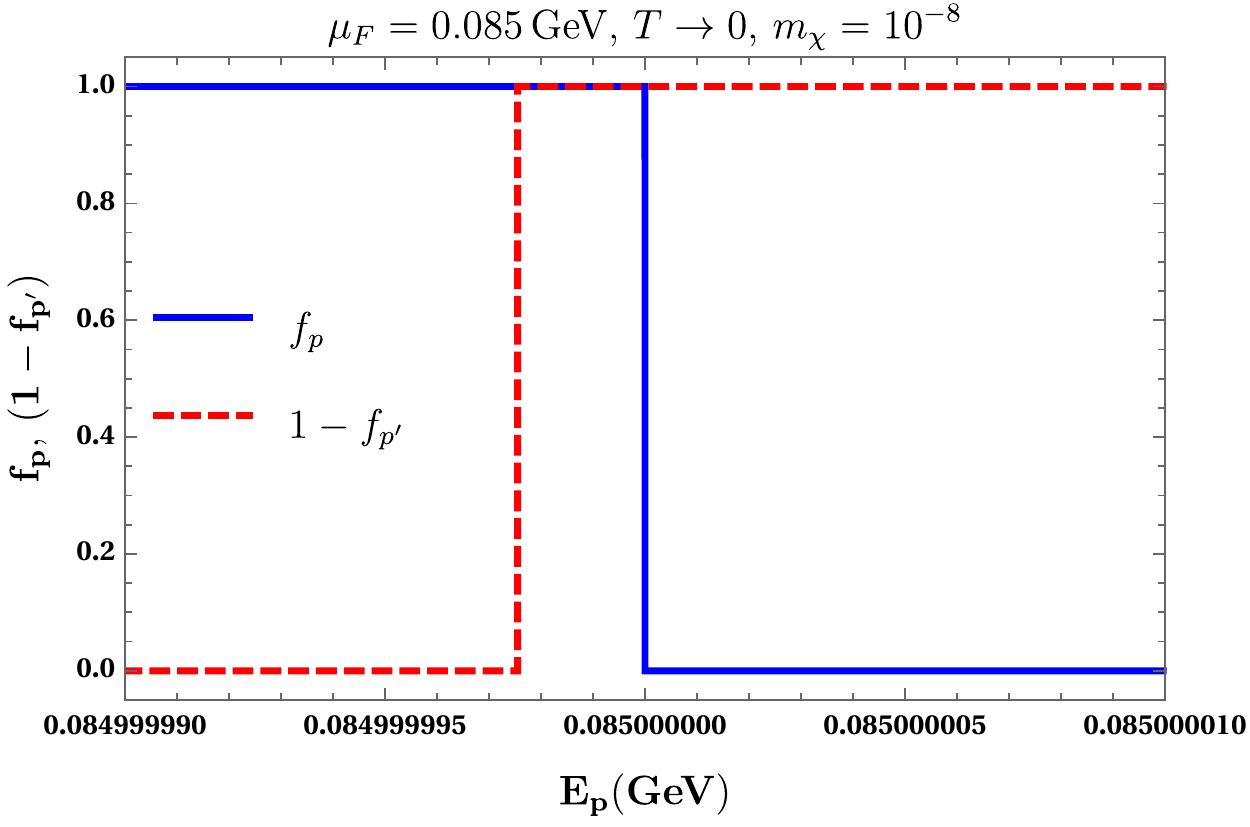}
	\end{center}
	\caption{Distribution functions of Eqs.~(\ref{eq:fp}) (blue line) and~(\ref{eq:fpp}) (red line) for $T=0$. Each panel corresponds to a different DM mass, $m_{\chi}=10^{-6,-7,-8}$ GeV from left to right.}
	\label{fig:distribzerotemp}
\end{figure}

\begin{figure}[htb]
	\begin{center}
	  \includegraphics[width=0.32\linewidth]{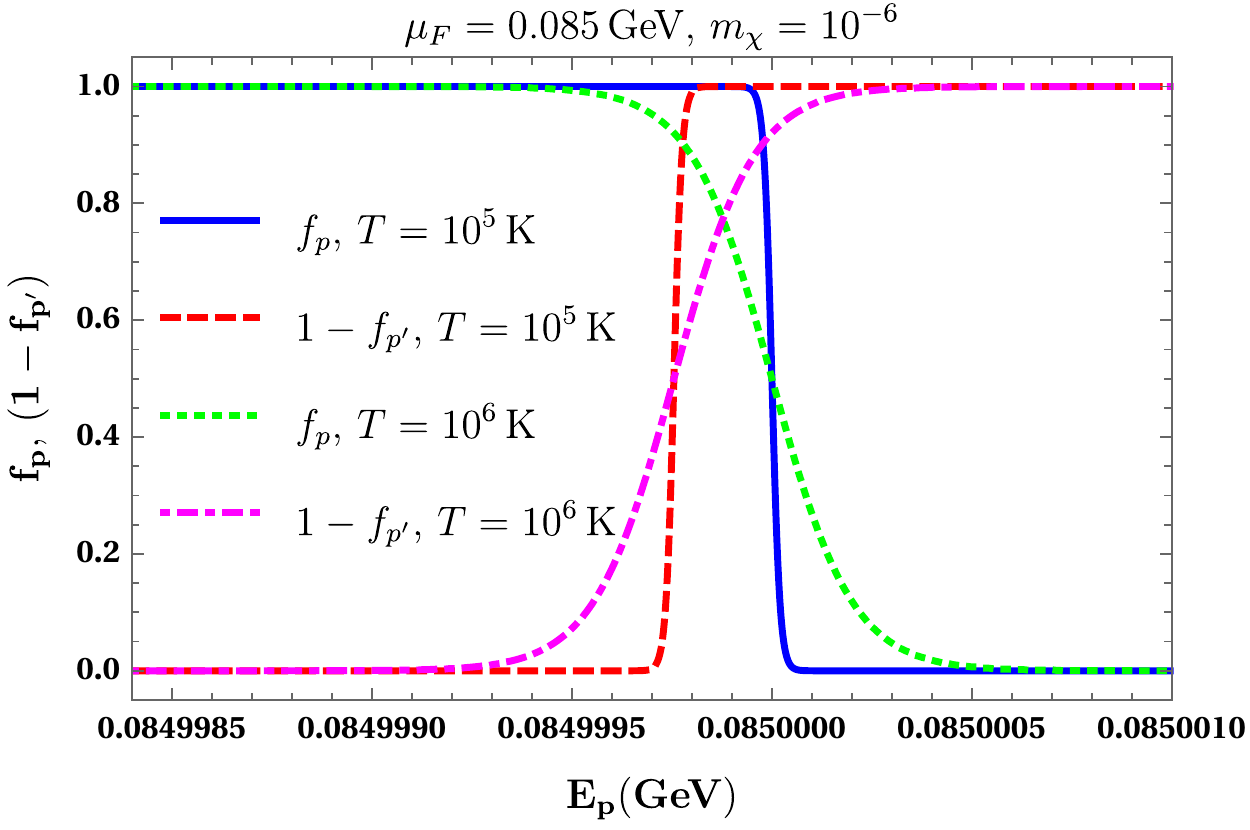}\quad
	   \includegraphics[width=0.32\linewidth]{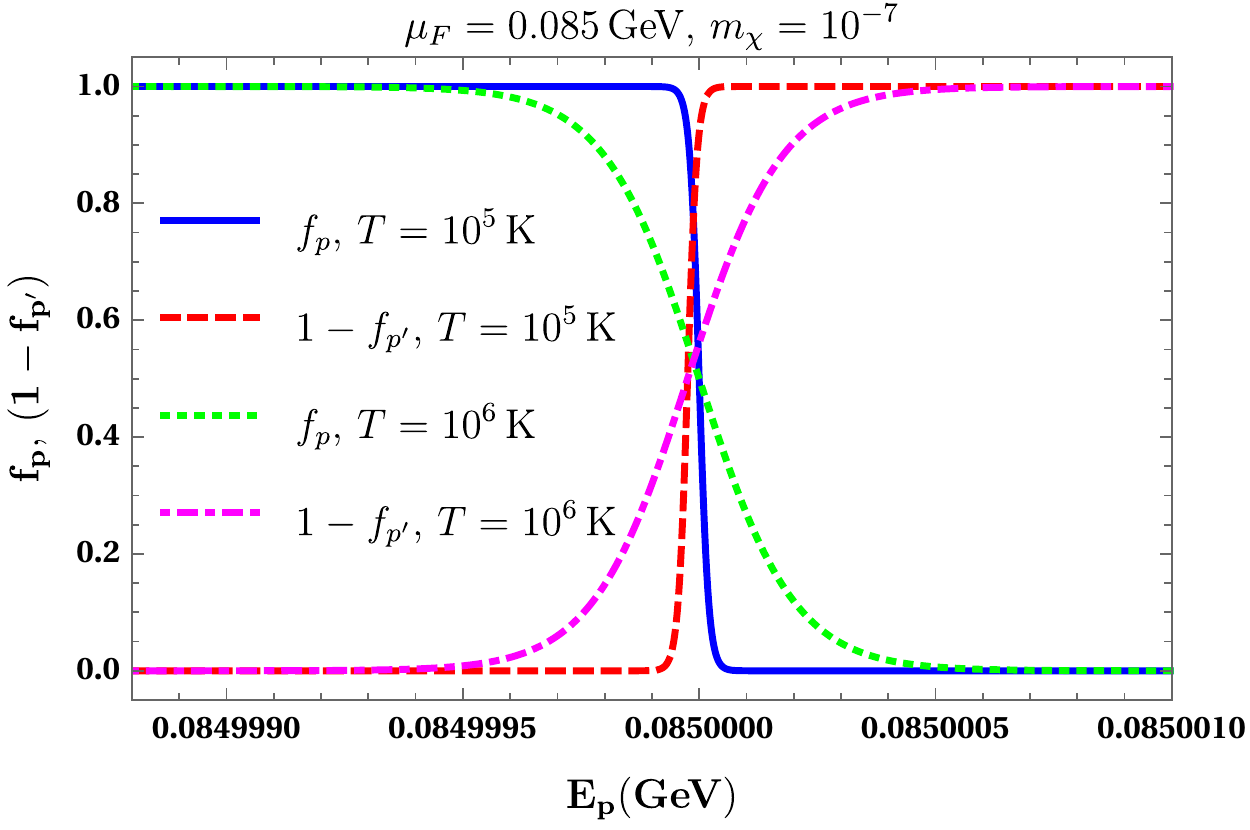}\quad
	   \includegraphics[width=0.32\linewidth]{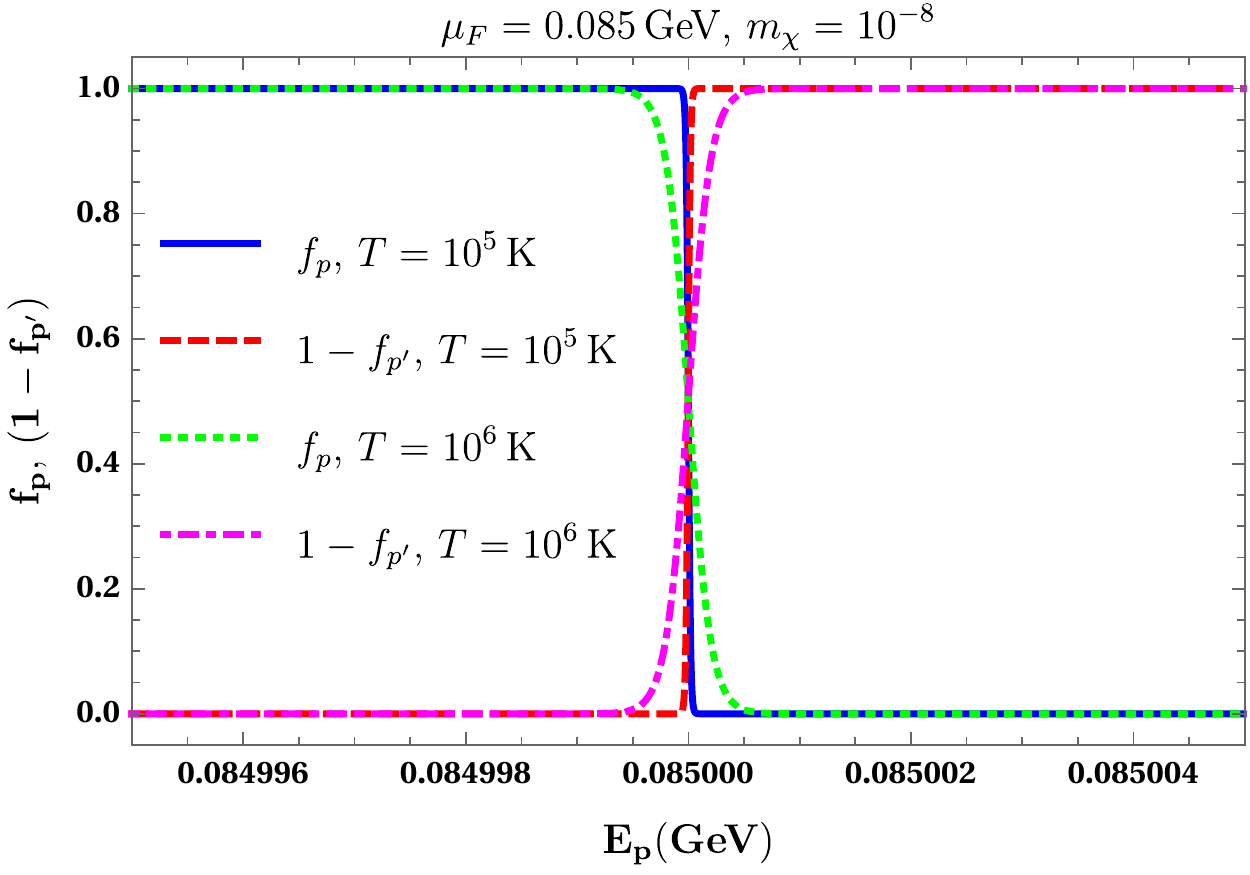}
	\end{center}
	\caption{Same legend as Fig.~\ref{fig:distribzerotemp} but for two typical NS temperature $10^{5}$ K and $10^{6}$ K.}
	\label{fig:distribnonzerotemp}
\end{figure}

\begin{figure}[htb]
	\begin{center}
	  \includegraphics[width=0.4\linewidth]{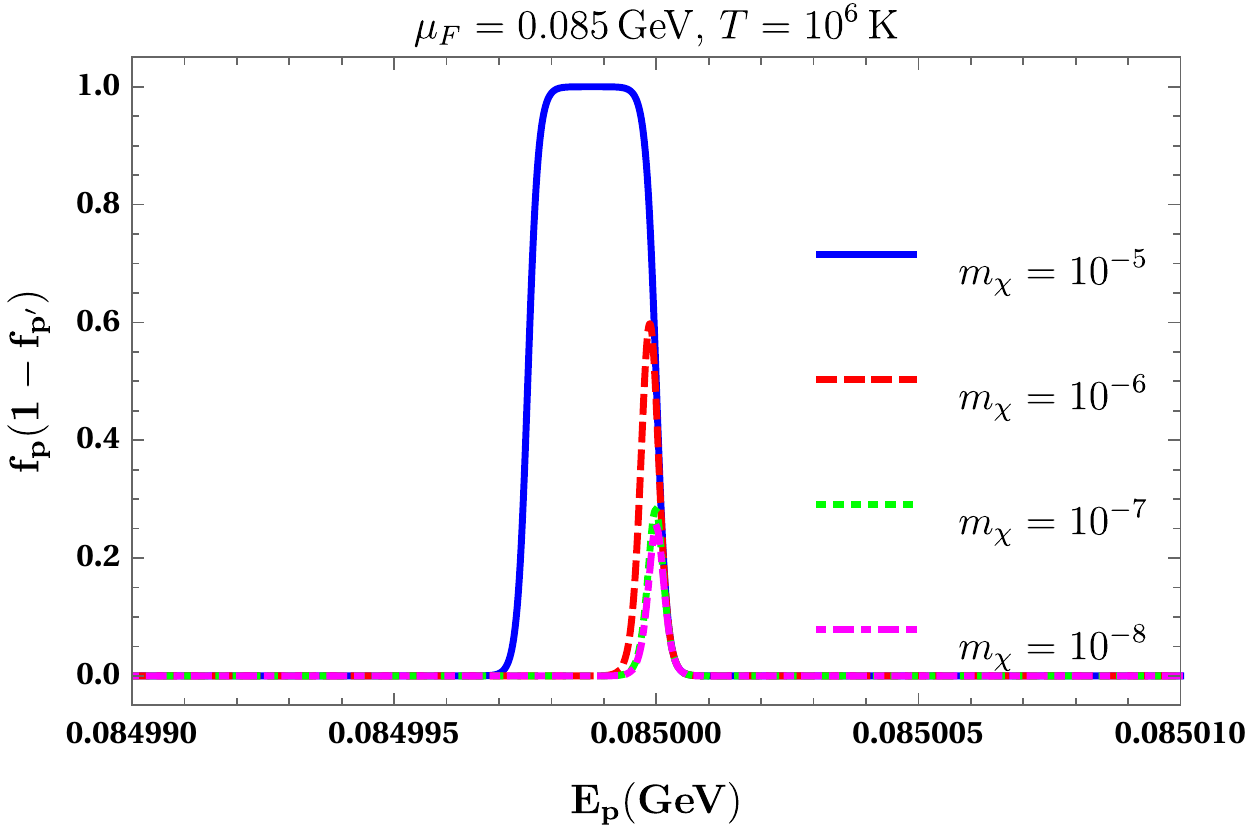}\quad
	   \includegraphics[width=0.4\linewidth]{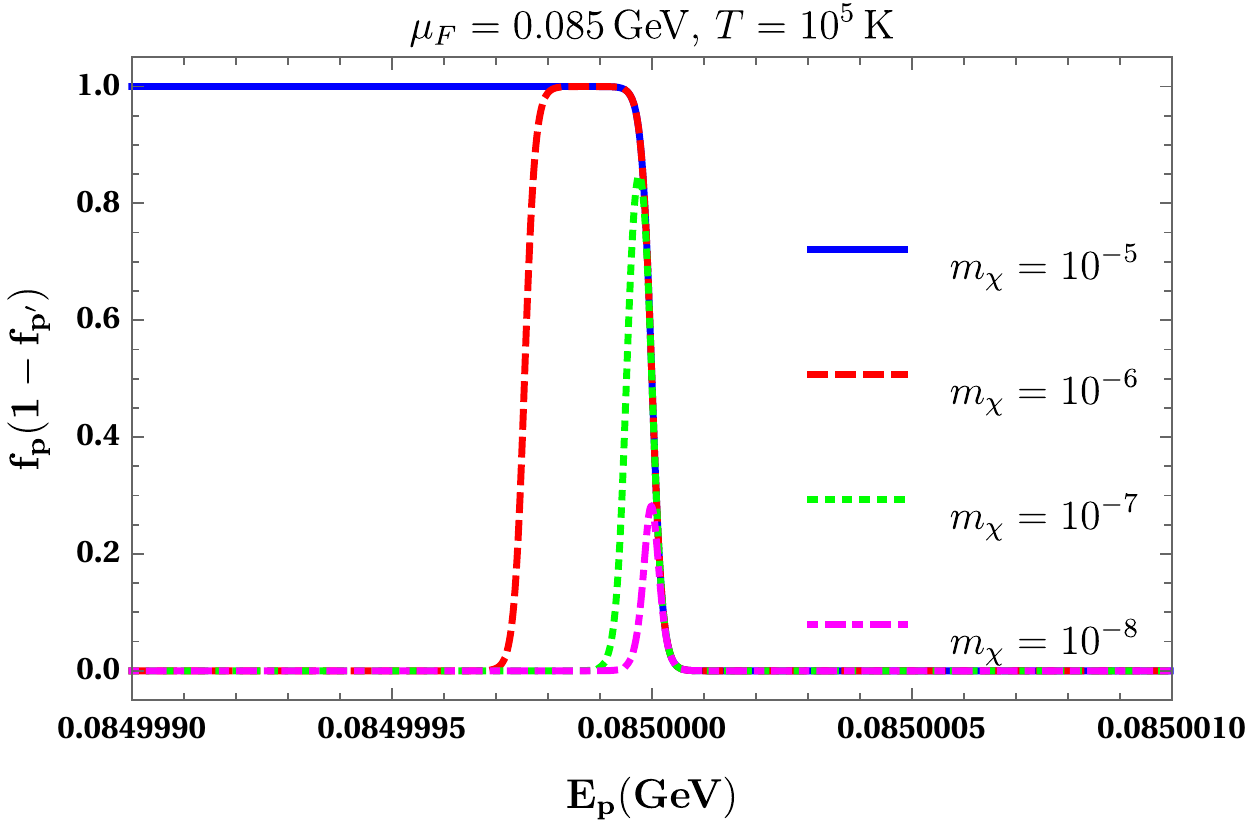}
	  
	\end{center}
	\caption{Overlap of the distribution functions (Eqs.~(\ref{eq:fp}) and~(\ref{eq:fpp})), i.e.~product of both the distribution functions which appears in Eq.~(\ref{eq:cap-diff}), for two typical NS temperatures $10^{5}$ K and $10^6$ K. Each curve corresponds to a different DM mass, from top to bottom $m_{\chi}=10^{-5,-6,-7,-8}$ GeV. }
	\label{fig:distribnonzerotempoverlap}
\end{figure}

\subsubsection{$T\rightarrow0$ Results: Regimes I and II}
It is not a surprise that Pauli blocking affects the capture rate for $m_{\chi}$ below 1 GeV. The capture rate is proportional to the integral of the product of the in-coming and out-going neutron distribution functions, i.e. proportional to the overlap of these two functions. For a vanishing temperature these distribution functions behave like theta functions,

\begin{eqnarray}
f_{p}(E_p-\mu_F)&=&\Theta(\mu_F-E_p),  \\
1-f_{p'}(E_p+q_0-\mu_F)&=&\Theta(E_p+q_0-\mu_F),
\end{eqnarray}
whose overlap occurs on an interval of energy equal to the energy gained by the neutron, $q_0$.
When $m_{\chi}$ is above $\sim$~GeV, the $1-f_{p'}(E_p+q_0-\mu_F)$ distribution function fully overlaps with the $f_{p}(E_p-\mu_F)$ function, i.e. the outgoing neutron is not Pauli blocked. However when $m_{\chi}$ is below $\sim$~GeV the overlap scales as $q_0$.
Thus we need to determine how $q_0$ scales with $m_{\chi}$, or more exactly how the distribution of $q_0$ which holds for a given $m_{\chi}$ scales with $m_{\chi}$. Here it is useful to recall that for a given value of $m_{\chi}$ and a given value of the initial DM particle velocity, $w$, an all range of outgoing DM particle velocity, $v$, is kinematically available, with different transition probabilities, resulting in a distribution of energy transfer $q_0$.
Hence, to understand the scaling in $m_{\chi}$ of the accretion rate, we start by writing the expression of $q_0$ for the elastic collision in terms of momentum transfer $\boldsymbol{q}$ and initial neutron momentum $\boldsymbol{p}$
\beq
\label{eq:q0}
q_0 = \frac{1}{2 m_n} \left( (\boldsymbol{p}+\boldsymbol{q})^2  - \boldsymbol{p}^2 \right) = \frac{1}{2 m_n} \left(\boldsymbol{q}^2 + 2 \,\boldsymbol{p} \cdot \boldsymbol{q}  \right)\;.
\eeq
Next, for fixed $v$ and $w$, we can integrate over both final neutron momentum angles.\footnote{Note that to derive the differential scattering rate we first integrate over the final state neutron phase space which is then followed by integration over the initial state neutron phase space in the non-relativistic limit. } Given the $\delta$ functions associated to conservation of energy and momentum, the angle between $\boldsymbol{q}$ and $\boldsymbol{p}$ is fixed to a single possible value, which fixes $q_0$ as a function of $w$, $v$ and $m_{\chi}$,
\beq
\label{eq:q0-int}
q_0 = \frac{1}{2} m_{\chi} \left(w^2 -v^2 \right).
\eeq
This shows that, for fixed initial velocity $w$, the distribution of $q_0$ is totally fixed by $m_{\chi}$ and the distribution of final velocity $v$.
When $m_{\chi} \ll m_n$, it can be shown that for fixed value of $w$, the distribution of $v$ is independent of the value of $m_{\chi}$. 
Thus the $q_0$ distribution scales as $m_{\chi}$,  which leads to an accretion rate independent of $m_{\chi}$ (given that the flux of DM particles scales as $1/m_{\chi}$).
This allows to understand \underline{regime II}, that is to say the mass scaling of the accretion rate between 0.1 GeV ($\sim \langle \mu_F \rangle$) and 2 T, see Fig.~\ref{fig:cap-bsk20A}. 

To better illustrate this dependence, we display in Fig.~\ref{fig:distribzerotemp} the Fermi distribution functions $f_{p}$ and the distribution of holes $1-f_{p'}$ for a vanishing NS temperature and for maximal recoil energy $ q_0^{max} = E_k^\chi$. As an example, we chose the values $\mu_F= 0.085$~GeV, $w=0.7$, $v=0$ and display the distribution functions for $m_{\chi}$ equals to $10^{-6}$ GeV, $10^{-7}$ GeV, and $10^{-8}$ GeV, from the left to the right panel, respectively. One observes that the overlap of both functions occurs on an interval equal to $q_0$ which scales as~$m_{\chi}$.

 The transition from \underline{regime II} to \underline{regime I} is observed when $m_{\chi} \sim m_n$ (or equivalently for $q_0 \sim \mu_F$). This can be understood from the fact that for such values the distribution of $v$ for fixed value of $w$ is not anymore independent of $m_{\chi}$. Note that we can obtain an analytical expression for the differential scattering rate in the limit of vanishing temperature, see Appendix~\ref{app:diff-rate}. After performing both $v_t$ and $v_s$ integrals in Eq.~(\ref{eq:rate-cm2}), the result we obtain contains many terms involving different $\Theta$ functions, corresponding to different kinematical conditions, see Eq.~(\ref{eq:r-factor}). This shows explicitly that a simple ansatz for Pauli blocking in Ref.~\cite{McDermott:2011jp} is not necessarily correct. Among these terms, there are two types of terms, the ones involving $\Theta$ functions linear in the velocities and the ones involving $\Theta$ functions which are quadratic in the velocities. It is the second class which turns out to be dominant for \underline{regime II} whereas it is the first class which is relevant for \underline{regime I}, i.e. the turnover which occurs around the neutron mass.
Note also that quantitatively, based on this mathematical definition of \underline{regimes I and II}, the latter implies $v_f^2 - m_{\chi}/m_n (w^2 -v^2) >0 $ ($\Lambda_- >0$) which means $ \mu_F > q_0$ (see Eqs.~(\ref{eq:r-factor}) and~(\ref{eq:variables-2})).

\subsubsection{Finite Temperature Effects: Saturation in Regime III}

For non vanishing temperature, as a result of thermal fluctuations, part of the in-coming neutrons have energies above $\mu_F$ while part of the out-going states with energies below $\mu_F$ are available. In this case, the Fermi distribution $f_{p}(E_p-\mu_F)$ and the distribution of holes $1-f_{p'}(E_p+q_0-\mu_F)$ can no longer be approximated by $\theta$ functions, but show a smooth tail above $\mu_F$ and below $\mu_F$, respectively. The typical spread of these tails is equal to the temperature $T$. Hence, the larger the temperature, the broader these tails. This behavior is illustrated in Fig.~\ref{fig:distribnonzerotemp} where both distribution functions are plotted for two values of temperatures, three values of $m_{\chi}$ ($10^{-6}$ GeV, $10^{-7}$ GeV, and $10^{-8}$ GeV, from left to right) and using the maximal recoil energy $ q_0^{max} = E_k^\chi$. For the same values, we display in Fig.~\ref{fig:distribnonzerotempoverlap} the overlap resulting from the product of both distribution functions. We observe that when $q_0$ is larger than the temperature, the overlap is close to a box function and so virtually the same as the one given in the $\theta$ function approximation. On the other hand, when $m_{\chi}$ is so small that $q_0$ reaches the thermal energy of the neutrons $E_n$, that is to say when
\begin{equation}
 q^{max}_0  = E_k^\chi  \simeq \langle E_n \rangle \simeq T\;,
\end{equation}
the overlap of the distribution functions quickly reaches an asymptotic value showing a peaked profile, which strongly differs from the overlap of two $\theta$ functions. This asymptotic behavior is reached when $m_\chi=10^{-7}$ GeV for $T=10^6$ K as shown in the left panel of Fig.~\ref{fig:distribnonzerotempoverlap}, and when $m_\chi=10^{-8}$ GeV for $T=10^5$ K as shown in the right panel of the same figure. The fact that the overlap does not go to 0 but towards this asymptotic area is due to the non-vanishing temperature. 

The typical width of the asymptotic area is actually $\sim 2 T$. Physically, once the transferred energy $q_0$ becomes smaller than $T$, ``thermal'' holes in the Fermi sea are available for the scattered neutron. In that regime, the overlap function is only determined by the temperature and $q_0$ becomes irrelevant. 
In practice, equating the maximal recoil energy $q_0^{max}$ to the temperature $T$ defines the typical mass 
\begin{equation}
m_{\chi}^* = \frac{2 \,T}{w^2} \approx 50\, \left(\frac{T}{\rm 10^5\,K}\right) \,\hbox{eV}\,,\label{eq:mchistar}
\end{equation}
below which the accretion rate increases again as $1/m_{\chi}$, similar to the case where $m_{\chi}>1$ GeV, scaling as the number density of DM particle. This precise behavior explains \underline{regime III} (see Fig.~\ref{fig:cap-bsk20A}), which according to the condition of Eq.~(\ref{eq:mchistar}) starts at smaller masses for smaller temperatures. 

Note that in Ref.~\cite{Kouvaris:2010vv,Kouvaris:2011fi,Kouvaris:2010jy}  the Fermi sea suppression was not considered, whereas in Ref.\cite{McDermott:2011jp} (or similarly in Ref.~\cite{Bell:2013xk}) it was estimated by multiplying the accretion rate obtained without Fermi suppression by a factor $\delta p/p$, where $p$ is the Fermi momentum and $\delta p$ is the recoil momentum in the neutron rest frame. The result obtained in this case is also shown in Fig.~\ref{fig:cap-bsk20A} (blue-dashed line). They result in an accretion rate which is $\mathcal{O}$(1) larger than what we get in \underline{regime I}, and up to an order of magnitude larger in \underline{regime II}. Also, as these estimations do not take into account any temperature dependence, they do not reproduce the \underline{regime III}.

%%%%%%%%%%%%%%%%%%%%%%%%%%%%%%%%%%%%%
\subsubsection{Finite Temperature Effects: DM Evaporation}

Evaporation is a thermal phenomenon where DM gains energy from scattering on neutrons, such that the resulting DM energy is larger than the escape energy. Thus DM is ejected from the NS. In general, this phenomenon is relevant when thermal energy is comparable to the escape energy. To our knowledge, for a Fermi degenerate medium such as in a NS, this effect has never been explicitly computed in the literature so far. The evaporation rate is given by~\cite{Gould:1987ju, Garani:2017jcj}, 
\begin{eqnarray}
\label{eq:evaporate1}
E_{\star} &=& \int_0^{R_\star} n_\chi(r) 4 \pi r^2 \dd r \int_0^{v_{esc}(r)}  f_{\chi}(w,r)\, 4 \pi w^2 \dd w \int_{v_{esc}(r)}^{\infty} R^+ (w \to v) \,  \dd v ~\;,
\end{eqnarray}
in which the expression of $R^+$ can be obtained from the one of $R^-$ in Eq.~(\ref{eq:cap-diff}), by replacing $ f_{p'}(E_p +q_0)$ by  $f_{p'}(E_p -q_0)$.
In Eq.~(\ref{eq:evaporate1}) we assume that the DM phase space density in the NS is time-independent, and can be factorized in two functions: the radial number density $n_\chi(r)$ and the velocity distribution $f_{\chi}(w,r)$ (for one DM particle).
Hence to compute evaporation, one needs a prior on these two distributions. As we will see below, in the case of a NS, DM is actually sensitive to evaporation for masses $m_\chi\lesssim$~keV. We will see in the next section (more specifically the remark regarding Eq.~(\ref{eq:mass_limit_T})) that, for those masses, after the first collision which leads to capture, one can assume DM to have thermalized. Hence, its velocity distribution $f_{\chi}(w,r)$, follows a Maxwell-Boltzmann distribution truncated at the escape velocity $v_{esc}(r)$ given by~\cite{Gould:1987ju,Garani:2017jcj}

\beq
\label{eq:fchi-mb}
f_\chi(w,r) = \frac{1}{\pi^{3/2}}\left(\frac{m_{\chi}}{2 T_\star}\right)^{3/2}\frac{e^{-\frac{m_\chi}{2 T_\star}w^2} \Theta(v_{esc}(r) -w)}{Erf\left(\sqrt{\frac{v^2_e(r)\,m_\chi}{2 T_\star}}\right) - \frac{2}{\sqrt{\pi}}\sqrt{\frac{v^2_{esc}(r)\,m_\chi}{2 T_\star}} \;e^{-m_\chi v_{esc}^2(r)/2T_\star} }~,
\eeq
where we have assumed the star to be isothermal with temperature $T_\star$. In this limit, we can also write the normalized radial distribution of DM~\cite{Gould:1987ju,Garani:2017jcj}:
\beq
\label{eq:nchi-iso}
n_\chi(r) = \frac{e^{-m_\chi \phi(r)/T_\star}}{\int_0^{R_\star} \dd^3r\, e^{-m_\chi \phi(r)/T_\star}}=\frac{4}{r^3_\chi \sqrt{\pi}} \exp{\left( -\frac{r^2}{r_\chi^2}\right)}\;,
\eeq
where in the last equation we introduce the gravitational potential of the star $\phi(r)$ which allows us to define a typical length $r_\chi = \left(\frac{3 T_{\star}}{2 \pi G \rho_\star m_\chi}\right)^{1/2}$, corresponding to the typical extension of the thermalized DM core.\\

\begin{figure}[t]
	\begin{center}
	  \includegraphics[width=0.5\linewidth]{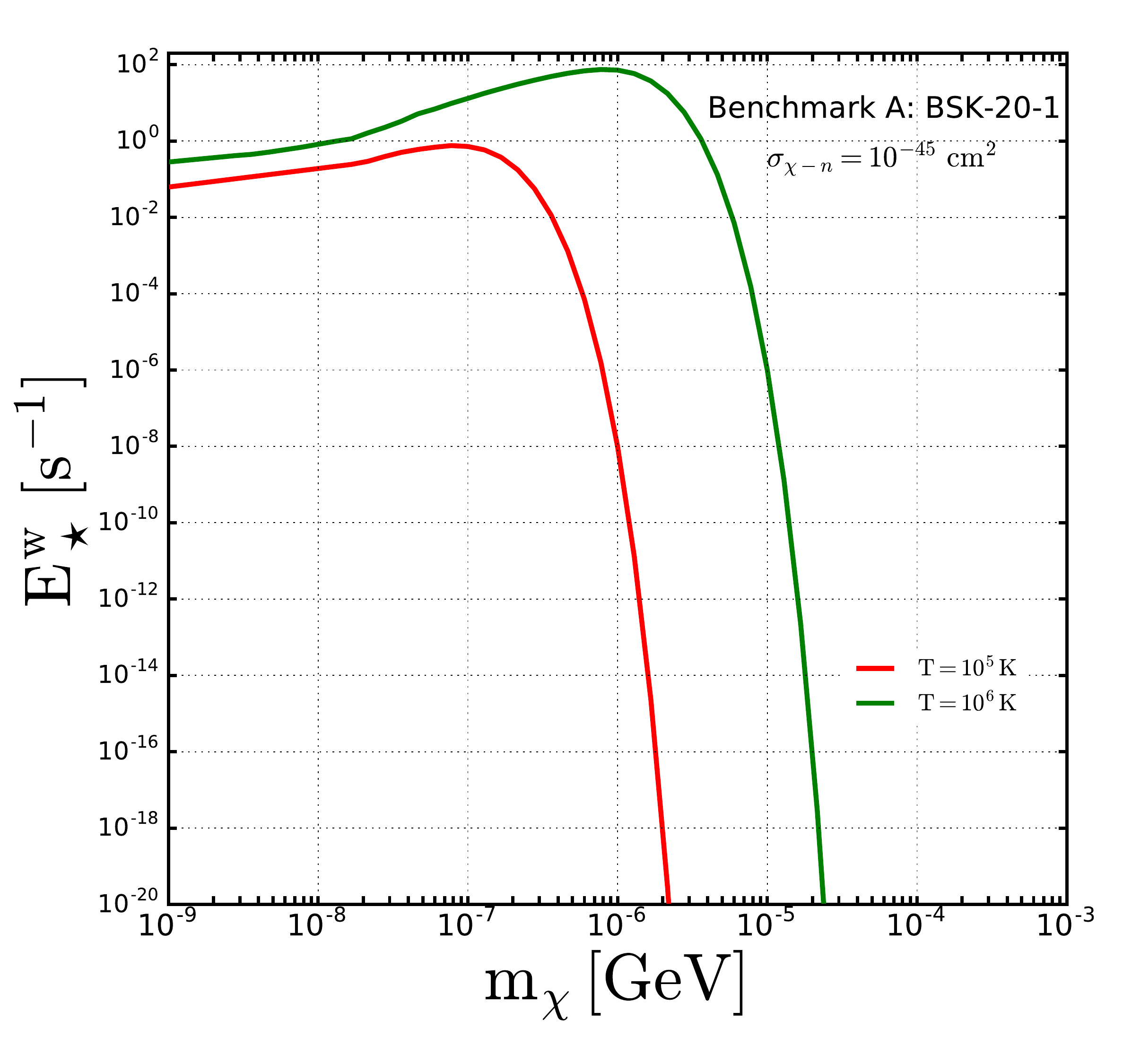}\\
	\end{center}
	\caption{Evaporation rate as a function of DM mass for benchmark Model A for typical NS temperatures.}
	\label{fig:evap-bsk20A}
\end{figure}

With these analytical handles on the DM radial distribution and the velocity distribution inside NS, we can now proceed to compute the evaporation rate. Since it is impossible to obtain a closed analytical expression for $R^+$, we present numerical results obtained for the evaporation rate, for benchmark NS model A in Fig.~\ref{fig:evap-bsk20A} for two different NS temperatures $10^5$ K (10 eV, in red) and $10^6$ K (100 eV,  in green), respectively. As the evaporation rate is $\propto e^{-m_\chi v_{esc}^2/2T_\star}$ we see a clear exponential suppression for DM masses above few T, i.e.~when $m_\chi v^2_{esc}/(2T_\chi)\gtrsim 1$. This feature is only quantitatively different with respect to the case where neutrons are described by a Maxwell-Boltzmann distribution (see e.g. Fig.~3 of~\cite{Garani:2017jcj}). 
Similar to the accretion term $R^-$, the evaporation term $R^+$ is also proportional to the product of the in-coming and out-going neutron distribution functions. However, for evaporation to proceed neutron has to lose momentum such that the final DM velocity exceeds $v_{esc}$. Thus, the overlap is now given by $f_{p}(E_p) (1- f_{p'}(E_p -q_0))$. This implies in practice that only those neutrons whose energy is larger than $\mu_F$ will participate in this process. 

\begin{figure}[t]
	\begin{center}
	  \includegraphics[width=0.5\linewidth]{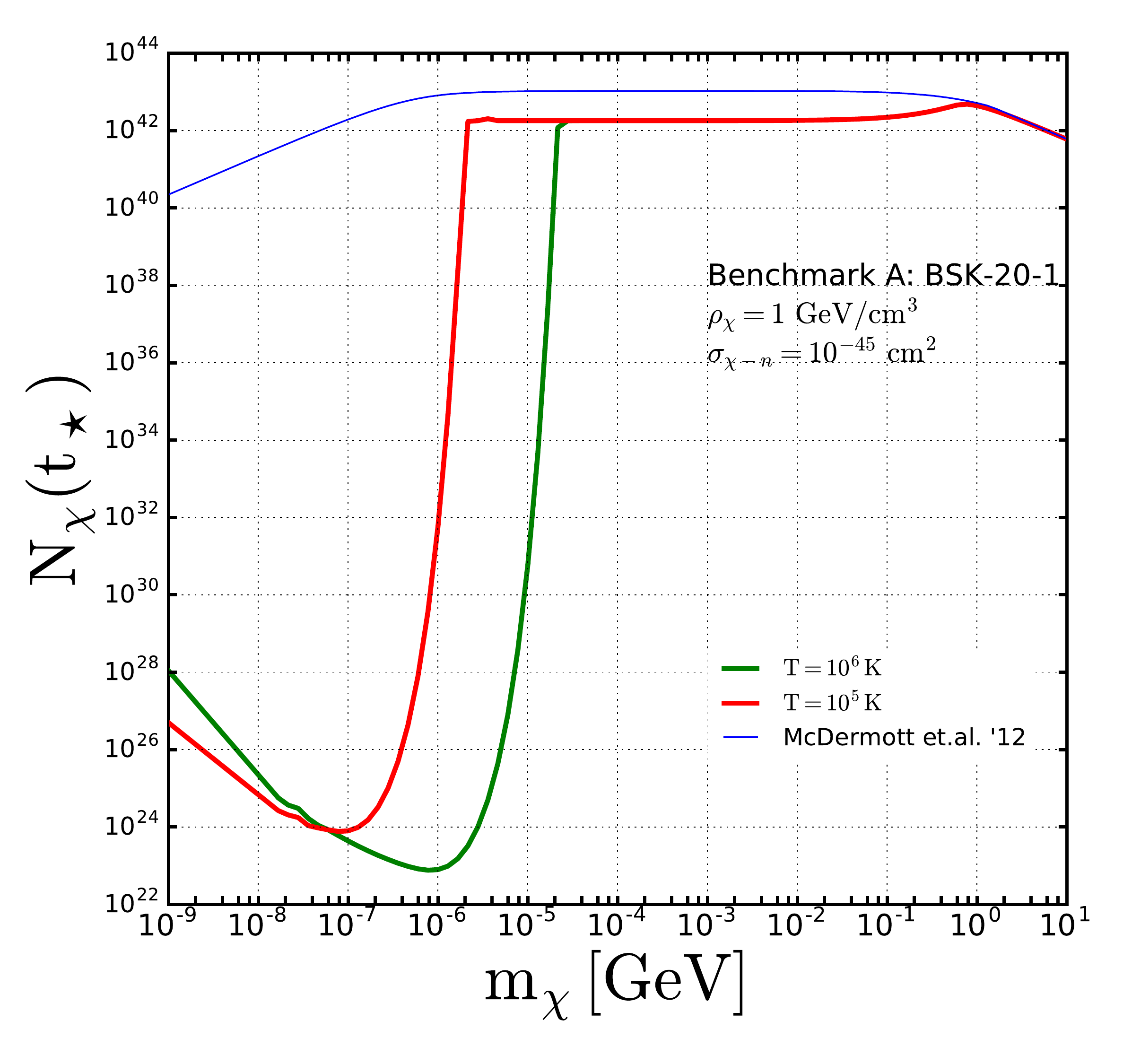}\\
	\end{center}
	\caption{Number of DM particles at $t= t_\star = 10 \,{\rm Gyr}$ as a function of DM mass for benchmark Model A, for DM density $\rho_\chi=1$ GeV/cm$^3$ and $\sigma_{\chi-n}=10^{-45}\rm cm^2$. The green and red lines result from our accurate treatment of the Fermi sea, for two typical NS temperatures, while the blue line is computed from the heuristic argument used in Ref.~\cite{McDermott:2011jp}.}
	\label{fig:nchi-bsk20A}
\end{figure}

%%%%%%%%%%%%%%%%%%%%%%%%%%%%%%%%%
\subsubsection{Putting Accretion and Evaporation Together}

In the ADM scenario the time evolution of number of DM particles is governed by Eq.~(\ref{eq:diffeq-3}), which results in the following solution
\begin{equation}
\label{eq:evolution-asym-sol-r}
N_\chi (t) =  C_{\star}^{\rm w} t_{\star} \left (\frac{1 - e^{- E_{\star}^{\rm w} t_\star}}{E_{\star}^{\rm w} t_\star} \right).
\end{equation}

With all the ingredients (capture and evaporation rate) already computed we now proceed to compute the number of accumulated DM particles at $t=t_\star = 10^{10}$ yr). In Fig.~\ref{fig:nchi-bsk20A} we present the results we get for $N_\chi(t_\star)$ for $T_\star= 10^5$ K (red curve) and for $T_\star= 10^6$ K (green curve). We find that the minimum DM mass for which DM particles are not largely evaporated away (evaporation mass) is $2 \cdot 10^{-6}$ GeV for $T_\star= 10^5$ K and  $2 \cdot 10^{-5}$ GeV for $T_\star= 10^6$ K, respectively. We find a heuristic expression for evaporation mass given by
\beq
\label{eq:evap-mass}
m_{\rm evap} \simeq 2 \cdot 10^{-6} \left(\frac{T_\star}{\rm 10^5\,{\rm K}} \right) {\rm GeV}.
\eeq
Thus the $1/m_{\chi}$ scaling of the accretion rate in \underline{regime III}, see Fig.~\ref{fig:cap-bsk20A}, is completely washed out by evaporation.
Note that, even if $E_\star^{\rm w}$ drops exponentially for $m_\chi$ above a few times $T$ (see Fig.~\ref{fig:evap-bsk20A}), the evaporation effect becomes important as soon as $m_\chi<m_{\rm evap}\sim 2$~keV, that is to say as soon as $m_\chi \lesssim 200\,m_{\rm evap}$.
This is due to the fact that the argument $E_{\star}^{\rm w} t_\star$ of the exponential in Eq.~(\ref{eq:evolution-asym-sol-r}) involves a large value of $t_\star$, and thus is of order one only when $E_\star^{\rm w}$ is as small as $1/t_\star \sim 10^{-17}$~sec$^{-1}$.\\  

%%%%%%%%%%%%%%%%%%%%%%%%%%%%%%%%%%%%%%%%%%%%%%%%%%%%%%
\section{Time evolution of DM density inside the neutron star and thermalization}
\label{sec:thermalization}

As computed previously, after the first interaction with NS matter, DM can be gravitationally trapped. As shown in Eq.~(\ref{eq:evap-mass}), for $m_{DM}\gtrsim $~keV evaporation is negligible and the number of particles collected is simply given by $C_\star^{\rm w} t_\star$.
After the first interaction, part of DM particles have orbits going out of the NS, whereas another (in general much smaller) part have orbits within the NS. Subsequently, DM loses further energy through scatterings, leading to smaller orbits, such that it moves towards the center of the NS. Eventually DM particles have lost so much energy that their energy reaches the neutrons temperature, at which point they thermalize with the neutrons.
All this dynamics is fully relevant for determining when DM could eventually gravitationally collapse to form a black hole.
So far the possibility of black hole formation has been considered only for those cases where DM has thermalized. In this case, to see whether a black hole forms one can just plug in the gravitational collapse condition (see below), the value of the "thermal radius" of the DM core.
The latter is just the radius of the orbit of a particle with energy equal to the temperature.
This makes sense if the characteristic "thermalization time", that a DM particle needs on average to thermalize, is smaller than the NS age. 
This "thermalization  time" is the sum of the time $"t_1$" during which the DM particles stay on orbits larger than the NS, and of the time $"t_2"$ which is subsequently needed for this particle to thermalize.

In this section we go beyond this usual approach in 3 steps:
\begin{itemize}
\item First, since the dynamics towards thermalization is based on the fact that a gravitationally trapped DM particle loses energy from subsequent elastic scattering with neutrons, we start in section~\ref{sec:orbits} by determining the energy loss rate, including Pauli blocking. This has been done in Ref.~\cite{Bertoni:2013bsa}, showing large differences with previous estimates. Although largely based on the formalism of this reference, our calculation here will differ from the fact that to calculate the energy a particle loses when it scatters, we take into account the variation of its energy along its orbital motion, and average over it.
\item Second, we will calculate in section~\ref{sec:estimtherm} the characteristic thermalization time $t_{th}=t_1+t_2$. Here too we will essentially proceed as in Ref.~\cite{Bertoni:2013bsa}, with the difference that we do not sum over the average time each scattering takes (i.e.~assuming that all particles undergo their n$^{th}$ scattering at the same time from the same average energy $E_{n-1}$ to the same final energy $E_n$). Instead, we compute this time directly from integrating over time the energy losses averaged along DM orbits. 

\item Third, after these preliminaries, we compute in section~\ref{sec:profile}, as a function of time, the energy distribution of DM particles. 
This is necessary for a proper treatment of the thermalization process and, to our knowledge, this has never been considered before.
This will allow us to determine at any time what are the numbers of accreted DM particles which have already, or have not yet, thermalized. Equivalently this will also allow us to determine, as a function of time and of the distance from the NS center, the distribution of DM particles in the NS.
Beyond the fact that to determine these distributions as a function of time is interesting in itself (as it is relevant for any phenomenological effect DM could induce), this will allow us to set more precise constraints on when a gravitational collapse would occur. In particular this allows us to determine if there is gravitational collapse also for cases where most of the DM particles would not have enough time to thermalize in the NS.
We will compare this result with the one obtained under the assumption of ``instantaneous'' thermalization of all DM particles after the characteristic ``thermalization time'' defined in section~\ref{sec:estimtherm}.  
\end{itemize}

%%%%%%%%%%%%%%%%%%%%%%%%%%%%%%%%%%%%%%%%%%%%%%%%%%%
\subsection{Average of DM Energy Loss Along Orbits\label{sec:orbits}}

Depending on the energy lost during the first collision, DM particle may have an orbit larger or smaller than the neutron star radius. As already mentioned above, two typical times are relevant: $t_1$, the typical time for a particle orbiting in and out of the NS to be fully contained in the NS, and $t_2$, the typical time for a particle orbiting within the NS radius to reach the thermal energy $E_{th}=3/2 k_B T$. The thermalization time is given by the sum $t_{th}=t_1+t_2$. Obviously these times depend on the rate of energy loss of DM particles. In this subsection we calculate these rates taking into account that a particle scatters at random positions along its orbit, thus with a kinetic energy which varies along this orbit. First we present \textit{(i)} how we parameterize the motion of DM, and secondly \textit{(ii)} how we compute the averaged energy losses along DM orbits.

\textit{(i) DM orbits in the neutron star:} We assume the orbits of particles to be eccentric  and approximate them as linear with radial extension $r_0$. Since particles are gravitationally trapped, there is a one to one correspondence between the orbital extent $r_0$ and the total energy $E_{\rm tot}$ of DM particles. One has
\begin{eqnarray}
	E_{\rm tot} = \begin{cases}- \displaystyle\frac{GM_\star m_\chi }{r_0} & \text{for }  r_0 > R_\star  \;\\
	- \displaystyle\frac{GM_\star m_\chi}{2R_\star} \left(3 - \frac{r_0^2}{R_\star^2}\right) & \text{for }  r_0 \leq R_\star\;. \end{cases}\label{eq:etot}
\end{eqnarray}
Assuming the NS to be a sphere of constant density, the kinetic energy $E_{\rm kin}$ at a time $t$ can be expressed as a function of $r(t)$. It follows that
\begin{eqnarray}
	E_{\rm kin} = \begin{cases} E_{\rm tot} + \displaystyle\frac{ G M_\star m_\chi }{r} & \text{for }  r > R_\star  \;,\\
	E_{\rm tot} + \displaystyle\frac{GM_\star m_\chi}{2R_\star} \left(3 - \frac{r^2}{R_\star^2}\right)		 & \text{for }  r \leq R_\star\,. \end{cases}\label{eq:ekin}
\end{eqnarray}
In the following, to characterize an orbit we use the maximal kinetic energy that we denote by $E$ (which is also the kinetic energy of particles crossing the center of the star),
\begin{eqnarray}
	E\equiv E_{\rm kin}^{\rm max} = \begin{cases} -\displaystyle\frac{GM_\star m_\chi }{r_0} +3\, \displaystyle\frac{GM_\star m_\chi}{2R_\star}  & \text{for }  r_0 > R_\star  \;,\\
				\displaystyle\frac{GM_\star m_\chi}{2R_\star} \left(\frac{r_0^2}{R_\star^2}\right)				 & \text{for }  r_0 \leq R_\star\,. \end{cases}\label{eq:ekin}
\end{eqnarray}
 
\textit{(ii) Averaged energy losses along DM orbit:} To compute the energy losses we start from the differential scattering rate defined in Eq.~(\ref{eq:rate-cm1}) which encodes the number of collisions a DM particle with velocity $w$ undergoes per unit time and per unit interval of out-going velocity, $v$. In this rate the velocities can be traded for the in-coming and out-going DM kinetic energies, $E_k$ and $E'_k$, leading to the following differential scattering rate in energy
\beq
\frac{d\Gamma}{d E_k'}=\sqrt{2 m_\chi E_k'}\left( R^-(E_k \rightarrow E_k')+R^+(E_k \rightarrow E_k') \right) \label{eq:diff_enl_our} \;.
\eeq
For simplicity, instead of using the numerical result we obtained in the previous section for this quantity, we will use the simpler analytic approximation obtained in Eq.~(22) of Ref.~\cite{Bertoni:2013bsa}, 
\beq
\frac{d\Gamma}{d E_k'}=\sigma_\chi \frac{ m_n^2 m_\chi}{2 \pi^2 m_r^2}\sqrt{\frac{E_k'}{E_k}}\;(E_k-E_k').\label{eq:diff_enl_berto} 
\eeq
We have checked the numerical compatibility of both results for values of $E_k$ above the thermal one, $E_{th}$. For those energies and for typical NS temperatures, the backreaction of neutrons encoded in the second term of the right-hand side of Eq.~(\ref{eq:diff_enl_our}) (i.e.~$R^+$) turns out to be negligible as long as $E>E_{th}$, and to very quickly become comparable to the $R^-$ term as soon as the energy becomes very close to $E_{th}$.
 
Thus, in the following we consider that as long as the DM particle has an energy above the thermal one, we can safely neglect the back reaction $R^+$ term, whereas as soon as it becomes smaller it follows a Maxwell-Boltzmann distribution, as given in Eq.~(\ref{eq:nchi-iso}).
Since we are interested in cross-section orders of magnitude smaller than the geometric cross-section ($\sim 10^{-45}$ cm$^2$), the successive scatterings are spread randomly along the orbits. Thus, for orbits with $r_0<R_\star$, the differential scattering rate for a DM particle moving along an initial orbit with maximum radius $r_0$ (corresponding to maximum kinetic energy $E$), going to an orbit with maximum radius $r'_0$ (corresponding to maximum kinetic energy $E'$), averaged over the initial orbit, is 
\bea
\frac{d{\Gamma_2}}{d E'}&=& \left\langle \displaystyle\frac{d{\Gamma}}{d E_k'} \right\rangle_{r_0} = \displaystyle\frac{\displaystyle\int_0^{\tau_0} \displaystyle\frac{d{\Gamma}}{d E_k'}[E_k(r_0,r(t))\rightarrow E_k'(r'_0,r(t)) ]\;dt}{\displaystyle\int_0^{\tau_0} dt}\\
&=&\sigma_\chi\frac{m_n^2 m_\chi}{2 \pi^2 m_r^2} \left(1-\sqrt{1-\displaystyle\frac{E'}{E}}\right)(E-E')\;.\label{eq:diff_el_averaged}
\ena 
Where $\tau_0$ is the period of the initial orbit. In this equation $E_k(r_0,r(t))$ and $E'_k(r'_0,r(t))$ are the kinetic energies of the in-coming and out-going DM particles when they lie at a distance $r(t)$ from the NS center along the orbit of extent $r_0$ and $r'_0$, respectively.
From this expression one can directly compute the average energy lost by a DM particle of energy $E$ per unit time:
\bea
b_2(E) &=& \int_0^E \frac{d{\Gamma_2}}{d E'} (E-E') d E' \,,\\
&=&\frac{\sigma_\chi }{42 \pi^2}\frac{m_n^2 m_\chi}{m_r^2}\; E^3\;.\label{eq:el_averaged2}
\ena 
Recall that the latter is valid for an initial orbit inside the NS, $r_0<R_\star$. For $r_0>R_\star$, similar average of $d\Gamma_1/d E'$ leads to:
\bea
b_1(E) &=& \int_0^E \frac{d{\Gamma_1}}{d E'} (E-E') d E' \nonumber \\
&=&\displaystyle\frac{2\sigma_\chi}{105 \pi^3}\frac{m_n^2 m_\chi}{m_r^2}\; E_\star^3 \times \frac{\eta(E)}{B^6} \times I(E)\;,\label{eq:el_averaged1}
\ena
with $\eta(E)$ the fraction of the period for which the DM is traveling inside the star, and $I(E)$ a non trivial function of energy (see Appendix \ref{app:fct_eq} for details). 

\begin{figure}[htb]
	\begin{center}
	  \includegraphics[width=0.49\linewidth]{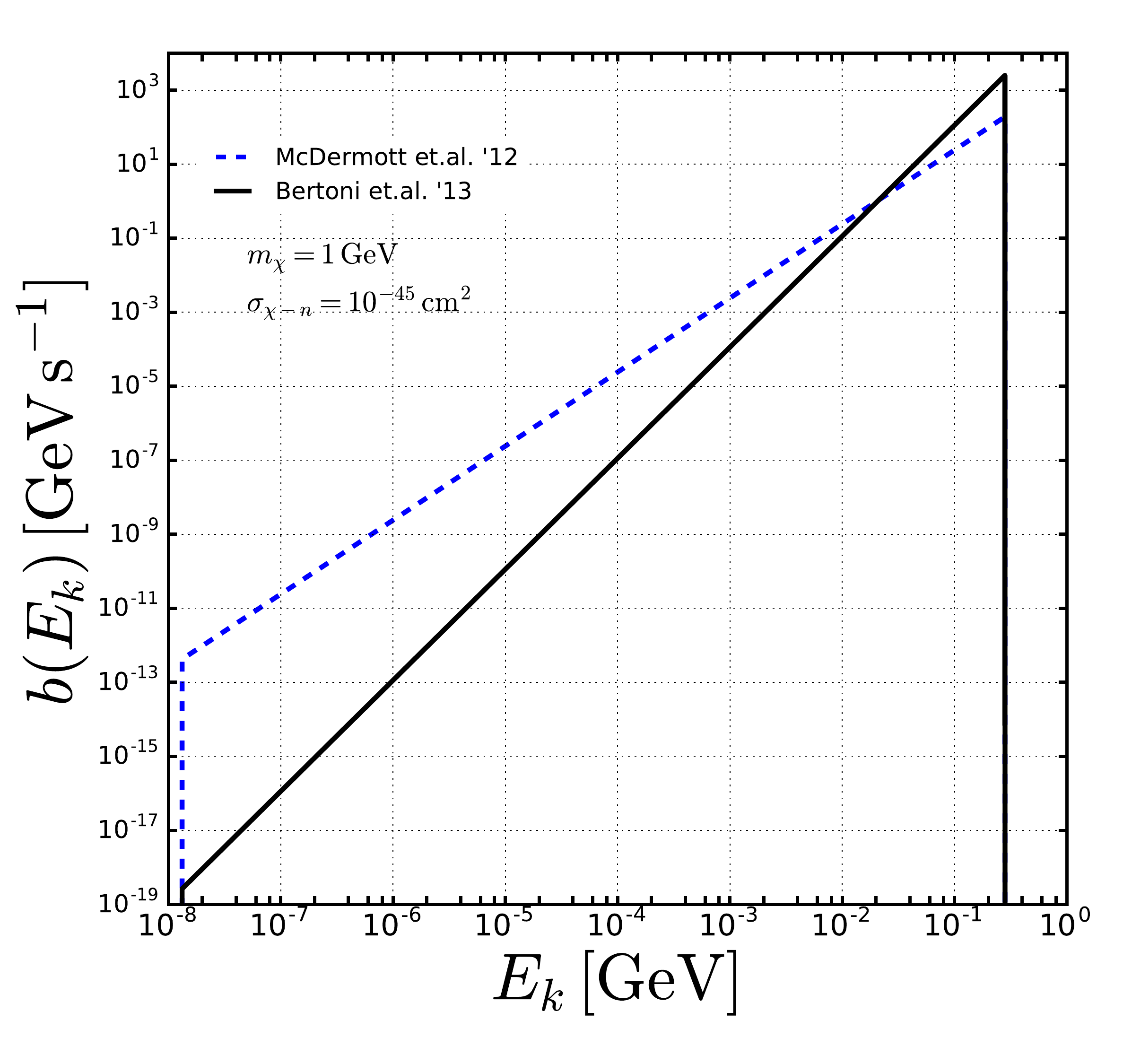}
	  \includegraphics[width=0.49\linewidth]{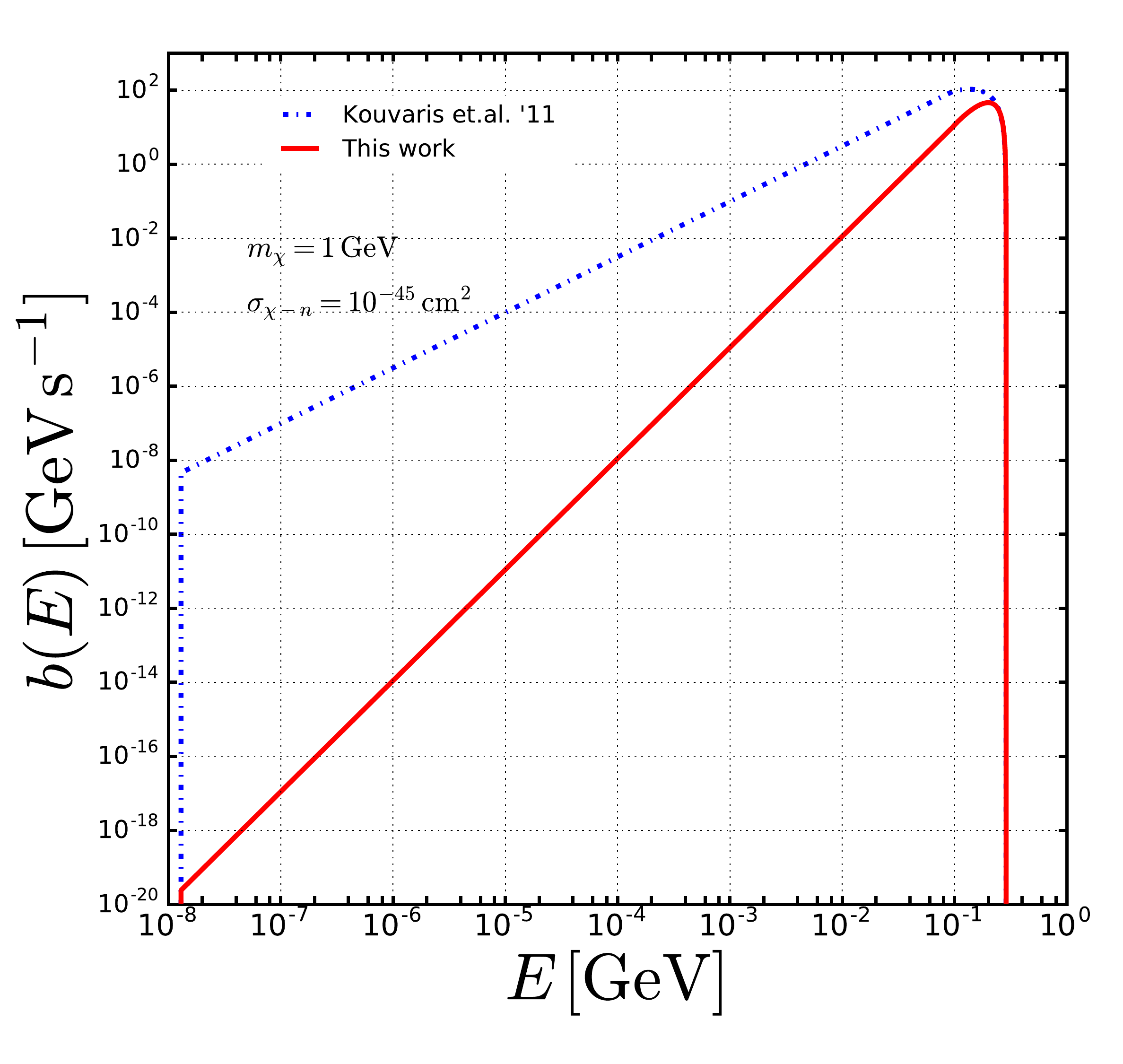}\\
	\end{center}
	\caption{Energy losses as a function of DM energy without (left) and with (right) averaging over orbits, for $m_\chi=1$~GeV and $\sigma_\chi=10^{-45}\rm cm^2$. The solid red line on the right panel is from Eqs.~(\ref{eq:el_averaged2}) and~(\ref{eq:el_averaged1}), and corresponds to adding the orbit average effect to the result of Ref.~\cite{Bertoni:2013bsa}, given by the black solid line on the left panel. The dashed blue line on the left panel is from Ref.~\cite{McDermott:2011jp}. Finally the dotted-dashed line on the right panel is from Ref.~\cite{Kouvaris:2010jy}. }
	\label{fig:energy_losses}
\end{figure}

On the left panel of Fig.~\ref{fig:energy_losses} we show the energy loss rate, $b(E)$, from Refs.~\cite{McDermott:2011jp} and~\cite{Bertoni:2013bsa}. Since these results are not averaged over orbits, we plot this rate as a function of kinetic energy $E_k$ which, in this case, monotonously decreases with time.
The energy scale is bounded from below by the thermal energy and from above by the kinetic energy corresponding to the escape velocity, $E_{esc}$. On the right panel of Fig.~\ref{fig:energy_losses}, we plot the energy loss rate we get averaging over the orbits, Eqs.~(\ref{eq:el_averaged2}) and~(\ref{eq:el_averaged1}). Not surprisingly, the more energy the DM particle has already lost, the slower it will subsequently lose energy, see in particular Eq.~(\ref{eq:el_averaged2}), which applies when the DM particle has an orbit already fully contained inside the NS. Note however, that for energies just below $E_{esc}$, the rate of energy loss increases as the energy decreases, leading to a maximum in the energy loss rate. This is due to the fact that, when $E$ becomes smaller than $E_{esc}$, the energy losses are sharply increasing as the fraction of the orbital period spent in the NS increases.
Eventually DM reaches the thermal energy, at which point the energy loss rate is set to zero. Note that, comparing the result of Ref.~\cite{Bertoni:2013bsa}
on the left panel to the one from Eqs.~(\ref{eq:el_averaged2}) and~\ref{eq:el_averaged1} on the right panel, one observes that
the typical effect of averaging over the orbits is to reduce the energy loss rate by a factor of 4, except for energies slightly below $E_{esc}$, where the effect is much larger.
Also shown on the right panel is the result of Ref.~\cite{Kouvaris:2010jy} which did an orbit average but did not consider the effect of Fermi sea suppression. A comparison of both lines in this panel shows that the Fermi sea effect is not only important in the accretion process but also in the thermalization process. This can be also seen partly from a comparison with the results of Ref.~\cite{McDermott:2011jp} in the left panel.

In the following we will also make use of the total interaction rate per unit time, averaged over the orbits:
\beq
\Gamma(E)=\int_0^{E} \frac{d{\Gamma}}{d E'} dE'\;.
\label{totalenergylossrate}
\eeq

%%%%%%%%%%%%%%%%%%%%%%%%%%%%%%%%%%%%%%%%%%%%%%%%%%%
\subsection{Estimate of the Thermalization Time\label{sec:estimtherm}}

Before considering the thermalization process more rigorously in the next subsection (directly from the evolution of the DM energy distribution), we begin by estimating the time $t=t_1+t_2$ it takes for DM to thermalize with the neutrons in the NS. To this end, we simply integrate over time the orbit averaged energy losses we obtained in the previous section.
The time $t_1$ can be obtained from the energy loss $b_1$, Eq.~(\ref{eq:el_averaged1}), by computing the time it takes for a DM with initial energy $E_0$ (just after the first collision) to reach the NS surface energy $E_{\rm surf}$ corresponding to $r_0=R_*$. Similarly, $t_2$ can be estimated from $b_2$, computing the time it takes for DM at $E_{\rm surf}$ to reach the thermal energy $E_{th}$. Hence,
\bea
t_1 &=& \int_{E_{0}}^{E_{\rm surf}}\frac{dE}{b_1(E)}\,,\\
t_2 &=& \int_{E_{\rm surf}}^{E_{\rm th}}\frac{dE}{b_2(E)}\;.
\ena 
The average energy $E_0$ that a DM particle has just after being gravitationally trapped can be estimated from the initial average kinetic energy $E_i$. From the average energy lost per collision,
\bea
\left\langle \Delta E \right\rangle = E_i-E_0 = \frac{\int_0^k d\Gamma(E_i) \,(E_i-E_f)}{\int_0^k d\Gamma(E_i)} = \frac{4}{7}E_i\,,\label{eq:delta_E}
\ena
we get $E_0 = 3/7\cdot E_i \approx 3/7 \cdot E_{esc}$, where $E_{esc}=\frac{3}{2}\frac{G m_\chi M_\star}{R_\star}$ is the escape kinetic energy for a particle at the center of the star.
Note that, if, after the first scattering, $E_{0}$ falls below the energy $E_{\rm surf}=\frac{1}{2}\frac{G m_\chi M_\star}{R_\star}$ corresponding to particles orbiting till the surface, then $t_1$ is irrelevant because the orbit will be enclosed inside the NS. But this does not happen on average since the average ratio is $E_0/E_{surf}=9/7$. However it can happen that 
the orbit size at thermal energy is larger than the NS size. 
This happens for DM masses when $E_{\rm surf}<E_{th}$, namely:
\beq
m_\chi<\frac{3R_\star T}{G M_\star}=1.33 \times 10^{-7} ~{\rm GeV}\left( \frac{T}{10^{5}\rm K}  \right) \;\label{eq:mass_limit_T} .
\eeq 
In the worst scenario for capture, DM can be so light that the thermal energy is actually larger than the escape energy. In the following we will not consider such small masses since they lead to very diffuse DM halos around the neutron star, which are not suitable for black hole formation. These considerations naturally lead to the notion of thermal radius $R_{th}$ corresponding to the typical size of the core obtained by equating gravitational energy, $-4\pi GR_{th}^2\rho_B m_\chi/3$, with the thermal one, $\frac{3}{2} k_B T$,
\bea
\label{eq:r-therm}
R_{th}=\sqrt{\frac{9T}{4\pi G \rho_B m_\chi}}=4.29~{\rm m } \left( \frac{T}{10^{5}\, \rm K}  \right)^{1/2}\left( \frac{1 \,\rm GeV}{ m_\chi}  \right)^{1/2} \;.
\ena
Note that $R_{th}$ defined in this way, differs from the scale height $r_\chi$  of the Maxwellian distribution, Eq.~(\ref{eq:nchi-iso}), by a factor $\sqrt{3/2}$, $R_{th}=\sqrt{3/2} \,r_\chi$. 

Coming back to the estimation of $t_1$ and $t_2$, while the first can only be computed numerically, $t_2$ can be approximated as:
\begin{eqnarray}
	t_2\approx \frac{21 \pi^2 m_r^2}{\sigma_{\chi}  m_n^2 m_\chi}\frac{1}{E_{th}^2}\approx 10700 ~{\rm yrs}\;\frac{\gamma}{(1+\gamma)^2} \left( \frac{10^{5}\,\rm K}{T}  \right)^{2} \left( \frac{10^{-45}\rm cm^{2}}{\sigma_\chi}  \right)  \;,\label{eq:t2}
\end{eqnarray}
with $\gamma=m_\chi/m_n$. Note that in Refs.~\cite{McDermott:2011jp} and~\cite{Kouvaris:2010jy}, the $m_\chi$ and $E_{th}$ dependence of $t_2$ are not the same as in Eq.~(\ref{eq:t2}), since they do not include a proper treatment of Pauli blocking. This leads to drastically different results. For example, for $m_\chi=1$~GeV, we obtain  values 4$\times 10^6$ and 1.5$\times 10^{10}$ times smaller than in Refs.~\cite{McDermott:2011jp} and~\cite{Kouvaris:2010jy}, respectively.  Computing $t_1$, it appears that it is always much smaller than $t_2$~\cite{Kouvaris:2010jy}, and account for less than 1\% of the time prior to thermalization for DM masses above $10^{-6}$ GeV. Thus, in the following we neglect this first step of the thermalization process, and approximate $t_{th} \approx t_2$.

%%%%%%%%%%%%%%%%%%%%%%%%%%%%%%%%%%%%%%%%%%%%  
\subsection{Evolution of the DM Energy Distribution Prior to Thermal Equilibrium\label{sec:profile}}  

The estimation of thermalization time ($t_1$ and $t_2$) basically means  that, for times below $t_{th}$, one assumes that none of the DM particles have reached $E_{th}$. This would be strictly true if the captured DM particles were all losing energy in the same continuous way. 
However, energy losses are not continuous but catastrophic: the jumps in energy are significant compared to the total energy a DM particle must lose to reach $E_{th}$. For example, according to Eq.~(\ref{eq:delta_E}), 1~GeV DM particle needs only 10 scatterings on average for its energy to go below $E_{th}$. Even smaller numbers of scatterings hold for lighter DM particles. Furthermore, the size of the energy jumps is stochastic, thus the final DM energy after $n$ number of scatterings is a random variable. 
Hence one should consider the evolution of DM energy distribution. Since we are dealing with large numbers of particles, to compute and to use this distribution makes sense. In this section we will compute this DM energy distribution as a function of time. We will perform this computation from the orbit averaged energy loss rates computed above. This will allow us to determine at any given time the number of DM particles which are in and out-of thermal equilibrium, as well as the spatial distribution of the DM particles. These are of importance to determine when a black hole could form.\\

To this aim we introduce the function which gives the number of DM particles that have an orbit with maximal kinetic energy $E$, per unit interval of $E$, $f_\chi(E,t)=\displaystyle\frac{d \,N_{\chi}(E,t)}{d \, E} $. This distribution is time dependent and follows the Boltzmann equation, 
\begin{eqnarray}
	\frac{\partial f_\chi}{\partial t}(E,t)= \int_{E}^{+\infty} dE' \frac{d \Gamma}{d E'} (E' \rightarrow E) f_\chi(E',t) - \Gamma(E)f_\chi(E,t) + q(E,t)\;,\label{eq:redistrib}
\end{eqnarray}
where $\Gamma(E)$ is the total energy loss rate defined in Eq.~(\ref{totalenergylossrate}). The source term $q(E,t)$, i.e.~the
number of particle accreted per unit time and per unit interval of energy $E$, is given by
\begin{align}
	q_\chi(E)&=  \int_0^{R_\star} 4 \pi r^2 \dd r \int_0^\infty \dd u_\chi \, \left(\frac{\rho_\chi}{\mx}\right) \, \frac{f_{v_\star}(u_\chi)}{u_\chi} \, w(r) \; \int d\, E_k'\;  \delta(E_k'-E_k'(E,r))  \sqrt{2 m_\chi E_k'} R_i^- (E_k \to E_k') \,, 
\end{align}
with $E_k'(E,r)$ the kinetic energy at position $r$ of a particle with a given $E$. This integral
is similar to the one given above for the capture rate, Eq.~(\ref{eq:capture1}), except that now we do not only count the number of particles that gets trapped but also keep track of their energies.
This equation can be solved semi-analytically by using a discrete grid in energy (see Appendix~\ref{app:num_sol} for details). In the following we adopt two simple approximations already discussed above: (i) first we only consider the evolution of the DM distribution inside the NS. This approximation is very good since we have seen above that $t_1$ is perfectly negligible in the DM mass range considered. Hence $\frac{d \Gamma}{d E'}=\frac{d \Gamma_2}{d E'}$ and $\Gamma=\Gamma_2$. (ii) Secondly, we assume that there is no neutron ``feedback'' as long as $E>E_{th}$ (see discussion above Eq.~(\ref{eq:diff_el_averaged})) and that all particles whose energy is less than $E_{th}$ are in thermal equilibrium with the neutrons, i.e.~follows a Maxwell distribution as given in Eq.~(\ref{eq:nchi-iso}).
Furthermore, as already mentioned above too, the strong energy dependence of the energy losses, Eq.~(\ref{eq:diff_enl_our}), implies that the larger the energy of DM particle, the faster it loses energy. Hence the shape of the source term $q(E)$ has very little impact on the solution $f_\chi(E,t)$. We have checked that, for values of cross sections such as the ones probed by neutron stars and for typical NS ages, the extreme source terms,
\beq 
q(E)=C_\star^{\rm w}\delta(E-E_{surf}) \quad \text{  and  } \quad q(E)=\frac{C_\star^{\rm w}}{E_{surf}-E_{th}}\theta(E-E_{th})\theta(E_{surf}-E)\;,
\label{source}
\eeq
give the same results for $f_\chi(E,t)$ (at the permille level) as soon as $E$ is below $E_{surf}$ by a small fraction. In the following we use the latter accretion term for convenience. 

\begin{figure}[htb]
	\begin{center}
	  	  \includegraphics[width=0.49\linewidth]{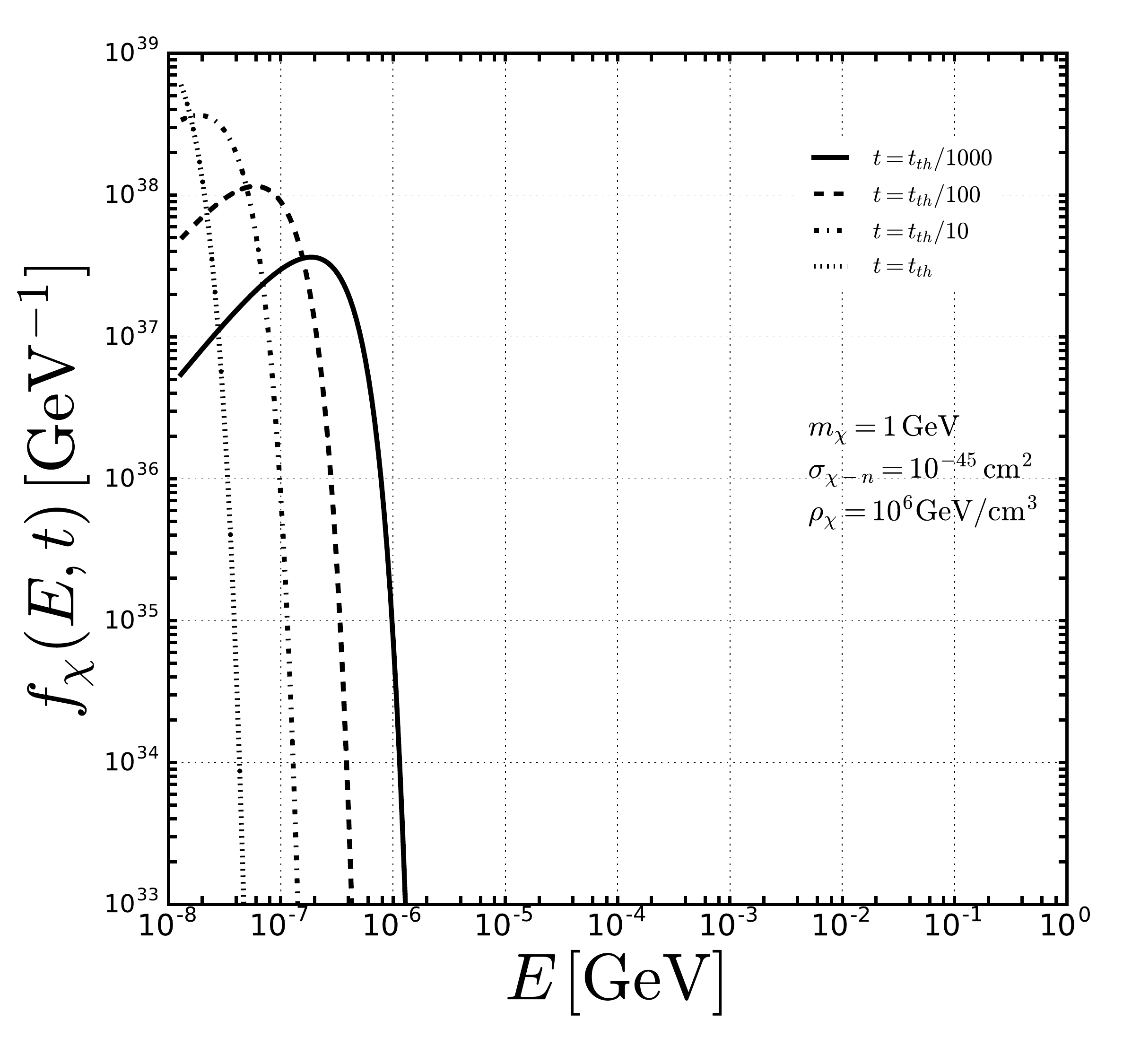}
		  \includegraphics[width=0.49\linewidth]{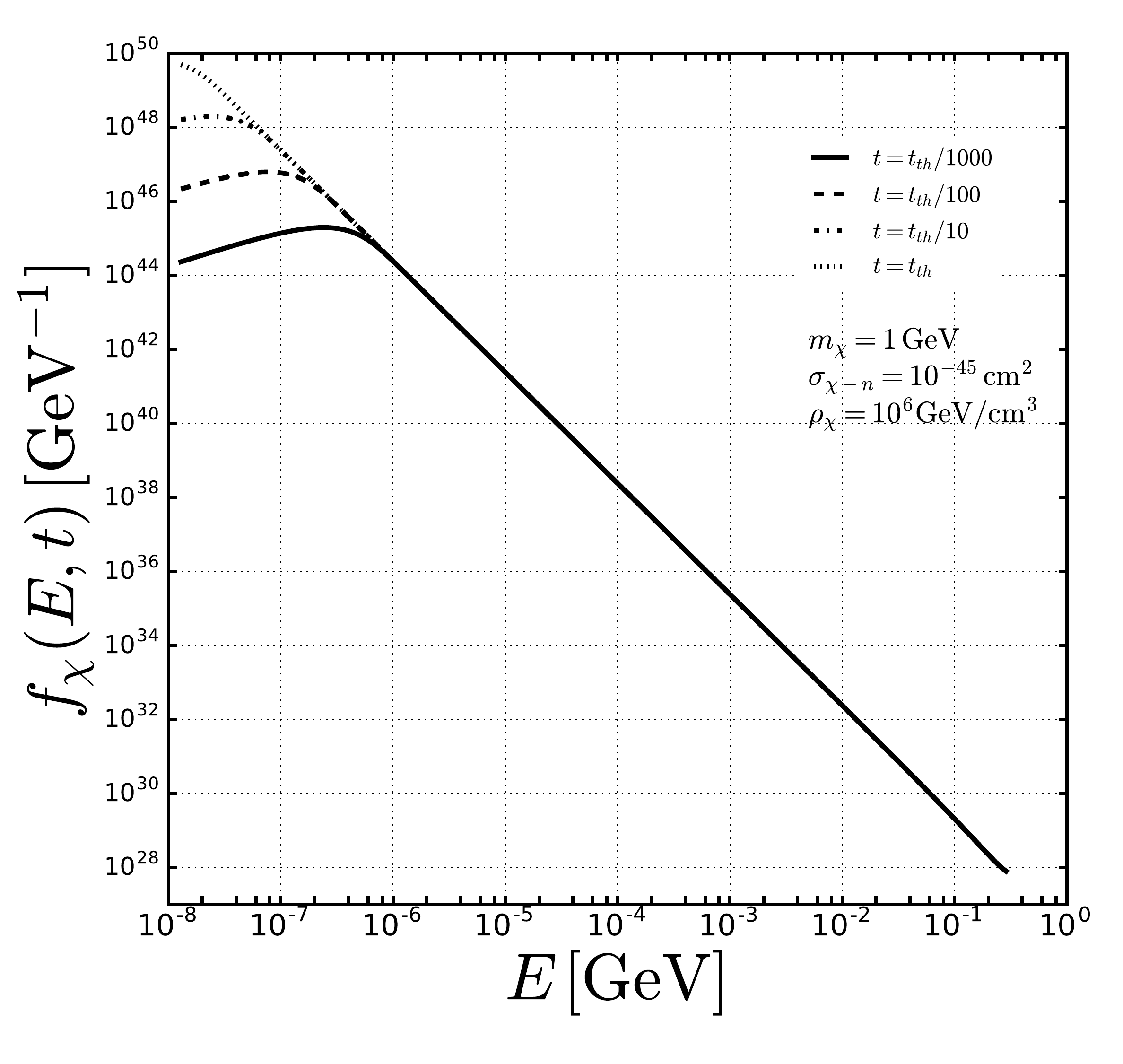}
	\end{center}
	\caption{\textit{Left:} The time evolution of DM energy distribution for a punctual source term  $q(E)\delta(t)$ for several times $t$, fractions of the thermalization time $t_{th}$.  \textit{Right:}  Same as the left panel, but using a time-independent source term $q(E)$.}    
	\label{fig:evolution_f_E}
\end{figure}

Before looking at what gives the source terms of Eq.~(\ref{source}), let us consider the case where all the DM particles would have been trapped at the same given time, i.e.~using a punctual source term in time $ q(E)\delta(t)$ with $q(E)$ any of the source terms in Eq.~(\ref{source}). This case is interesting because it gives the distribution $f_\chi(E,t)$ which, up to a normalization factor is proportional to the number of DM particles. This is nothing but the probability that a single particle trapped at $t=0$ with energy $E=E_{surf}$, ends up with an energy $E$ at time $t$.
The results for this case are shown in the left panel of Fig.~\ref{fig:evolution_f_E}, where we show the evolution of $f_\chi(E,t)$ as a function of time, using the solution of Eq.~(\ref{eq:redistrib}) given in App.~\ref{app:num_sol}.
Because of the large energy loss rate for large energies, the high energy tail of $f_\chi(E,t)$ at a given time is very steep and the time spent at those energies is short. For a given time $t$, the distribution peaks around the energy $\sim E^*$, given by the equation
\begin{eqnarray}
	t = \int^{E^*}_{E_{surf}} \,\frac{d E'}{b_2(E')}\;. 
\end{eqnarray}
Note that for $E^*=E_{th}$, $t=t_2$ as we expect. Below $E^*$, the tail of the distribution behaves as given by Eq.~(\ref{eq:diff_el_averaged}), proportional to $E'$ for small values of $E'$. As times goes on, the distribution is peaked at energies closer and closer to the thermal energy. 

\begin{figure}[htb]
	\begin{center}
	\includegraphics[width=0.49\linewidth]{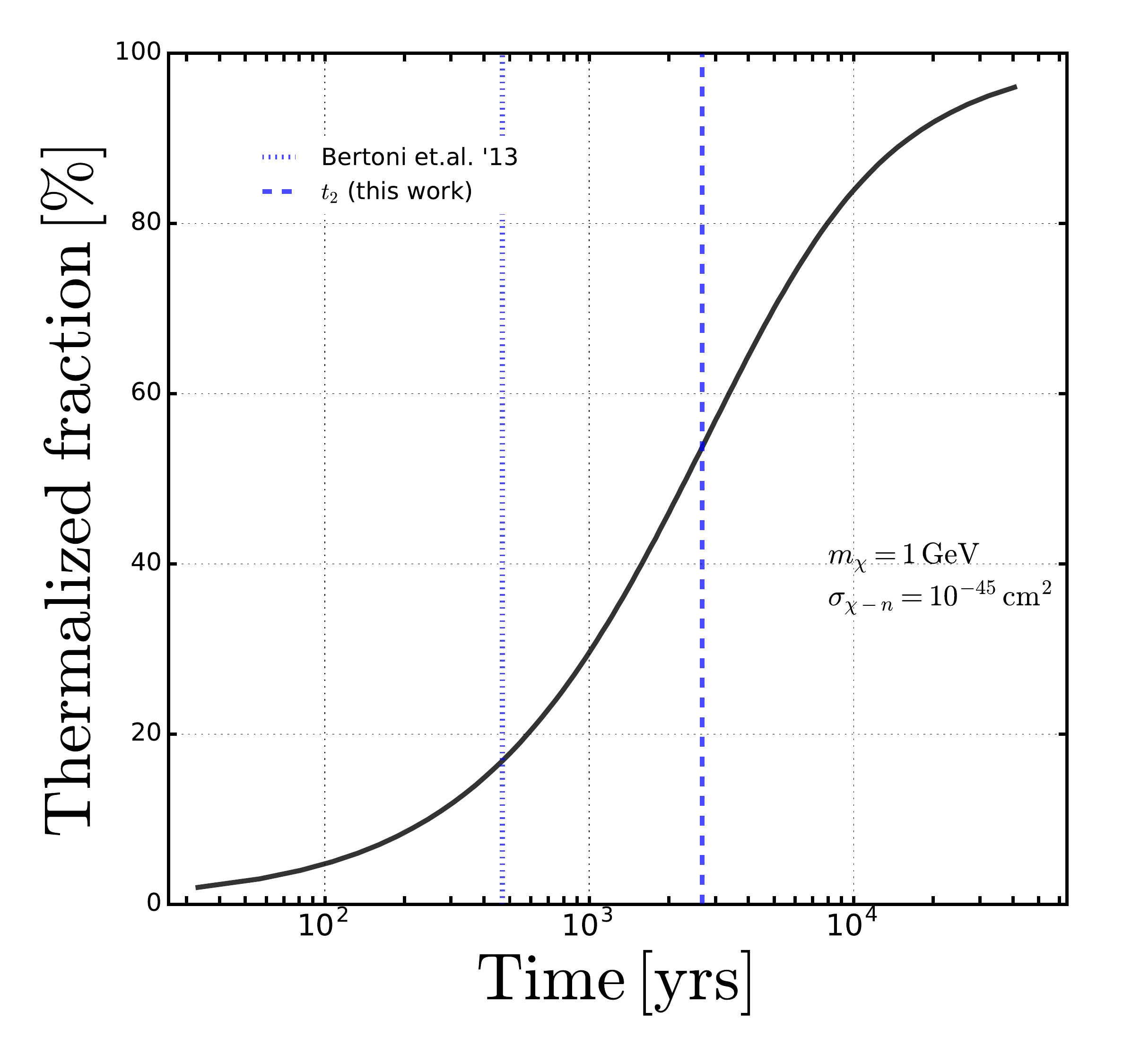}
	  \includegraphics[width=0.49\linewidth]{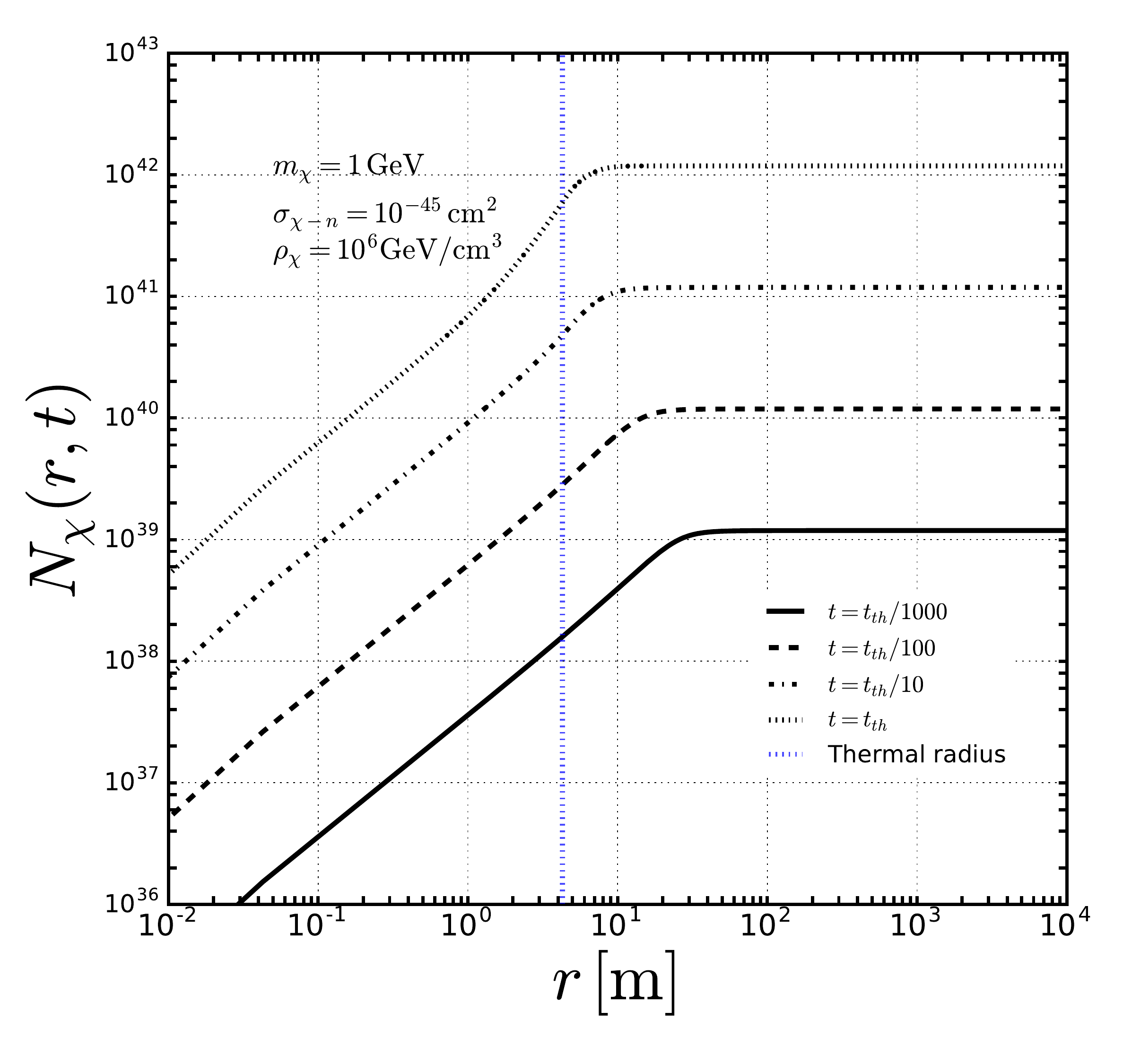}
	\end{center}
	\caption{\textit{Left:} Thermalized fraction $f=N^{th}_\chi/N^{tot}_\chi$ as a function of time for 1 GeV DM mass and $\sigma_{\chi-n}=10^{-45}\, \rm cm^2$. The vertical lines correspond to the times $t_2$ defined Eq.~(\ref{eq:t2}) (dashed line), and the discrete estimate from Ref.\cite{Bertoni:2013bsa} (dotted line). \textit{Right:} Evolution of number of DM particles within a sphere of radius $r$. The several times '$t$' considered are fractions of the thermalization time $t_{th}$.}   
	\label{fig:evolution_N_DM}
\end{figure}

On the right panel of Fig.~\ref{fig:evolution_f_E} we show the evolution of $f_\chi(E,t)$ using the continuous source term of Eq.~(\ref{source}). We observe the same behavior at low energy: $f_\chi(E,t)$ is proportional to $E$ and peaks at $E^*$. Above $E^*$ the distribution decreases as $C_\star^{\rm w}/b_2(E)\propto E^{-3}$, the time independent solution of equation Eq.~(\ref{eq:redistrib}).  
From this distribution, one can deduce the number of particles which have thermalized as the difference between the total number of particles accreted $N^{tot}_\chi$ and the number of those which are still cooling down:
\beq
\label{eq:ntherm}
N^{th}_\chi(t)=C_\star^{\rm w}\times t-\int_{E_{th}}^{E_{surf}} f(E',t) dE'.
\eeq
The left panel of Fig.~\ref{fig:evolution_N_DM} shows as a function of time, the fraction $f=N^{th}_\chi/N^{tot}_\chi$ of particles which have thermalized with respect to the total number of particles which have been gravitationally trapped.
This fraction is given for a cross section $\sigma_\chi^0=10^{-45}\rm cm^2$ and for $m^0_{\chi}= 1$~GeV. Note that the result can be obtained from any other values of these 2 parameters by simply rescaling the time axis by the factor $\alpha=\frac{\sigma_\chi^0}{\sigma_\chi}\frac{m_\chi}{m_\chi^0}\frac{(1+\gamma_0)^2}{(1+\gamma)^2}$, with $\gamma=m_\chi/m_n$ and $\gamma_0=m_\chi^0/m_n$. 
Obviously if one waits long enough this fraction tends to unity. Thus, depending on the fraction of DM particles one requires to consider that DM has thermalized, the thermalization time varies. For example, the time required to get a thermalized fraction of 90\% is 8 times larger than to get a thermalized fraction of 50\%.

For comparison, in Fig.~\ref{fig:evolution_N_DM} we highlight with vertical lines the typical thermalization time $t_{th}=t_2$ we got above in Eq.~(\ref{eq:t2}), and the thermalization time obtained from a discrete estimate of Ref.\cite{Bertoni:2013bsa}. These times differ by a factor 6 and the corresponding thermalized fraction $f$ obtained for these times are 54\% and 17\%, respectively.\footnote{Thus thermalization at the 90\% level requires about 38 times more time than to get a thermalization at the 17\% level.}
 
Conversely, from the rescaling relation above, one can compute for a given time $t$ the corresponding cross-section $\sigma_{\chi-n}$ required to yield a fraction $f$ of thermalized DM particles. The results for $t=10^{10}$ yrs are shown as a function of DM mass in Fig.~\ref{fig:cap-exclusion-neutrons}. The solid, dotted, dotted-dashed green lines correspond  to values of $f$ of 90\%, 70\% and 50\%, respectively. The areas under each of these lines correspond to the region of the parameter space where less than $f$\% of the DM particles have actually thermalized. The area for $f=90\%$ is shaded in green. In the same figure, we display for comparison the line defined in~\cite{Bertoni:2013bsa} as the frontier between thermalized and not thermalized DM. This gives a cross section a factor of 6 and 38 smaller than the ones we get for 50\% and 90\% thermalized fraction. Below we will see what are the fractions of thermalized DM which are relevant for deriving the bounds on the parameters from the requirement of having no black hole formation. These percentages can be quite low in some cases.

Note that to determine thermalization times corresponding to given fractions of thermalized DM, as we just did, is instructive to give an idea of the time scale involved, but is useless for all practical purposes. The fundamental physical quantity from which all relevant constraints below can be obtained is the energy profile $f_\chi(E,t)$ or equivalently the DM density profile $\rho_\chi(r,t)$ as a function of the distance from the center of the NS. The equivalence between both profiles stems from the one-to-one correspondence between the kinetic energy that the DM particle has when it crosses the center of the NS and the orbit it follows (as long as we make the approximation of linear orbits as we do in this work).

To get $\rho_\chi(r,t)$ from $f_\chi(E,t)$ we first use the correspondence between $E$ and $r_0$, the radius of DM orbit, see Eq.~(\ref{eq:ekin}). There is a one to one correspondence between the energy distribution $f_\chi(E,t)$ and orbit-size distribution of the DM particles, which we denote as $g_\chi(r_0,t)$,
\begin{eqnarray}
	f_\chi(E,t)=\frac{d \,N_{\chi}(E,t)}{d \, E} \; \Leftrightarrow \; g_\chi(r_0,t)=\frac{d \,N_{\chi}(r_0,t)}{d \, r_0}\;. 
\end{eqnarray}
Next, from $g_\chi(r_0,t)$, one can get the number of particles in a given sphere of radius $r<R_\star$,
\bea
N_\chi(r,t)=\int_{R_{th}}^{R_\star} dr_0\, g_\chi(r_0,t)\tau(r,r_0)+N^{th}_\chi(t)\times \int_0^{r} n_\chi(r') \,4\pi r'^2\,\dd r' \, ,\label{eq:num_radius}
\ena
with $\tau(r,r_0)$ the fraction of time spent within the radius $r$ by a particle with orbital extent $r_0$. 
If $r_0<r$, the relation $\tau(r,r_0)=1$ obviously holds, whereas, if $r<r_0<R_\star$, one gets $\tau(r,r_0)=2/\pi \arcsin{(r/r_0)}$. The second term of Eq.~(\ref{eq:num_radius}) corresponds to thermalized particles which are distributed following $n_\chi(r)$ (Eq.~(\ref{eq:nchi-iso})). In Fig.~\ref{fig:evolution_N_DM} we display $N_\chi(r,t)$ as a function of $r$ for the same times as the ones considered in Fig.~\ref{fig:evolution_f_E}. We see that the number of particles increase with $r$, and flattens at a peculiar radius $r_0^*(t)$ which is the typical orbit size of particles with energy $E^*(t)$. For times $t$ smaller than $t_{th}$, the density increases up to $r_0^*(t)$ which is larger than $R_{th}$. For times $t$ larger than $t_{th}$, $r_0^*(t)$ saturates to the radius $R_{th}$ (vertical line of the plot), below which $N_\chi(r)$ increases as $r^3$. Finally the DM number density $\rho_\chi(r,t)$ is given by
\bea
\rho_\chi(r,t) &=& \displaystyle\frac{m_\chi}{4\pi r^2}\frac{\partial N_\chi}{\partial r } = \displaystyle\frac{m_\chi}{4\pi r^2} \int_r^{R_\star}dr_0 \,g_\chi(r_0,t) \frac{\partial \tau}{\partial r}   +N^{th}_\chi(t)\times n_\chi(r) \,4\pi r^2\,.
\ena
One can show that the first term, which gives the contribution from those DM particles which have still not thermalized, diverges as $r$ tends to $0$. This peaked density profile is actually not physical, since it comes from the linear orbit approximation made here. To be more realistic, the central density should be averaged within a sphere of typical radius, the mean ellipticity of the bulk of DM particles at time $t$.\footnote{In practice however this has no effect on the bounds we will get below on the number of particles accreted, or equivalently on the cross section, from the requirement of no black hole formation.
This is due to the fact that these bounds are set to a very good approximation by the particles which have thermalized, and not by the ones which have still not thermalized.}

\section{Constraints on DM from Black Hole Formation}\label{sec:blackholeconstraints}

For a flux of DM particles scattering off NS matter, we have so far determined the number of DM particles accreted, their energy density  profiles, and the related number of DM particles which have thermalized. Eventually DM particles accumulate so much towards the center of the NS that the resulting DM core contracts to a black hole (BH) through gravo-thermal processes. This will be the case when the system becomes self gravitating and satisfies Chandrasekhar criteria. The formed black hole can eventually consume the neutron star, thus destroying it. With the mere observation of neutron stars (i.e. they are not destroyed due to DM) in various DM backgrounds we can place stringent constraints on DM elastic cross section, assuming DM is asymmetric. On the basis of the results obtained in the previous sections, in this section we re-evaluate constraints on DM-neutron elastic cross section. We will also determine the constraints which hold on the DM-proton and DM-muon elastic cross sections.

\subsection{Gravitational Collapse: Chandrasekhar Limit and Self Gravitation}

As abundantly discussed in the literature, 
a black hole can form provided the DM core contains large enough number of DM particles \cite{Goldman:1989nd,Bell:2013xk,McDermott:2011jp,Kouvaris:2010jy,Kouvaris:2010vv,Kouvaris:2011fi,Bramante:2013nma,Bramante:2013hn}.
As already discussed above, the DM particles which have thermalized form a core of radius $R_{th}$ given by Eq.~(\ref{eq:r-therm}), as a result of the balance between the gravitational energy, $-4\pi R^2_{th}\rho_Bm_\chi/3$ and the thermal energy, $\frac{3}{2} k_B T$. In this core, we assume that particles are uniformly distributed and non-interacting, the kinetic energy of order $T$, which means that they are highly non-relativistic (unless the mass is tiny, below $T\sim eV$).
The first condition to have gravitational collapse is that this stable thermal radius configuration resulting from the balance between the "thermal pressure" and the gravitational potential induced by the baryons is destabilized. 
This will be the case if DM particles begin to self-gravitate. The self gravitation condition is $\rho_{\chi}\gtrsim \rho_b$, which means
\beq
\label{eq:self-gr}
 \frac{G N m_\chi^2}{R}  \gtrsim \frac{4 \pi}{3} G \rho_b m_\chi R^2 \,.
\eeq
If DM thermalizes with NS matter, the above equation can be re-written as
\beq
\frac{N_{\chi} m_{\chi} }{ 4/3 \pi R_{th}^3} \gtrsim \rho_b,
\label{eq:self-grav}
\eeq
Inserting the expression for thermal radius Eq.~(\ref{eq:r-therm}) we get
\beq
N^{\rm self}  \simeq  4.8 \times 10^{41} \left(\frac{100 \,{\rm GeV}}{m_{\rm DM}} \right)^{5/2} \left(\frac{T_{NS}}{10^5 {\rm K}} \right)^{3/2}.
\label{eq:N-self-grav}
\eeq
If this condition is satisfied, the ``thermal pressure'' cannot counteract the effect of self-gravitation because the self-gravitation potential in $1/R$ varies faster with respect to $R$ than the energy associated with the thermal pressure, and with the gravity induced by baryons.
From this time, as $R$ decreases, DM is more and more confined and as a result of the Heisenberg uncertainty principle DM momentum increases.
For a bosonic DM particle confined within a radius $R$, the momentum simply goes as $1/R$. 
Thus as long as the DM particle remains non-relativistic the total energy of a bosonic DM particle is
\beq
\label{eq:boson-nonrelch}
E_{\rm tot} \sim - \frac{G N m_\chi^2}{R}  +\frac{1}{2 m R^2} \,,
\eeq
where the last term is the kinetic energy, $E_{kin}\sim p^2/2m_\chi$.
This configuration has a stable minimum for
\begin{equation}
\label{eq:minR}
R\sim \frac{1}{G N m_\chi^3} \,,
\end{equation}
so that it does not lead the system towards $R=0$.
However, if $N$ is large enough, the minimum with respect to $R$ is so small that DM particles do not remain non-relativistic anymore, in this case the total energy per particle is
\beq
\label{eq:boson-nch}
E_{\rm tot} \sim - \frac{G N m_\chi^2}{R}  +\frac{1}{R}\,.
\eeq
The latter has no stable minimum for a finite value of $R$ but, provided the first term dominates over the second one, it has a singular minimum at $R=0$.\footnote{In practice one can check that, for $N$ as large as the one needed to satisfy the self-gravitation criteria, when the Heisenberg uncertainty principle becomes saturated, the minimum, Eq.~(\ref{eq:minR}), lies in a value of $R$ so small that at this radius the particle is already relativistic. Thus the particle becomes relativistic before reaching this non-relativistic minimum and there is not much of a period where Eq.~(\ref{eq:boson-nonrelch}) applies.}
This condition that the first term dominates over the second one is nothing but the 
Chandrasekhar condition, which is satisfied if 
\beq
\label{eq:chandra-bos}
N^{\rm ch}_{\rm bosons}  \simeq 1.5 \times 10^{34} \left (\frac{100 \,{\rm GeV}}{m_{\chi}} \right)^2.
\eeq
Thus there is black hole formation if the number of thermalized particles is larger than ${\rm Max (N^{\rm self},N^{\rm ch} ) }$.
As well known, comparing both numbers, one observes that in the bosonic case, if the self gravitation condition is satisfied, the Chandrasekhar condition is satisfied, unless the DM mass is huge, above $\sim 10^{17} \, {\rm GeV} (T_{NS}/10^5 {\rm K})^3$ ~\cite{McDermott:2011jp}. 

As well known too, for fermions, to satisfy the Chandrasekhar condition requires many more particles, as a result of the Pauli exclusion principle which implies that inside the DM core the fermions are confined within a distance $\sim R/N^{1/3}$. This, for relativistic fermions, gives 
\beq
\label{eq:fermion-nch}
E_{\rm tot} \sim - \frac{G N m_\chi^2}{R}  + \left(\frac{N}{g_f}\right)^{1/3}\frac{1}{R}\,,
\eeq
where $g_f$ is the number of relativistic degrees of freedom, leading to a minimum at $R=0$ if the number of fermions is larger than
\beq
\label{eq:chandra-ferm}
N^{\rm ch}_{\rm fermions}  \simeq 1.8 \times 10^{51} g^{-1/2}_f \left (\frac{100 \,{\rm GeV}}{m_{\chi}} \right)^3.
\eeq
Thus for fermions the Chandrasekhar condition is more demanding than the self-gravitation one.

Note that in the above, to set the conditions for black hole formation,
we have taken into account only those DM particles which have thermalized. For situations where only a small fraction of DM has thermalized, one could wonder if all the particles which have still not thermalized change this condition. It can be checked that the effect on the bounds on DM cross section we will find below is small (because once the thermalized core begins to self-gravitate and collapses, the non-thermalized particles will still not self-gravitate and thus will not collapse as fast). Thus, to determine the black hole formation constraint, the shape of DM profile beyond the thermal radius doesn't enter into play. But we need to know what is the percentage of accreted DM which has thermalized at a given time, information which requires us to calculate the time evolution of the energy profile, as done above (see previous section).

%%%%%%%%%%%%%%%%%%%%%%%%%%%%%%%%%%%%%%
\subsection{Bose-Einstein Condensation}
For extremely large densities, and small temperatures, a gas of bosons can form a Bose-Einstein condensate (BEC), stemming from the fact that, for zero temperature, a Bose gas has a fundamental energy state whose energy vanishes~\cite{Landau:1980mil}. A BEC will form if the temperature of the gas (with $g_s$ degrees of freedom) is below the critical temperature $T_c$  
\begin{equation}
T_c =\frac{2 \pi}{m_{\chi}}\left(\frac{N/V}{g_s \, \zeta(3/2)}\right)^{2/3}.
\end{equation}
This will be the case if $N$ is large enough.
In this case, assuming a non-interacting system in a 3D box~\cite{McDermott:2011jp,Kouvaris:2010vv}, the number of particles in the ground state and the radius of the BEC can be estimated to the values
\bea
\label{eq:nchi0}
N_\chi^0 = N_\chi \left[ 1- \left( \frac{T}{T_c} \right)^{3/2} \right], \quad r_{BEC}= \left(\frac{3}{8 \pi G \rho_b m_\chi^2}\right)^{1/4}.
\ena
The last quantity is obtained equating the kinetic energy $p^2/2m_{\chi}$ (non-relativistic in this case) with the gravitational energy, $4\pi G \rho_b r^2_{BEC}m_{\chi}/3$. In Eq.~(\ref{eq:nchi0}), $N_\chi$ is the number of DM particles which have thermalized.
Thus the critical number of DM particles for a BEC to form a black hole is
\beq
\label{eq:self-grav-bec}
N^{BEC} \simeq N^{ch}_{\rm bosons} + g_s \,10^{36} \, \left(\frac{T_{NS}}{10^5 {\rm K}} \right)^{3} \,,
\eeq
where $N^{ch}_{\rm bosons}$ is given by Eq.~(\ref{eq:chandra-bos}). So far we didn't incorporate the fact that the DM gas is not a non-interacting system but experiences the gravitational potential of NS. For a potential of the form $V(r) = 2 \pi/3 G \rho_b m_{\chi} r^2 $, it was noted in~\cite{1987PhRvA..35.4354B} that the number of particles in ground state and the critical temperature for BEC formation are modified
\bea
\label{eq:nchi-exact}
N_\chi^0 = N_\chi \left[ 1- \left( \frac{T}{T_c} \right)^{3} \right] \quad \text{with:  } T_c = \left(\frac{N}{\pi^3 g_s \zeta(3) }\right)^{1/3} \left(\frac{ \pi}{3} G \rho_b\right)^{1/2}.
\ena
Next these expressions are modified further by considering a GR background, as discussed in~\cite{Jamison:2013yya}, with the prescription $\rho_b \rightarrow \rho_b + 3 P_b $. For the benchmark NS model A we have $\rho_b(0) = 4.31 \times 10^{-3} \,{\rm GeV^4}$ and $P(0) = 7.51\times 10^{-4} \,{\rm GeV^4}$, respectively. We obtain the following estimate for the number of accreted DM particles above which a BEC collapses into a black hole
\beq
\label{eq:self-grav-bec}
N^{BEC} \simeq 1.5 \times 10^{34} \left(\frac{100 \,{\rm GeV}}{m_{\chi}} \right)^{2} +  3.07 \times 10^{35} g_s \left(\frac{T_{NS}}{10^5 {\rm K}} \right)^{3}.
\eeq

%%%%%%%%%%%%%%%%%%%%%%%%%%%%%%%
\subsection{Black Hole Mass and Evaporation}

Once the black holes are formed they can continuously consume the neutron star, thus destroying it. However, if the formed black hole evaporates before the 'consumption', the neutron star can still survive, and BH constraints are alleviated. 

To take into account this effect we begin the discussion by noting that the initial black hole mass is proportional to the number of self gravitating DM particles times DM mass, $M_{BH} \sim m_{\chi} \, {\rm Max(N^{self}_{\chi},N^{ch}_{\chi})}$. In this work we assume that the time evolution of BH mass is governed by the following differential equation~\cite{McDermott:2011jp,Kouvaris:2010vv,Bell:2013xk}
\beq
\label{eq:bh-accretion}
\frac{\dd M_{BH}}{\dd t} = \frac{\dd M_{BH}}{\dd t}|_{\rm NS}  +  \frac{\dd M_{BH}}{\dd t}|_{\rm DM} + \frac{\dd M_{BH}}{\dd t}|_{\rm Hawking}.
\eeq
The various terms on the RHS of the above equation are:
\begin{itemize}
\item BH accretion of NS matter: The first term above captures the effect of NS matter accreting onto the BH. Assuming that matter falls into BH isotropically, the BH gains mass through Bondi-Hoyle accretion process~\cite{Bondi:1944jm,Bondi:1952ni}. However it is well known that most neutron stars have large angular momentum, and hence the in-falling matter could carry angular momentum, invalidating the constraints. This has been addressed in~\cite{Kouvaris:2013kra} which concluded that one recovers conditions for Bondi accretion if the effect of viscosity of nuclear matter in the core is considered. Thus the BH accretion of NS matter for a static spherically symmetric BH is given by~\cite{Shapiro:1983du}
\beq
\label{eq:bondi-accr}
\frac{\dd M_{BH}}{\dd t}\big|_{\rm NS} = \frac{4 \pi \lambda \rho_b G^2}{c_s^3} M^2_{BH},
\eeq
with $c_s = 0.33 c$ and $\lambda = 0.25 $ ~\cite{McDermott:2011jp}, respectively.
 
\item BH accretion of DM: The second term in Eq.~\ref{eq:bh-accretion} is DM accretion rate onto the BH. For the case of bosonic DM (without BEC formation)  and fermionic DM, assuming that DM is collisionless, the accretion is spherical, and the rate is given by~\cite{Shapiro:1983du}
  \beq
  \frac{\dd M_{BH}}{\dd t}\big|_{\rm DM} =  4 \pi (\frac{2 G M_{BH}}{v_\chi})^2  m_{\chi} n_\chi v_{\chi}.
  \eeq

For bosonic DM with BEC formation the right hand side is equal to the capture rate times mass. Once a BEC is formed all new particles that are captured go to the ground state~\cite{McDermott:2011jp,Kouvaris:2010jy}.

\item Hawking Evaporation: The last term in Eq.~(\ref{eq:bh-accretion}) is the rate at which BH loses masses through hawking evaporation

  \beq
  \frac{\dd M_{BH}}{\dd t}|_{\rm Hawking} = - \frac{1}{15360 \pi G^2 M_{BH}^2}.
  \eeq

\end{itemize}
From the above one finds that the black hole formed \underline{does evaporate} when
%following condition on DM mass for which constraints from BH formation are alleviated 
\begin{eqnarray}
  m_\chi &\gtrsim& 3 \times 10^6\, {\rm GeV} \quad {\rm for\, bosons \,with \,no-BEC}    \\
  m_\chi &\gtrsim& 16 \,{\rm GeV} \quad\quad\quad\, {\rm for \,bosons\, with \,BEC}  \\
  m_\chi &\gtrsim&  10^{10} \, {\rm GeV} \quad \quad\,\,{\rm for \,fermions}. 
  \end{eqnarray}
It is not impossible that these bounds may be slightly relaxed from the fact that when the BH evaporates there might be also at this stage a Fermi sea suppression effect at work~\cite{Autzen:2014tza}, a possibility we will not look at.

%%%%%%%%%%%%%%%%%%%%%%%%%%%%%%%%%%%%%%%%
\subsection{Exclusion Curves for Old Neutron Stars}

In this section we present the black hole formation exclusion curves we find for the DM-target (neutron, proton and muon) elastic scattering cross section as a function of DM mass, for bosonic as well as for fermionic DM. These curves depend on mainly three astrophysical parameters, the age of the NS ($\tau_{\star}$), the local DM density ($\rho_{\chi}$), and the NS temperature ($T_\star$). We start by discussing the values of these three inputs applying to the neutron stars which have been observed so far, and subsequently will mention possibilities of future observations of neutron stars in extreme DM density environment which can lead to more stringent constraints.

\underline{NS observation as of today}. In our galaxy numerous pulsars with age of order several billions of years have been observed~\cite{Manchester:2004bp}. Thus for the exclusion plots we will consider the reference value $\tau_\star = 10$ Gyrs.
Note that the uncertainties on the slowing process (technically on the braking index) and corrections due to secular motion affect the precise determination of NS age, however to adopt such a reference value is still reliable.
Most of these identified pulsars are located in our local neighborhood (within 1 kpc from Earth), for which the DM matter density corresponds to $\rho_\chi=0.3\,$GeV/cm$^3$.\footnote{Dependence on halo velocity profile is also relevant. We consider the standard Maxwellian velocity profile in the galactic rest frame. Using a more realistic profile derived from N-body simulations can lead to a difference of up to $\sim 20\%$ in the DM accretion rate with respect to the standard Maxwellian, similar to the case of DM accretion in the Sun~\cite{Choi:2013eda}.} 
For what concerns the NS temperature, a NS is thought to have formed as one of the possible end points of supernovae explosions, with temperatures $\sim 10^{11}$ K at birth. It is well known that NS surface temperature rapidly cools down to temperatures below $10^6$ K within 10$^5$ yrs, via neutrino emission through direct Urca process~\cite{Page:2005fq, Yakovlev:2004iq}. For neutron stars with ages larger than $10^5$ yrs the cooling is mostly driven by photon emission from the surface and other sub-dominant processes. 
Numerous old pulsars have been detected through their radio emission. However, measurement of thermal spectra is challenging, hence there are uncertainties in the deduced surface temperature. For the exclusion plots we will consider a uniform radial temperature profile and the reference value $T=10^5$~K.

For illustrative purposes, let us consider the examples of two old pulsars which have been observed nearby and for which thermal emission have been measured with small uncertainties: PSR J0437-4715 and PSR J2124-3358. These have a spin down age of $\tau_\star = 6.64 \times 10^9 $ yrs and $\tau_\star = 1.07 \times10^{10} $ yrs (corrected for the secular motion)~\cite{Manchester:2004bp} and lies at a distance equal to 139 pc and 270 pc from Earth, respectively~\cite{Kargaltsev:2003eb}. For PSR J0437-4715 the surface temperature $T^{surf}_\star$ is within $1.25-3.5\times 10^5$ K~\cite{2012ApJ...746....6D,Kargaltsev:2003eb}, whereas for PSR J2124-3358 $T^{surf}_\star<4.6\times 10^5$ K. According to a simple non-magnetic iron atmospheric model~\cite{1983ApJ...272..286G}, one obtains an inner temperature of $T_\star=2.1\times 10^6$K for PSR J0437-4715 and $T_\star<2.5\times 10^7$K for PSR J2124-3358. 
Note that pulsars close to the galactic center have been observed, such as J1745-2900, which is located 0.1 pc from the galactic center. This is interesting because they experience a much denser DM environment, $\rho_\chi\sim10^2\,$GeV/cm$^3$ (assuming a NFW DM halo profile). However, its age is much shorter, only $\sim 3.4$~kyr for this pulsar according to~\cite{Manchester:2004bp}. Computing DM constraints with this pulsar leads to less stringent constraints, as the increase in the local DM density does not compensate for its younger age.

\underline{Prospects for future NS observations} In the future no better constraints are expected from the observation of even older NS since the age we consider is already close to the age of the Universe. However it is possible that the limits improve largely from the observation of neutron stars experiencing larger DM local densities. In the exclusion plots we will also present our results for large DM densities, equal to $\rho_\chi=10^3\,$GeV/cm$^3$ and $\rho_\chi=10^6\,$GeV/cm$^3$. The latter value is quoted only for the purpose of a hypothetical far future or experimental breakthrough. The former value, instead, could perhaps be relevant in a not too far future. Actually, millisecond pulsars (MSPs), which are believed to be old recycled pulsars, have already been discovered in DM rich environment like globular clusters~\cite{Ozel:2016oaf} (where $\rho_\chi \sim 10^3\,{\rm GeV/cm^3}$~\cite{Bertone:2007ae,Kouvaris:2010vv}), with properties still under debate.

Different choices of values for $\rho_\chi$ and $\tau_\star$ imply a direct rescaling of the constraints (see below). A change of temperature, instead, implies a more complicated rescaling. Such a change does not only affect the accretion and evaporation rates (relevant for small DM masses, see Sec.~\ref{sec:accretion}), but also the conditions for self gravitation and more importantly on the condition for BEC formation, see Eqs.~(\ref{eq:N-self-grav}) and~(\ref{eq:self-grav-bec}), respectively. As for the local DM density we assume that it has not changed significantly during the evolution of NS.\footnote{In presenting the exclusion curves we have assumed that the NS does not inherit DM particles accreted by its progenitor. Inclusion of the stellar history of progenitor can at most double the number of DM particles in NS, according to Ref.~\cite{Kouvaris:2010vv}} 

The black hole formation exclusion contours we get for these input values are presented for bosonic DM in Fig.~\ref{fig:cap-exclusion-neutrons} (for neutrons), Fig.~\ref{fig:cap-exclusion-protons} (for protons), and Fig.~\ref{fig:cap-exclusion-muons} (for muons). In all these figures, the case without BEC formation are shown in the left panels and the case with BEC formation are shown in the right panels. The corresponding exclusion curves for fermionic DM is presented in Fig.~\ref{fig:cap-exclusion-fermions}.

\begin{figure}[htb]
	\begin{center}
	  \includegraphics[width=0.45\linewidth]{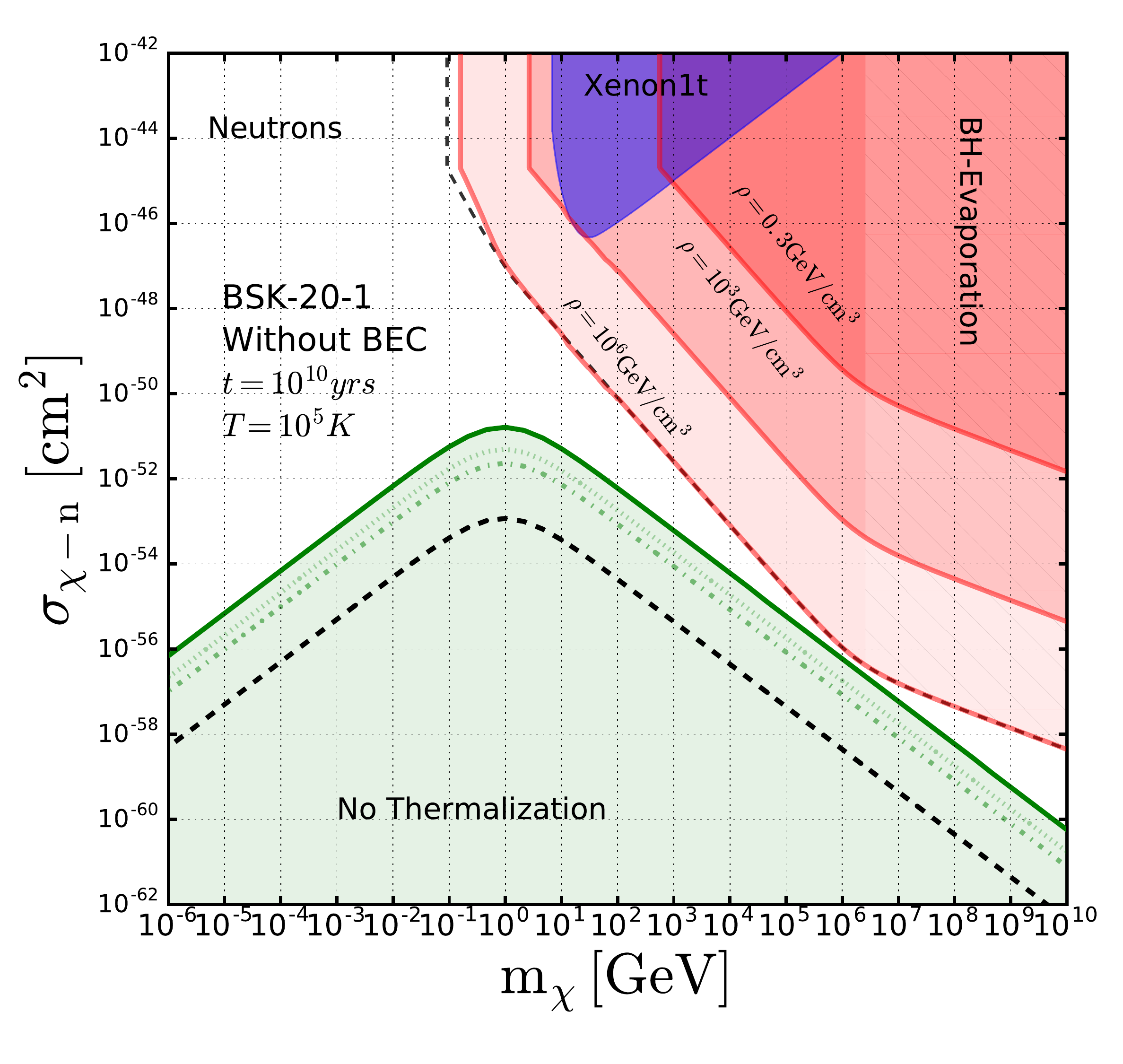}   \includegraphics[width=0.45\linewidth]{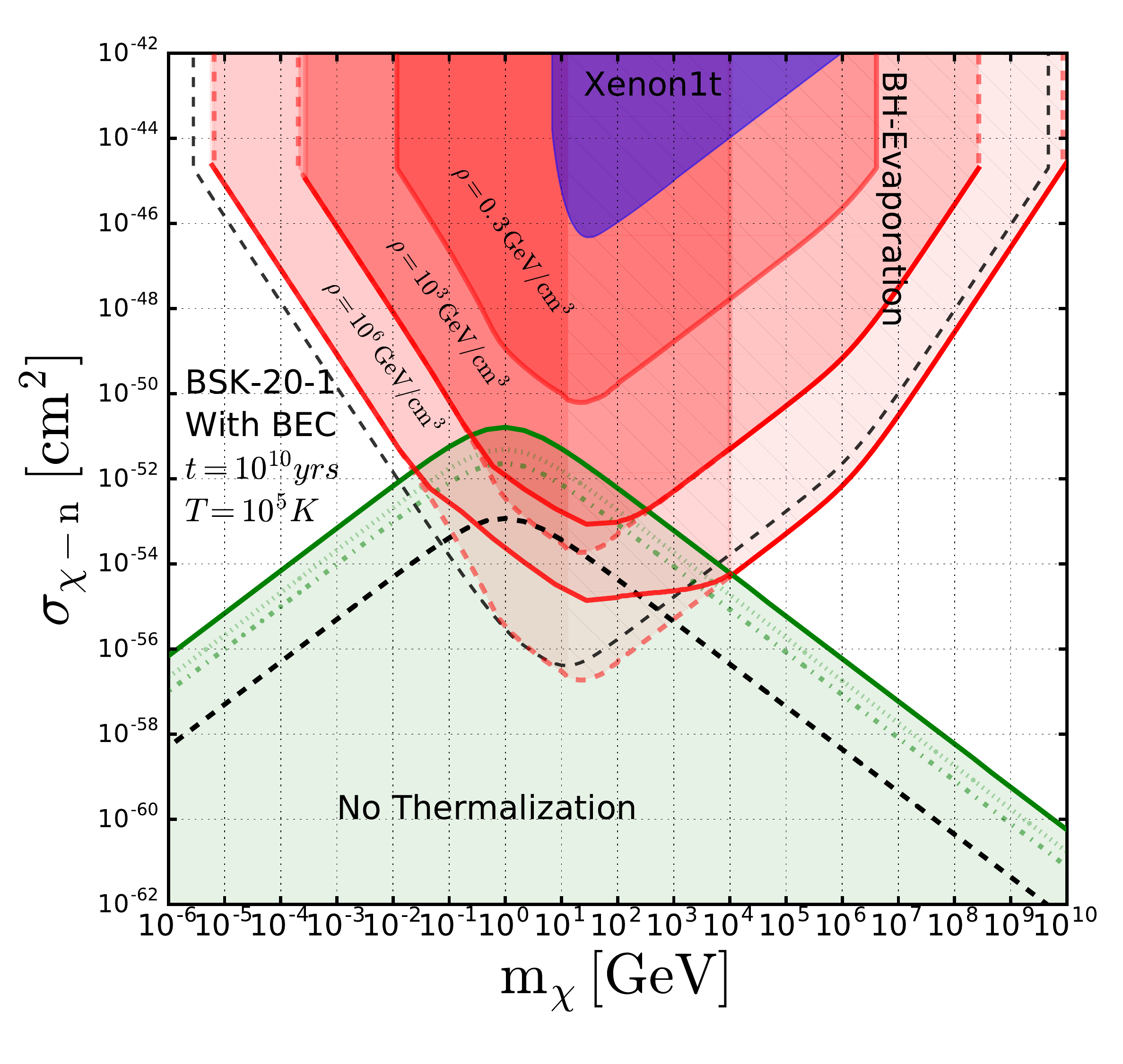}\\
	\end{center}
	\caption{\textbf{\textit{Exclusion plots for bosonic ADM scattering off neutrons:} } Left panel (gravitational collapse condition applied to the thermalized DM core and Right panel (gravitational collapse condition applied to the BEC): In the shaded red regions accumulated DM forms a BH thus constraining the parameter space. We present exclusion curves for $\rho_\chi = 0.3 \,{\rm GeV/cm^3}$ (dark red regions), $\rho_\chi = 10^3\, {\rm GeV/cm^3}$ (mild red regions) and $\rho_\chi = 10^6\, {\rm GeV/cm^3}$ (light red regions). The black dashed contour along the light red regions are the exclusion limits we obtain using the $\delta p/p_f$ approximation of~\cite{McDermott:2011jp}. In the green shaded regions DM does not thermalize with neutrons. The solid, dotted, dotted-dashed green lines correspond to thermalization of the DM particles at the level of 90\%, 70\% and 50\% respectively. The black dashed contour within the green shaded region are those obtained from~\cite{Bertoni:2013bsa}. In the hatched regions the formed BH evaporates before destroying NS, hence constraints are relaxed. The latest Xenon1t constraint on DM-neutron cross sections~\cite{Aprile:2018dbl} is shown in purple shaded region. For right panel only: the solid red contours deep in the ``No thermalization'' region is our result for the exclusion curve using Eq.~(\ref{eq:num_radius}), whereas the dashed red line is obtained still assuming that 100\% of DM has thermalized, even though this doesn't make any sense in this ``No thermalization'' region.}
	\label{fig:cap-exclusion-neutrons}
\end{figure}

%%%%%%%%%%%%%%%%%%%%%%%%%%%%%%%%%%%%%%%%%%%%%%
\subsubsection{Scattering off Neutrons (Fig.~\ref{fig:cap-exclusion-neutrons})}

The dark red region in Fig.~\ref{fig:cap-exclusion-neutrons} gives the excluded region for local density equal to the one expected in the solar system, $\rho_\chi = 0.3 \, { \rm GeV/cm^3}$. The mild-red region is for DM density $\rho_\chi = 10^3\, { \rm GeV/cm^3}$ and the light-red region is for an extreme DM density $\rho_\chi = 10^6 \, { \rm GeV/cm^3}$.
We also show the exclusion curve obtained using estimation of DM capture rate in Ref.~\cite{McDermott:2011jp} (thin dashed black line) applied to NS model A ($\rho_\chi = 10^6 \, { \rm GeV/cm^3}$). The hatched area correspond to regions where the formed BH evaporates before the destruction of NS, hence constraints do not hold in those regions. We also present in these figures the parameter space regions where DM does not thermalize with NS medium, these are shaded in green. The dotted, dashed and solid green curves correspond to the cases where $50\%$, $70\%$ and $90\%$ of DM particles have thermalized, respectively.
For comparison we also show the thermalization curve from~\cite{Bertoni:2013bsa} (thick dashed black line, deep in the green shaded region), obtained by calculating a characteristic thermalization time (see above).

First let us consider the case where the possibility of BEC formation is not taken into account (left panel). For $m_\chi > 1$ GeV there is no effect of the Fermi sea. For these masses and as long as $m_\chi \lesssim 10^6$~GeV, the effect of BH evaporation is negligible, and the number of DM particles accreted from the halo is $C^{\rm w}_\star \, \tau_\star$.  Using Eqs.~(\ref{eq:N-self-grav}) and~(\ref{eq:capture1}) one can read off from the plot the following upper limit on the cross section:
  \beq
  \label{eq:sig-nbec-ex>1}
  \sigma_{\chi-n} < 2.2 \times 10^{-47} {\rm cm^2} \left(\frac{0.3\, {\rm GeV/cm^3}}{\rho} \right) \left(\frac{10^{10}\,{\rm  years}}{t_\star} \right) \left(\frac{10^4 \, {\rm GeV}}{m_\chi} \right)^{3/2} \left(\frac{T_\star}{10^5 \,{\rm K}} \right)^{3/2}.
  \eeq

  For $m_\chi < 1 \,{\rm GeV}$ the effect of Pauli blocking becomes important as the phase space for scattering is greatly reduced (see Sec.~\ref{sec:accretion}). As noted already, BH formation in this case is governed by the self gravitation condition of Eq.~(\ref{eq:N-self-grav}). For masses smaller than $\sim 100$ MeV, for a given cross section, DM accretion is inefficient and can never accrete enough DM particles such that they self gravitate. For mass below 200 MeV we do not find any constraint on the cross section because for such masses even a cross section of the order of the geometric one doesn't lead to black hole formation.
For the range $200\,\hbox{MeV} \lesssim m_\chi \lesssim 1\,\hbox{GeV}$, and when the cross section is smaller than the geometric cross section, we find the upper limit to be
    \beq
  \label{eq:sig-nbec-ex<1}
  \sigma_{\chi-n} < 5.94 \times 10^{-47} {\rm cm^2} \left(\frac{10^6\, {\rm GeV/cm^3}}{\rho} \right) \left(\frac{10^{10}\,{\rm  years}}{t_\star} \right) \left(\frac{0.5 \, {\rm GeV}}{m_\chi} \right)^{5/2} \left(\frac{T_\star}{10^5 \,{\rm K}} \right)^{3/2}.
  \eeq
This limit is a factor 1 to 8 times smaller than in previous studies (compare the red curve with thin dashed black curve). These differences mainly stem from the improved evaluation of the capture rate. For $m_\chi \gtrsim 10^6\, {\rm GeV}$, the efficiency of capture decreases due to velocity suppression~\cite{McDermott:2011jp} and scales as $\propto m_\chi^{-2}$. Hence the change of slope around those masses. More importantly, for such masses, Hawking evaporation of BH dominates and the formed BHs are efficiently evaporated away, thus constraints from BH formation are relaxed (hatched regions).

Next, let us consider the case when the BEC formation is taken into account (right panel of Fig.~\ref{fig:cap-exclusion-neutrons}). As argued above, in this case the conditions for BH formation becomes more stringent.
This is visible from comparing the left and right panels of Fig.~\ref{fig:cap-exclusion-neutrons}.
As a result, regions where there is no thermalization (or little level of thermalization) now touch the regions excluded by the requirement of no black hole formation.
As discussed above, in the literature the ``no thermalization'' region (delimited by the dashed black line~\cite{Bertoni:2013bsa} in
Fig.~\ref{fig:cap-exclusion-neutrons})
is obtained from calculating a "typical" thermalization time (see above). Beyond (before) this time 100\% of the accreted DM is supposed to have (have not) thermalized. Thus in the thermalized region obtained in this way, the BEC black hole condition is obtained considering a BEC stemming from this fully thermalized DM population (i.e.~$N_\chi$ in Eq.~(\ref{eq:nchi-exact}) is the total number of DM particles accreted by the NS). In the no thermalization region instead, it is either said that the BEC black hole condition should relax (implying implicitly
that all of the no thermalization region is not excluded by any black hole formation), or still an excluded region is obtained by assuming a BEC
stemming from a fully thermalized DM population, even though one knows one lies in the no thermalization region. This latter condition, which makes no sense, gives the red dashed line in Fig.~\ref{fig:cap-exclusion-neutrons}).
Here instead, knowing from Eq.~(\ref{eq:ntherm}) the number of DM particles which have actually thermalized at all time, $N_\chi^{th}$, we can determine the number
of DM particles there are in the BEC at all times, and compare at all times this number with the number of DM particles
one needs in the BEC to form a black hole, Eq.~(\ref{eq:self-grav-bec}). This allows us to derive robust black hole formation exclusion regions, given by the solid red curve, especially in the region where only a small fraction of the accreted DM particles have thermalized. We find that this discussion turns out to be in fact irrelevant for a DM density of the order of the one in our vicinity, $\rho_\chi=0.3\,$GeV/cm$^3$, because in this case the exclusion curves lie in the region where the level of thermalization is high. But this discussion is fully relevant for
larger DM densities, for example within the $100 \,\hbox{MeV}\lesssim m_\chi\lesssim 1$~TeV mass range for $\rho_\chi=10^3\,$GeV/cm$^3$.
For this case Fig.~\ref{fig:cap-exclusion-neutrons} shows that the level of thermalization from which the black hole forms may largely vary. For example for $m_\chi=10$~GeV this level is as low as 25~\% whereas for other masses it is higher. For the extreme case $\rho_\chi=10^6\,$GeV/cm$^3$ and same DM mass the level is as low as 0.9\%.

The cross sections probed are much smaller than the ones probed by direct detection experiments. For example, for  $m_\chi = 1 $ GeV we find that no black hole formation requires $\sigma_{\chi-n} < 2 \times 10^{-54} \, {\rm cm^2}$ in environments very close to the galactic center ($\rho_\chi= 10^6\, {\rm GeV/cm^3}$) and $\sigma_{\chi-n} < 1\times 10^{-52} \, {\rm cm^2}$ for $\rho_\chi= 10^3\, {\rm GeV/cm^3}$, respectively. Our results give constraints  on $\sigma_{\chi-n}$  which may differ from previous constraints, by up to $\sim 2-3$ orders of magnitudes, especially in the region where only a little proportion of the accreted DM particles has thermalized.

As already noted in the literature, with BEC formation, the requirement of no black hole formation constrains $\sigma_{\chi-n}$ in ADM models with DM mass down to $\sim 2$~keV. 
Here we find that it does it down to $\sim 6$~keV (for $\rho_\chi= 10^6\, {\rm GeV/cm^3}$), whereas for $\rho_\chi= 10^3\, {\rm GeV/cm^3}$ and $\rho_\chi= 0.3\, {\rm GeV/cm^3}$
it does it down to 100 keV and 7~MeV respectively.

Note that the exclusion curve scales as $m_\chi^{-2}$ in regions where DM largely thermalizes with NS matter, for $m_\chi \lesssim 0.1 \,{\rm GeV}$. This scaling is due to the fact that DM accretion rate is independent of $m_\chi$ and the BEC condition (Eq.~(\ref{eq:self-grav-bec})) is driven by the term $\propto m_\chi^{-2}$. Empirically we find the following limit on the cross section
  \beq
  \label{eq:sig-nbec-ex<1}
  \sigma_{\chi-n} < 1.6 \times 10^{-47} {\rm cm^2} \left(\frac{0.3\, {\rm GeV/cm^3}}{\rho} \right) \left(\frac{10^{10}\,{\rm  years}}{t_\star} \right) \left(\frac{0.1 \, {\rm GeV}}{m_\chi} \right)^{2} .
  \eeq

The constraints derived here are factor $\sim 8$ smaller than those derived in previous studies for $m_\chi \lesssim 0.1 \,{\rm GeV}$.
For masses larger than $\sim 10-15$ GeV, there are no constraints due to evaporation of the formed black holes. 
Finally note that in Appendix~\ref{app:combinedresults} we combine into a single plot the constraints of both panels of Fig.~\ref{fig:cap-exclusion-neutrons}, i.e.~ black hole formation 
constraints from the BEC below 15 GeV and black hole formation constraints from the DM thermal core above 15 GeV. In this Appendix we also show these plots for temperature equal to $10^6$~K rather than $10^5$~K.

\begin{figure}[H]
	\begin{center}
	  \includegraphics[width=0.45\linewidth]{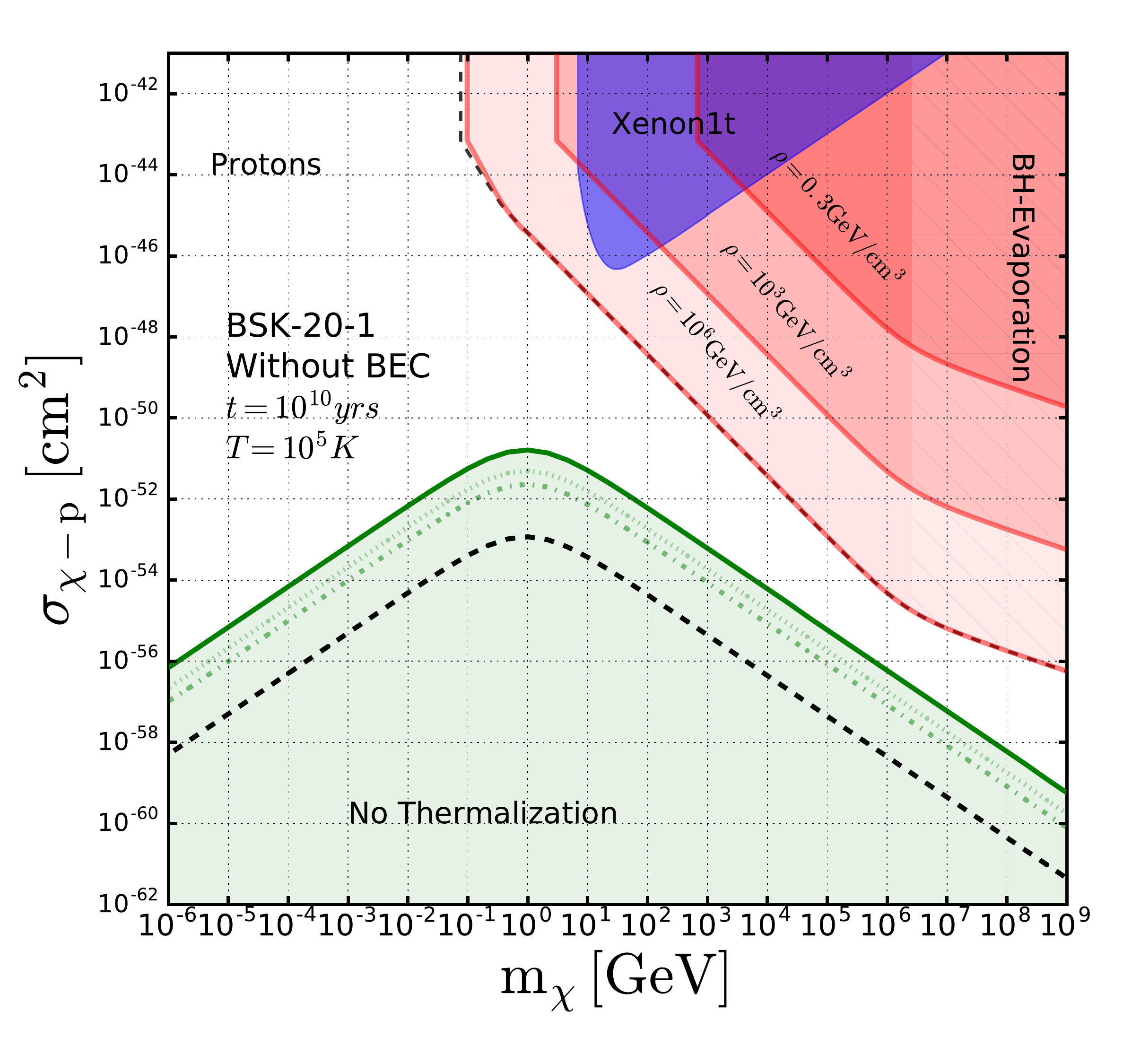}   \includegraphics[width=0.45\linewidth]{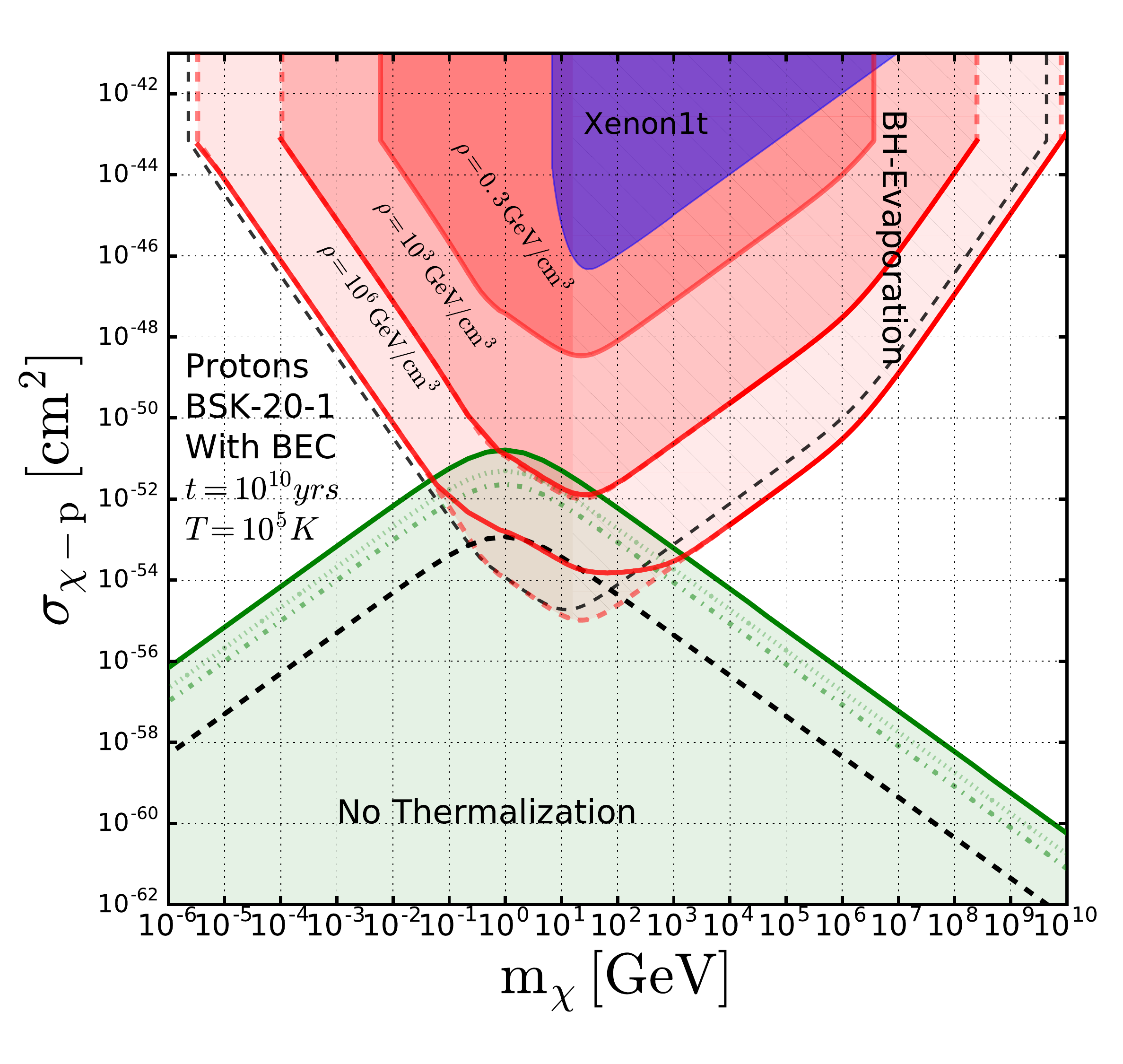}\\
	\end{center}
	\caption{\textbf{\textit{Exclusion plots for bosonic DM scattering off protons:} } Same as Fig.~\ref{fig:cap-exclusion-neutrons} for DM scattering off protons.
 }
	\label{fig:cap-exclusion-protons}
\end{figure}

%%%%%%%%%%%%%%%%%%%%%%%%%%%%%%%%%%%%%%%%%%%%  
\subsubsection{Scattering off Protons (Fig.~\ref{fig:cap-exclusion-protons})}

Assuming that DM scatter off protons of the NS one can also get constraints on the DM-proton cross section, given in
Fig.~\ref{fig:cap-exclusion-protons}. The most important difference with respect to neutrons is that there are far fewer protons than neutrons, but with similar chemical potentials. Thus, quantitatively the accretion rate and the resulting exclusion curves are simply rescaled by the number density of protons. For NS model A, the averaged fraction of free protons that DM can scatter off is approximately $2.7\%$ of the total number of baryons and the chemical potential is smaller than that of neutrons by factor 2 (see Appendix~\ref{app-profiles} for a discussion about NS profiles). Another point to note is that the geometric cross section per proton is larger than that of neutrons by factor $\sim 35$. Thus the saturation cross section is 35 times the critical geometric cross section for neutrons.  
  
  For the case without BEC formation, for $m_\chi >1 {\rm GeV}$ we find the following upper limit on the cross section
  \beq
  \label{eq:sig-nbec-p-ex>1}
  \sigma_{\chi-p} < 1.1 \times 10^{-45} {\rm cm^2} \left(\frac{0.3\, {\rm GeV/cm^3}}{\rho} \right) \left(\frac{10^{10}\,{\rm  years}}{t_\star} \right) \left(\frac{10^4 \, {\rm GeV}}{m_\chi} \right)^{3/2} \left(\frac{T_\star}{10^5 \,{\rm K}} \right)^{3/2}.
  \eeq
  
  For $m_\chi < 1 \,{\rm GeV}$, the impact of the Fermi sea suppression is large, similar to the case of neutrons. 
Here, we find the following upper limit on the cross section
  \beq
  \label{eq:sig-nbec-p-ex<1}
  \sigma_{\chi-p} < 1.0 \times 10^{-45} {\rm cm^2} \left(\frac{10^6\, {\rm GeV/cm^3}}{\rho} \right) \left(\frac{10^{10}\,{\rm  years}}{t_\star} \right) \left(\frac{0.5 \, {\rm GeV}}{m_\chi} \right)^{5/2} \left(\frac{T_\star}{10^5 \,{\rm K}} \right)^{3/2}.
  \eeq

For the case where BEC formation is taken into account (note that the BEC condition (Eq.~\ref{eq:self-grav-bec}) is independent of the target particle), the exclusion plot differ from that of neutrons only due to quantitative changes in the capture and thermalization rates. Again using the refined treatment of thermalization we can robustly estimate the bounds even in regions where DM does not thermalize. Here we require $N^{th} < N^{BEC}$ (solid red curves). For $m_\chi < 0.1 \, {\rm GeV}$ we obtain the following limit 

   \beq
  \label{eq:sig-bec-p-ex<1}
  \sigma_{\chi-p} < 1.8 \times 10^{-46} {\rm cm^2} \left(\frac{0.3\, {\rm GeV/cm^3}}{\rho} \right) \left(\frac{10^{10}\,{\rm  years}}{t_\star} \right) \left(\frac{0.1 \, {\rm GeV}}{m_\chi} \right)^{2}.
  \eeq
  
For DM scattering off protons (for the BEC case) we exclude DM masses down to $2\,{\rm keV}$. Clearly NS probe a large region of the parameter space much better than direct detection experiments, similar to the case of neutrons. Note nevertheless that 
these constraints on $ \sigma_{\chi-p}$ are interesting only if one assume a large violation of isospin symmetry. If we assume that, as expected, isospin is a good symmetry, these limits on $ \sigma_{\chi-p}$ are largely superseded by the constraints on $ \sigma_{\chi-n}$ of Fig.~\ref{fig:cap-exclusion-neutrons} (except for $m_\chi$ of order a few keV). Note also that the constraints on $ \sigma_{\chi-p}$ we plot in Fig.~\ref{fig:cap-exclusion-protons} from direct detection (purple region) are assuming isospin symmetry.

\begin{figure}[H]
  \begin{center}
    	  \includegraphics[width=0.45\linewidth]{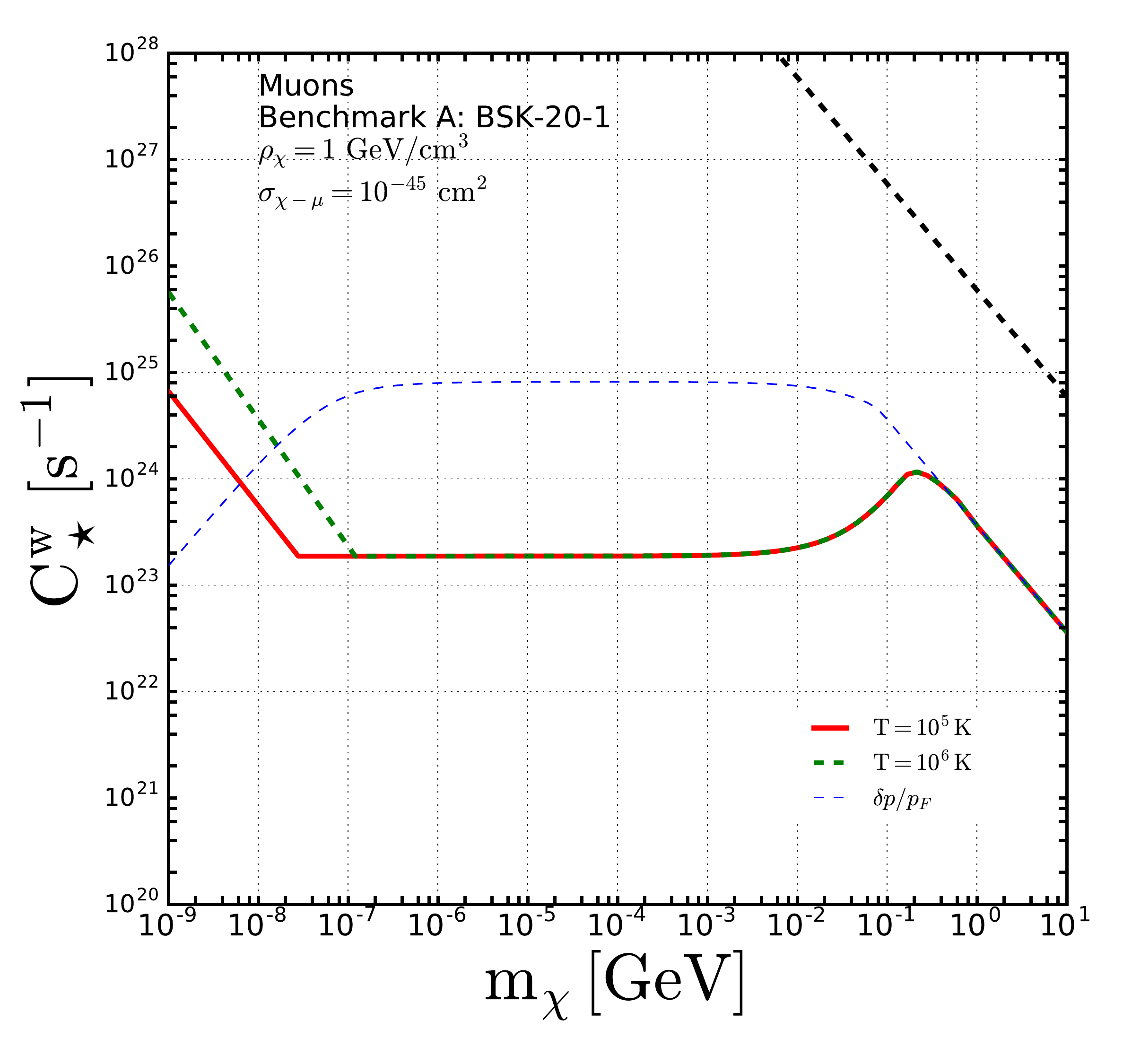}\\
	  \includegraphics[width=0.45\linewidth]{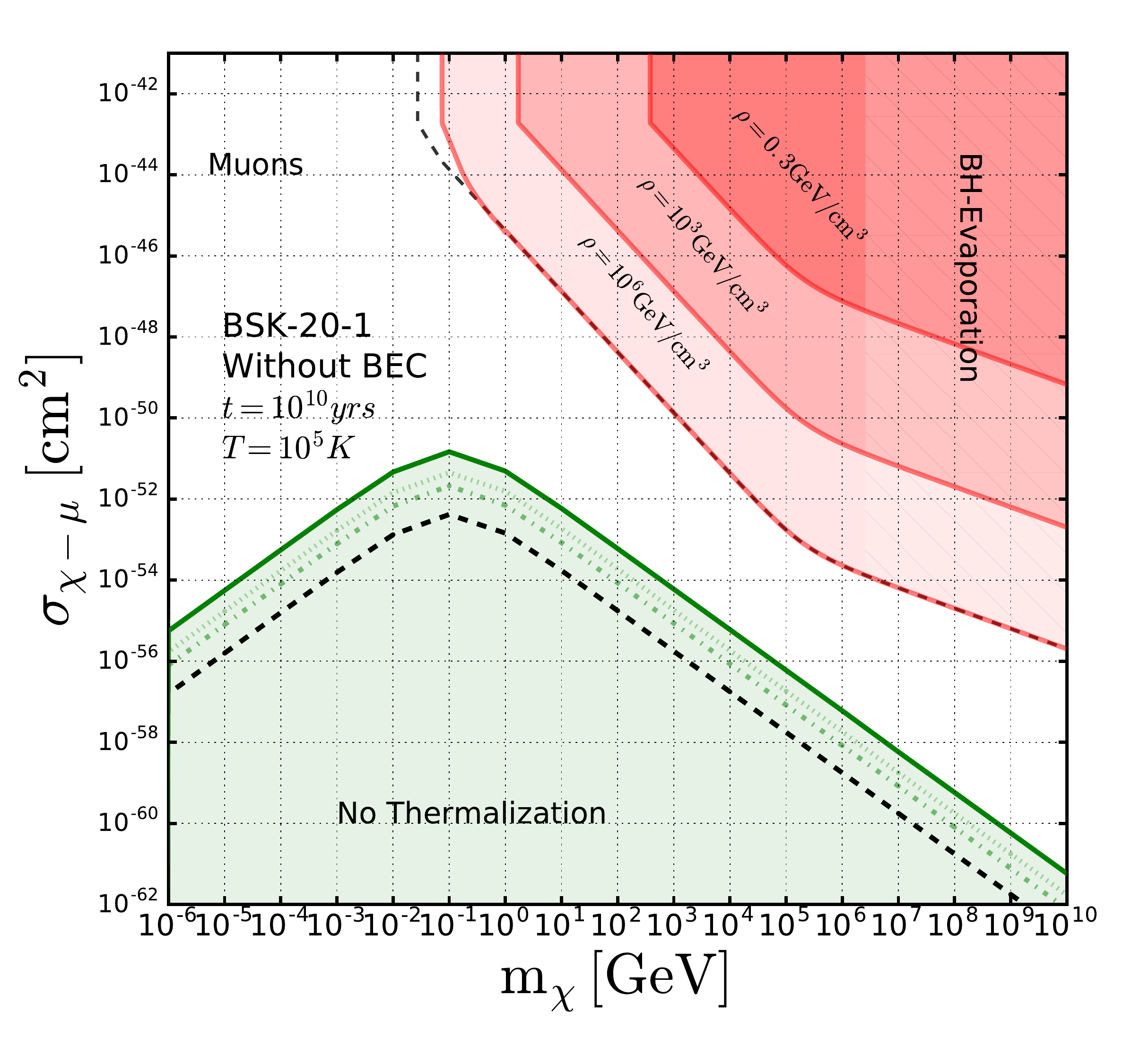}   \includegraphics[width=0.45\linewidth]{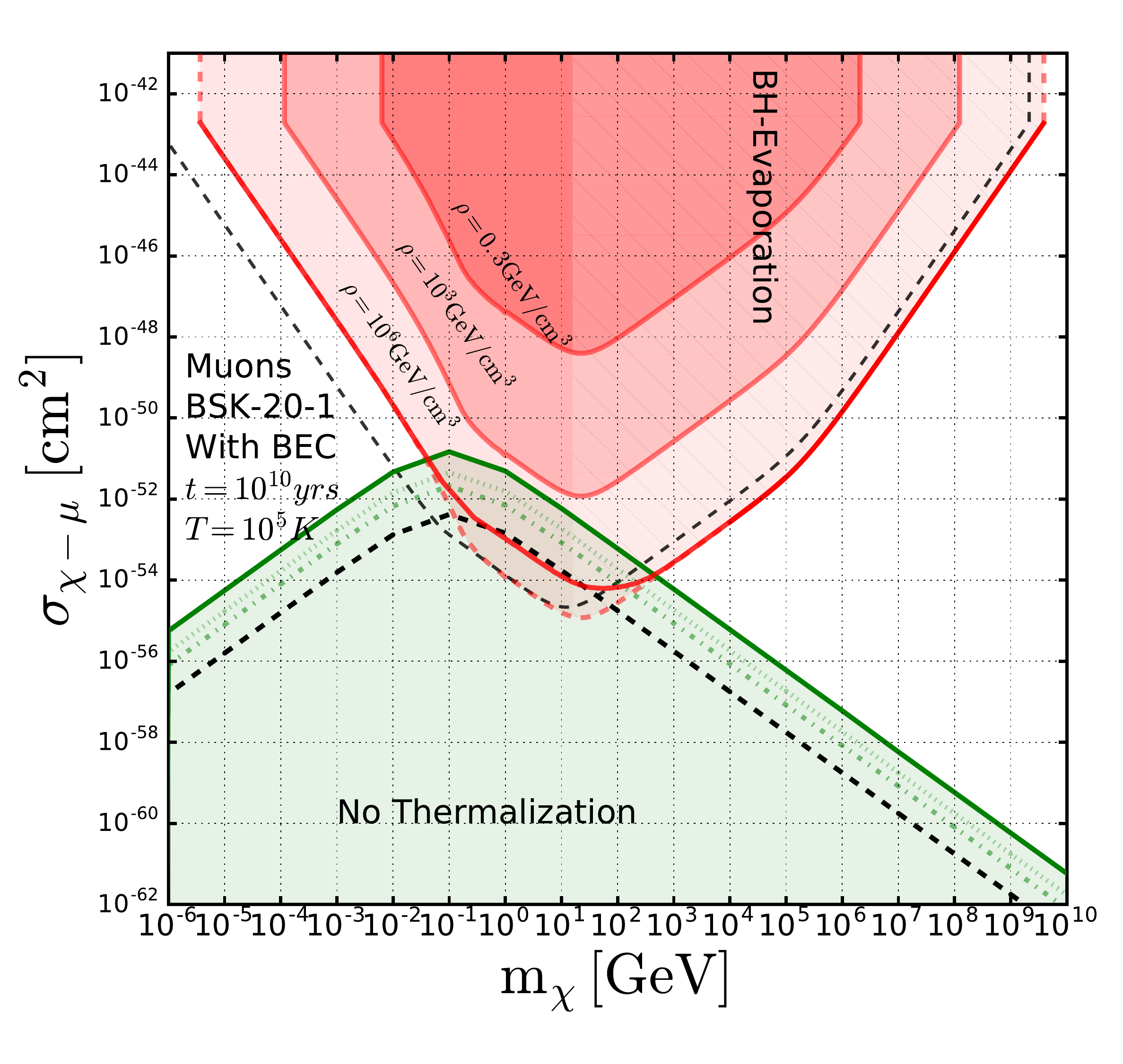}\\
	\end{center}
	\caption{\textbf{\textit{Upper panel}}: Capture rate for DM scattering off muons: same legend as Fig.~\ref{fig:cap-bsk20A}. \textbf{\textit{Lower panels}}: Exclusion plot for bosonic DM scattering off muons: same legend as Fig.~\ref{fig:cap-exclusion-neutrons}.}
 
	\label{fig:cap-exclusion-muons}
\end{figure}

%%%%%%%%%%%%%%%%%%%%%%%%%%%%%%%%%%%%%%%%%%  
\subsubsection{Scattering off Muons (Fig.\ref{fig:cap-exclusion-muons})} 

The possibility of the existence of muons in a NS was realized soon after the discovery of the first pulsars. When the Fermi momentum of electrons exceeds the muon mass, it becomes energetically favorable for electrons at the edge of the Fermi sphere to be converted to muons, through thermal fluctuations~\cite{Weinberg:1972kfs}. However, the existence of muons in NS have not been exploited to constrain properties of DM. Here we present the exclusion limits for DM$-\mu$ elastic scattering cross section from observation of old NS for the first time. For a discussion about the NS profile, equation of state and muon content, see Appendix~\ref{app-profiles}. Note that in, for instance, ``quarkophobic'' DM models, these constraints would be basically the only relevant ones.

In the upper panel of Fig.~\ref{fig:cap-exclusion-muons} we show the capture rate we obtain for DM scattering on muons, similar to Fig.~\ref{fig:cap-bsk20A} for the neutron case. 
Similar to protons and neutrons, muons are also non-relativistic in NS ($\mu_F = 0.018 \,{\rm GeV}$ for model A). Thus the formalism of capture and thermalization remains the same with few quantitative changes. 
 There are 3 main differences: the mass, the number density or fraction of muons and finally the value of the chemical potential. For the NS model A, the average number of free muons that DM can scatter off is approximately $1.16\%$ of the total number of baryons. Thus the geometric saturation cross section for muons can be estimated by rescaling the neutron critical cross section with the muon fraction. This yields the critical cross section for muons to be $4.9\times 10^{-43}\,{\rm cm^2}$, which is a factor 80 larger than the neutron critical cross section, similar to that of protons. When below the critical cross section, the DM accretion rate for scattering off muons peaks at the muon mass (whereas for neutrons it peaks  around the neutron mass, see above), and, for large masses, velocity suppression becomes relevant for $m_\chi \gtrsim 10^5\,{\rm GeV}$ (for neutrons $m_\chi \gtrsim 10^6\,{\rm GeV}$). Note also that for muons and $m_\chi\lesssim m_\mu$, the result we get from the $\delta p/p$ approximation are a factor 40 larger than the results we obtain.
  
In the lower panels of Fig.~\ref{fig:cap-exclusion-muons} we show the exclusion plots for the muon case, similar to Fig.~\ref{fig:cap-exclusion-neutrons} for the neutron case.
 For the case without BEC formation (left panel) we obtain the following limit on the cross section for $m_\chi \gtrsim 0.1\, {\rm GeV}$ (note the change of slope at $m_\chi \sim 10^5\,{\rm GeV}$)
  \beq
  \label{eq:sig-nbec-mu-ex>1}
  \sigma_{\chi-\mu} < 1.3 \times 10^{-45} {\rm cm^2} \left(\frac{0.3\, {\rm GeV/cm^3}}{\rho} \right) \left(\frac{10^{10}\,{\rm  years}}{t_\star} \right) \left(\frac{10^6 \, {\rm GeV}}{m_\chi} \right)^{3/2} \left(\frac{T_\star}{10^5 \,{\rm K}} \right)^{3/2}.
  \eeq
For $m_\chi \lesssim 0.1\, {\rm GeV}$, similar to the other cases, the best constraints are obtained for NS in large DM density environments, we get
    \beq
  \label{eq:sig-nbec-mu-ex<1}
  \sigma_{\chi-\mu} < 8.6 \times 10^{-45} {\rm cm^2} \left(\frac{10^6\, {\rm GeV/cm^3}}{\rho} \right) \left(\frac{10^{10}\,{\rm  years}}{t_\star} \right) \left(\frac{0.1 \, {\rm GeV}}{m_\chi} \right)^{5/2} \left(\frac{T_\star}{10^5 \,{\rm K}} \right)^{3/2}.
  \eeq
For the case with BEC formation (right panel), NS can probe DM masses down to $6\, {\rm keV}$ (for $\rho_\chi \sim 10^6\,{\rm GeV/cm^3}$). For neutron stars in our local neighborhood we obtain the upper limit for $m_\chi \lesssim 0.1\, {\rm GeV}$
    \beq
  \label{eq:sig-bec-mu-ex<1}
  \sigma_{\chi-\mu} < 1.6 \times 10^{-46} {\rm cm^2} \left(\frac{0.3\, {\rm GeV/cm^3}}{\rho} \right) \left(\frac{10^{10}\,{\rm  years}}{t_\star} \right) \left(\frac{0.1 \, {\rm GeV}}{m_\chi} \right)^{2}.
  \eeq
In Appendix~\ref{app:combinedresults} we present the plot combining the constraints from both panels of Fig.~\ref{fig:cap-exclusion-muons}, as well as the results for $T=10^6$~K rather than $T=10^5$~K.
  
\subsubsection{Fermion DM scattering off NS matter}

Similar to the case of bosonic DM we can obtain exclusion curves for fermionic DM scattering off NS matter. However, it is well known that the constraints in this case are much weaker than that of the bosonic case due to Pauli exclusion principle.  Parametric dependence of the critical number of particles on DM mass is different than that of bosons and scales $\propto m_\chi^{-3}$, as given by Eq.(\ref{eq:chandra-ferm}). In this case, Chandrasekhar limit (Eq.(\ref{eq:chandra-ferm})) is more stringent than self-gravitation condition (Eq.(\ref{eq:N-self-grav})) for all DM masses considered here. As we do not consider DM self interactions we do not consider various subtleties that might arise from self interactions. Another point to note is that Hawking evaporation of the formed black holes is only relevant for DM masses above $10^{10}$ GeV, for the fermion DM case (see above).

\begin{figure}[H]
  \begin{center}
    	  \includegraphics[width=0.45\linewidth]{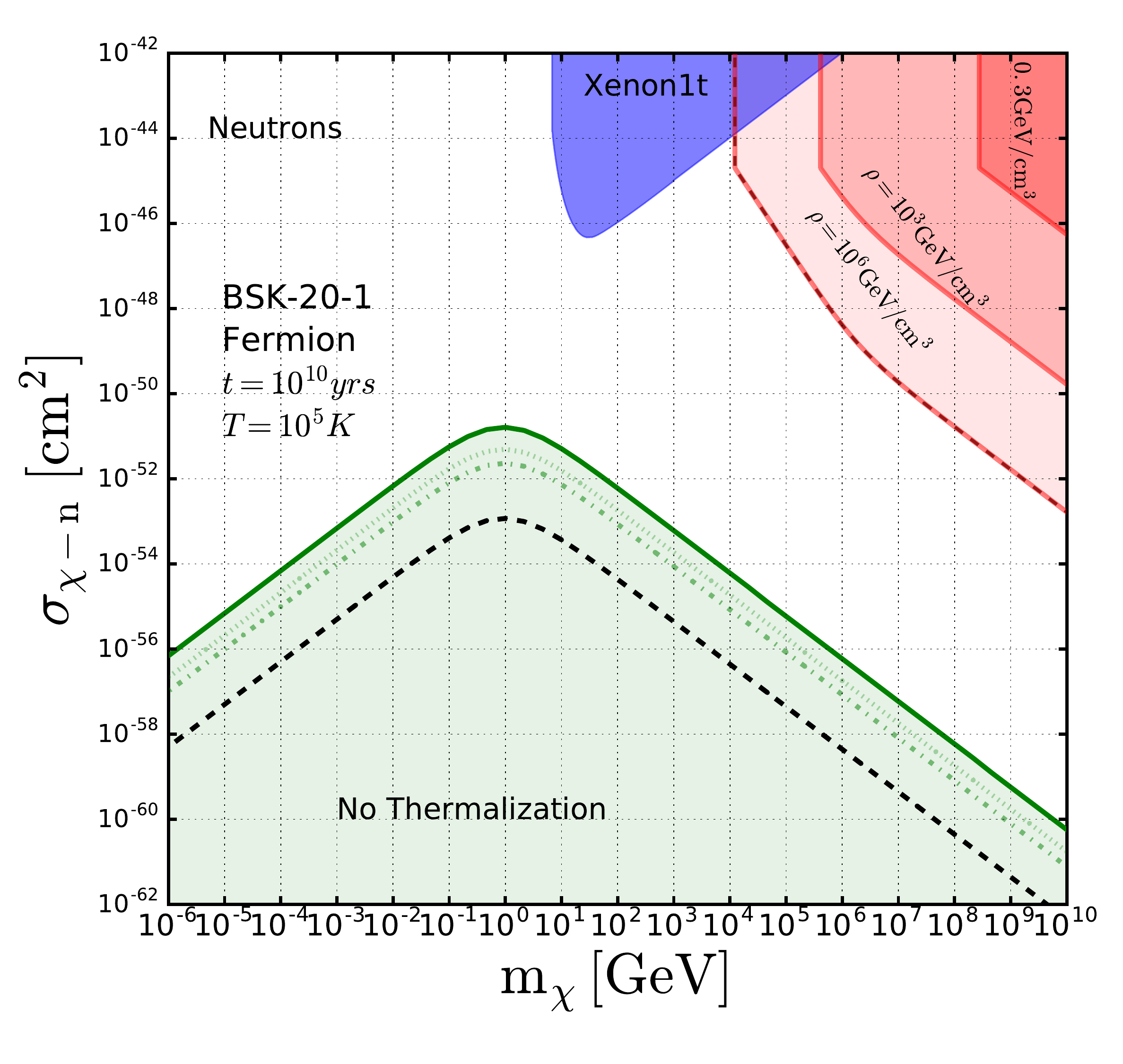}\\
	  \includegraphics[width=0.45\linewidth]{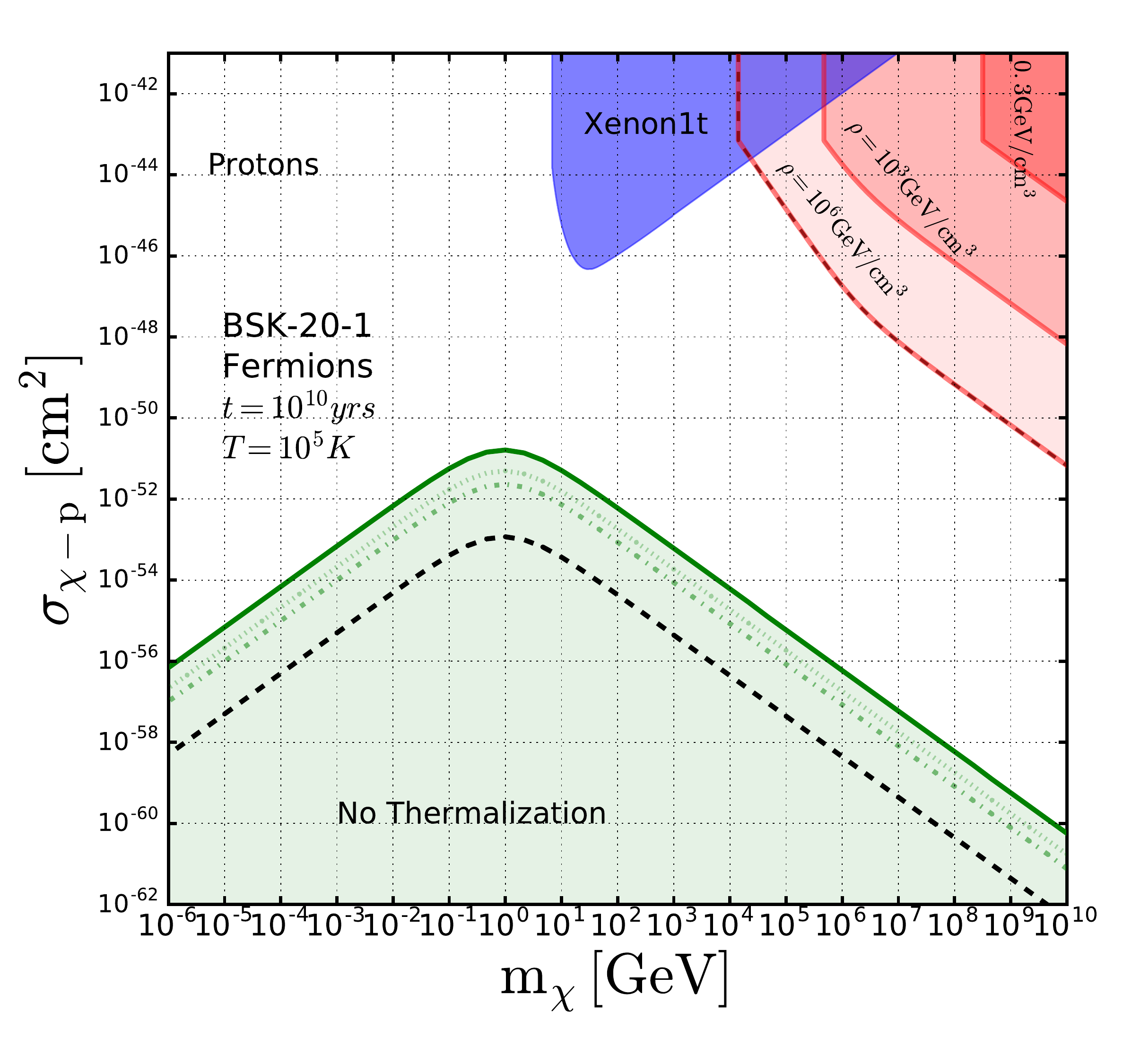}   \includegraphics[width=0.45\linewidth]{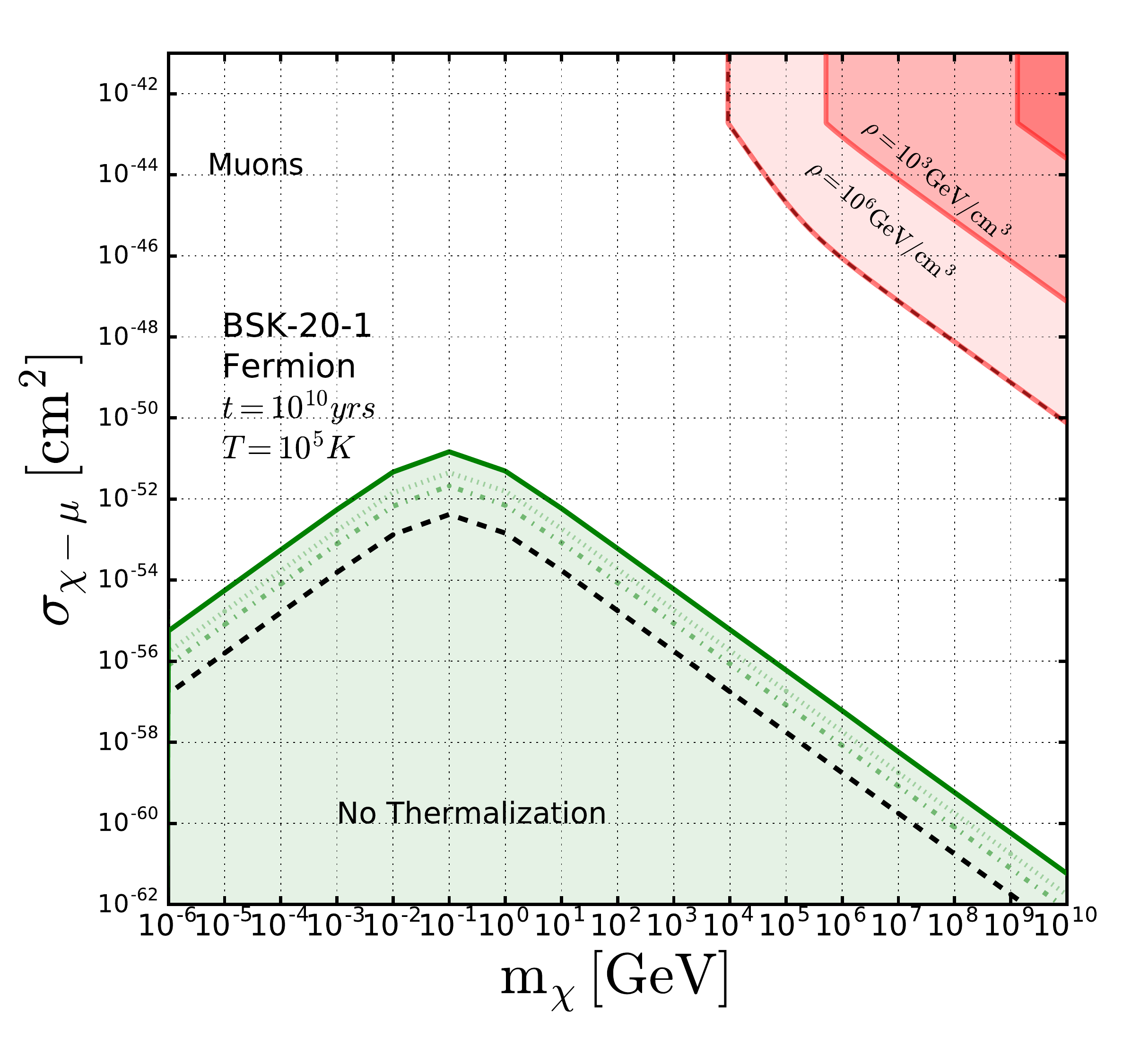}\\
	\end{center}
  \caption{\textbf{\textit{Exclusion plot for fermionic DM scattering off neutrons, protons and muons:} } Color code is the same as Fig.~\ref{fig:cap-exclusion-neutrons}, for fermion DM scattering off neutrons (upper panel), protons (lower left panel) and muons (lower right panel).}
	\label{fig:cap-exclusion-fermions}
\end{figure}

The exclusion plots for fermion DM scattering off neutrons (upper panel), protons (lower left panel), and muons (lower right panel) are shown in Fig.~\ref{fig:cap-exclusion-fermions}, respectively. The color code is the same as in Fig.~\ref{fig:cap-exclusion-neutrons}. Again, similar to the previous cases, the best constraints are obtained when NS in extreme DM density environments are considered.

For neutrons we find the following limit on the cross section (when below the geometric limit) for $m_\chi < 10^6$ GeV,
\beq
\label{eq:sig-ferm-neutron}
\sigma_{\chi-n} < 2.2 \times 10^{-47}\, {\rm cm^2} \left(\frac{10^6 \,{\rm GeV/ cm^3}}{\rho}\right) \left(\frac{10^{10} {\rm years}}{t_\star}\right) \left(\frac{10^5\, {\rm GeV}}{m_\chi} \right)^2,
\eeq
and for $m_\chi > 10^6$ GeV

\beq
\label{eq:sig-ferm-neutron-l}
\sigma_{\chi-n} < 5.4 \times 10^{-46}\, {\rm cm^2} \left(\frac{0.3 \,{\rm GeV/ cm^3}}{\rho}\right) \left(\frac{10^{10} {\rm years}}{t_\star}\right) \left(\frac{10^9\, {\rm GeV}}{m_\chi} \right).
\eeq
For DM scattering off protons we find the following limit (for $m_\chi < 10^6$ GeV)
\beq
\label{eq:sig-ferm-proton}
\sigma_{\chi-p} <  1.0 \times 10^{-45}\, {\rm cm^2} \left(\frac{10^6 \,{\rm GeV/ cm^3}}{\rho}\right) \left(\frac{10^{10} {\rm years}}{t_\star}\right) \left(\frac{10^5\, {\rm GeV}}{m_\chi} \right)^2,
\eeq
and for $m_\chi > 10^6$ GeV
\beq
\label{eq:sig-ferm-proton-l}
\sigma_{\chi-n} <  2.2\times 10^{-44}\, {\rm cm^2} \left(\frac{0.3 \,{\rm GeV/ cm^3}}{\rho}\right) \left(\frac{10^{10} {\rm years}}{t_\star}\right) \left(\frac{10^7\, {\rm GeV}}{m_\chi} \right).
\eeq
For DM scattering off muons we have
\beq
\label{eq:sig-ferm-muon}
\sigma_{\chi-\mu} < 1.5\times 10^{-43}\, {\rm cm^2} \left(\frac{10^6 \,{\rm GeV/ cm^3}}{\rho}\right) \left(\frac{10^{10} {\rm years}}{t_\star}\right)\left(\frac{1.3 \times10^9\, {\rm GeV}}{m_\chi} \right),
\eeq
and for $m_\chi < 10^5$ GeV
\beq
\label{eq:sig-ferm-muon-l}
\sigma_{\chi-\mu} < 1.7\times 10^{-45}\, {\rm cm^2} \left(\frac{0.3 \,{\rm GeV/ cm^3}}{\rho}\right) \left(\frac{10^{10} {\rm years}}{t_\star}\right)\left(\frac{10^5\, {\rm GeV}}{m_\chi} \right).
\eeq

%%%%%%%%%%%%%%%%%%%%%%%%
\section{Summary and Discussion}
It is well known that neutron stars, from their high density, low temperature and oldness, provide stringent constraints on the cross section between (asymmetric) Dark Matter (DM) particle and neutrons. Multiple steps are involved in obtaining these constraints: gravitational trapping of DM particles that intercept the neutron star; orbital trajectories these particles subsequently follow outside and inside the NS; formation of a thermalized DM core with or without the formation of a Bose-Einstein condensate; collapse of the DM core (or of the central BEC part of the core) into a black hole when DM starts to self-gravitate (with or without subsequent evaporation of the black hole).
In this context we have improved and refined several steps, and determined new constraints on DM elastic scattering cross section with SM particles which were not considered before. 

In computing the capture rate, instead of considering a uniform NS, we have considered a realistic Equation of State (EOS) for the neutron star called BSK~\cite{Potekhin:2013qqa}, with characteristic radial distributions for number densities and chemical potentials of neutrons, protons and muons. We have considered one representative benchmark model BSK20-1 throughout the text. We also determine and compare how the results vary when one considers a different EOS in Appendix~\ref{app-profiles}.

For DM accretion and for DM thermalization with neutron star medium, we perform a thorough calculation considering the fact that neutrons form a highly degenerate medium in the neutron star. This had been done for the thermalization process~\cite{Bertoni:2013bsa} but not in computing the number of accreted DM particles. We do it in a consistent way for both the accretion and the thermalization processes, going beyond some of the approximations made in~\cite{Bertoni:2013bsa}. 

For the accretion rate, Sec.~\ref{sec:accretion}, we discuss the effect of Fermi degenerate medium in detail and derive analytical expression (in Appendix~\ref{app:diff-rate}) for the differential scattering rate for DM scattering off non-relativistic particles, such as neutrons (protons and muons). We have also taken into account the fact that the temperature of the NS is non-vanishing ($\sim 10\, {\rm eV}$). This induces two effects, which were not considered before, and turns out to be crucial for small DM mass, i.e.~for $m_{\chi}$ below $\sim $~keV: saturation of the Pauli blocking suppression and evaporation of DM from scattering off neutrons.
Numerically, for such masses these effects are huge. For larger masses and up to $\sim 0.1$~GeV we find that the suppression due to the Fermi sea is very large for the accretion rate, and is about 6-8 times lower than the estimate in~\cite{McDermott:2011jp}.

For the thermalization process, Sec.~\ref{sec:thermalization}, instead of computing a typical thermalization time and assuming that after (before) this time all particles have (have not) thermalized, we compute the evolution of the energy density profile, or equivalently (in the linear orbit approximation) the radial DM number density profile.
Beyond the fact that the knowledge of these profiles is in principle relevant for any phenomenological study of DM properties in a NS, this allows us to determine at any time the number of DM particles which have already, or have not still, thermalized.
This information is the one needed for deriving more reliable constraints due to black hole formation, and allows to set such constraints also when the level of thermalization is low.
To determine these profiles we take into account the fact that a DM particle has a varying velocity along its orbit, in the linear orbit approximation. To go beyond the linear orbit approximation would certainly be quite interesting, but it basically requires simulations which we leave for future work.

With all the ingredients in hand, in Sec.~\ref{sec:blackholeconstraints} we present exclusion curves in the cross section vs mass plane for ADM scattering off non-relativistic neutrons, see Fig.~\ref{fig:cap-exclusion-neutrons} for the bosonic DM case with and without formation of BEC and Fig.~\ref{fig:cap-exclusion-fermions} for the case of fermionic DM, respectively.
Depending on the value of the parameters (particularly DM densities) considered, the black hole constraints we obtain 
may differ from previous estimates by factors of order unity or by orders of magnitudes.

For fermionic DM, as well as for bosonic DM with no BEC formation, the upper bounds we find on the elastic cross section on neutron are similar to the ones of~\cite{McDermott:2011jp}. The upper bounds may differ by up to a factor 3-4. For bosonic DM with no BEC formation we find that our upper bounds are always obtained for cases where the level of thermalization is very close to 100\%.  For the bosonic case where we do take into account the fact that a BEC can form, the constraints are more stringent. For a DM local density similar to the one of the solar system, $\rho_\chi=0.3\,$GeV/cm$^3$, the upper bounds may differ again by factors of order 3-4. For larger DM density, the results differ by orders of magnitudes, due to the fact that in this case the gravitational collapse may occur when the level of thermalization is still low. For instance for $\rho_\chi=10^6\,$GeV/cm$^3$ the results differ by up to 2-3 orders of magnitudes, precisely where the black hole constraints are the most stringent.

Beyond the bounds on the elastic cross section off neutrons, we have also determined the amount of DM accreted from scattering with protons and with muons.
The proton case is of limited interest as it gives less stringent bounds than the neutron case (except for $m_\chi$ of order few keV), hence these proton bounds are irrelevant, unless one assumes that isospin symmetry is largely violated. We stress, however, that there are many muons in a neutron star, and that consequently quite interesting bounds can be obtained on the elastic cross section between the DM and muons. These bounds, which we give in Figs.~\ref{fig:cap-exclusion-muons} and Fig.~\ref{fig:cap-exclusion-fermions}, could be the best one can get in many models, in particular in leptophilic models. As for the bounds on the cross section with electrons, we leave their determination for future work.
Electron in a neutron star are to a large extent relativistic and thus, unlike the other components, cannot be considered in the non-relativistic formalism we have considered here.
\section*{Acknowledgments}
We warmly thank Nicolas Chamel for providing us the equations of state for old neutron stars, and for several discussions, within the framework of joint particle physics and astrophysics Brussels Lab. of the Universe. We thank Petr Tinyakov for discussions and comments. The work of R.G. is supported by the "Probing DM with Neutrinos" ULB-ARC grant. The work of Y.G. is supported by the IISN Belgian fund. This work is also supported by the F.R.S.-FNRS and the  Excellence of Science grant (EOS) convention 30820817.

\bibliographystyle{unsrt}
\bibliography{biblio}
\appendix

\section{Uncertainties from Neutron Star Equations of States}\label{app-profiles}

As mentioned in the introduction, the capture rate depends on the properties of NS. Its precise computation does not only depend on the NS mass and radius, but also on the radial profile of the number density and chemical potential of the target material considered. For a realistic estimation of the capture rate, we have used four NS benchmark profiles. These four models stem from two different equations of state (EOS) called BSK20 and BSK21, described in Refs.~\cite{Potekhin:2013qqa,Goriely:2013xba}. All regions of the star are treated in a unified and consistent way using the nuclear energy density functional theory. The functionals underlying these two EOS were fitted to microscopic neutron matter EOS that differ in their degree of stiffness~\cite{Goriely:2013xba}. {For} each EOS we consider two extreme profiles corresponding to low and high "mass configuration" of NS (see the corresponding radius and mass values in Tab.~\ref{tab:numbers}). Hence the four cases: BSK20-1 (model A), BSK20-2 (model B), BSK21-1 (model C) and BSK21-2 (model D).
In the {discussion} above we only refer to BSK20-1 (model A), since the results do not qualitatively differ {from} the other cases, but quantitatively by ${\cal O} (1)$ factors. Before discussing the dependence of our results on the profiles, note that these EOS are still allowed by the latest constraints from LIGO observation of a binary NS merger (see e.g. Fig.1 of~\cite{Most:2018hfd} and Fig.3 of~\cite{Abbott:2018exr}). {Moreover} note that these EOS consider only ordinary {baryonic and leptonic} matter (i.e. protons $p$, neutrons $n$, electrons $e$, muons $\mu$) without ``exotic'' particles {such has hyperons which contain strange quarks.}

In the top panel of Fig.~\ref{fig:chem-pot}, we show the number density of baryons (neutrons + protons) {as a function of NS radius} for the four benchmark models considered. In all cases, we notice that the number density is relatively constant in the inner regions of the NS and starts to drop in the outer regions, from about two thirds of the NS radius. In the {following} four plots of Fig.~\ref{fig:chem-pot} we show the chemical potential of {neutrons for the four models mentioned above (right middle panel)} and, we show the chemical potentials of neutrons, protons, electrons and muons for BSK20-1 (middle left panel). We also show neutron fraction as a function of radius (lower right panel) for the four NS models and, the NS composition for BSK20-1 (lower right panel). All these figures show that the values of chemical potentials and number densities of target particles are not constant within the NS. Hence, as a result of these inhomogeneities,  DM capture is also inhomegeneous (see below). \\

\textit{NS models and DM constraints}:\\

{For each model the NS profile we consider, we give in Tab.~\ref{tab:numbers} the total number of particles $N$ of each species, relative to the numbers which hold for BSK20-1.} For example, one observes that while the number of neutrons varies by 30\%, the (free) proton number can differ by as much as factor 6. The capture rate increases linearly with this number $N$, and {depends on} the NS mass through $v_{esc}$ (see Eq.~(\ref{eq:capturegeom})).

\begin{figure}[!htb]
	\begin{center}
	  \includegraphics[width=0.4\linewidth]{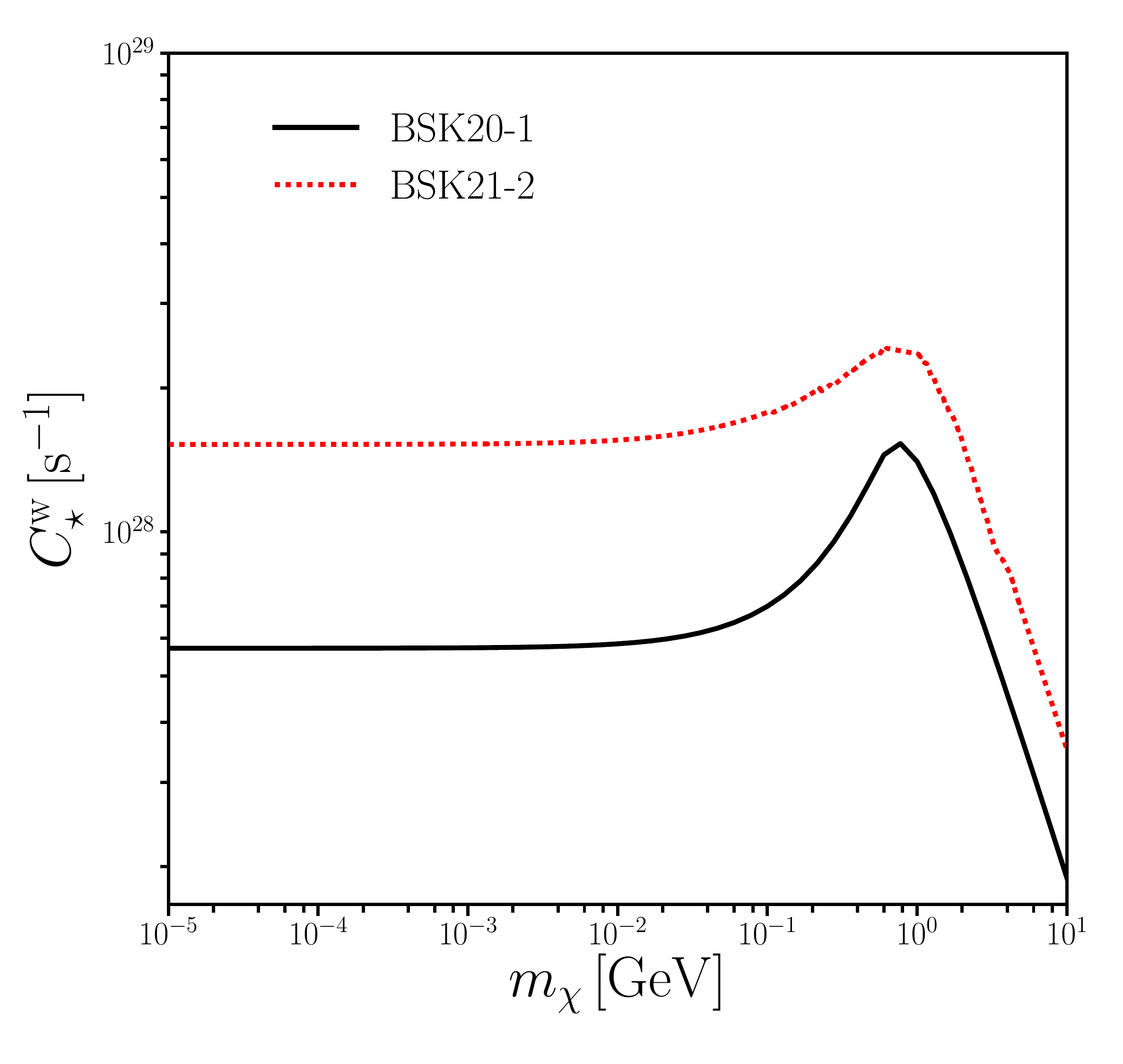}\\
          		\includegraphics[width=0.4\linewidth]{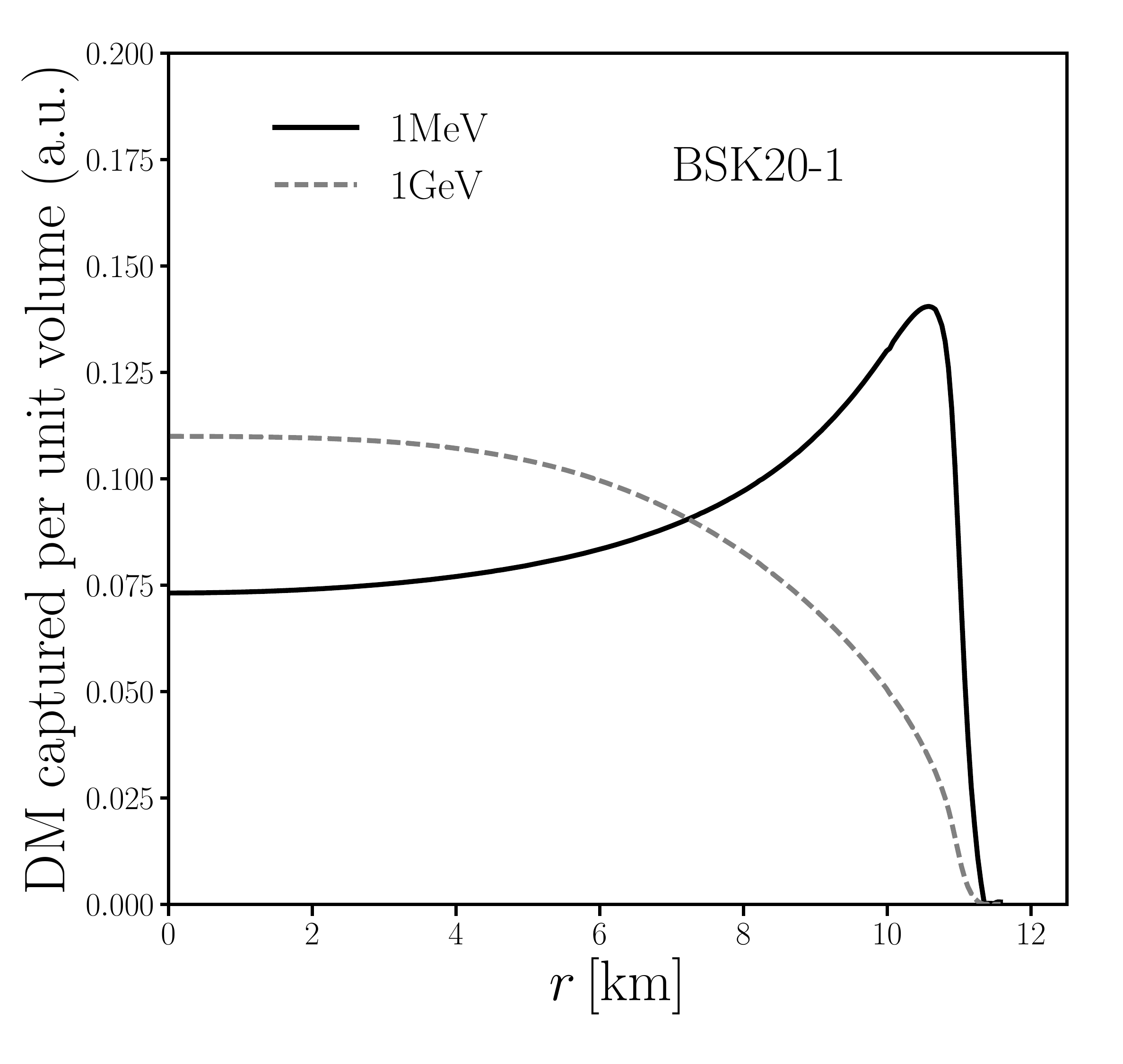}
		\includegraphics[width=0.4\linewidth]{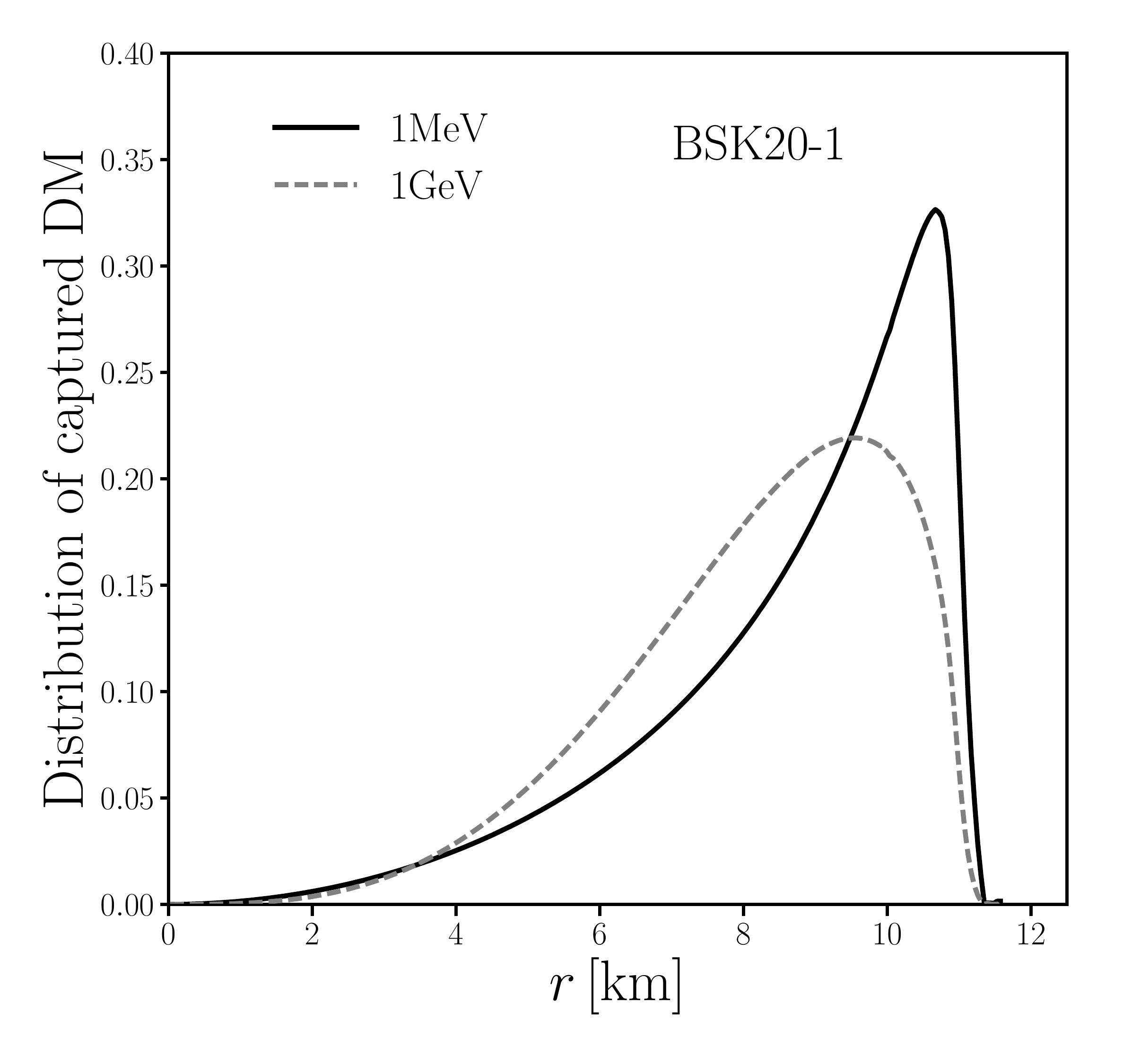}
	\end{center}
	\caption{\textit{Upper panel:} Capture rates (from DM scattering off neutrons) with respect to DM mass for the two models: BSK20-1 and BSK21-2. \textit{Lower left panel:} Capture rates per unit volume for BSK20-1, for $m_\chi =1\,$MeV and 1 GeV. The rates are normalized to make the comparison easier. \textit{Lower right panel:} Distribution of DM particles captured as a function of NS radius, for DM masses of 1 GeV and 1 MeV for BSK20-1.}  \label{fig:cap_two_mod}
\end{figure}

\begin{figure}[H]
	\begin{center}
		\includegraphics[width=0.45\linewidth]{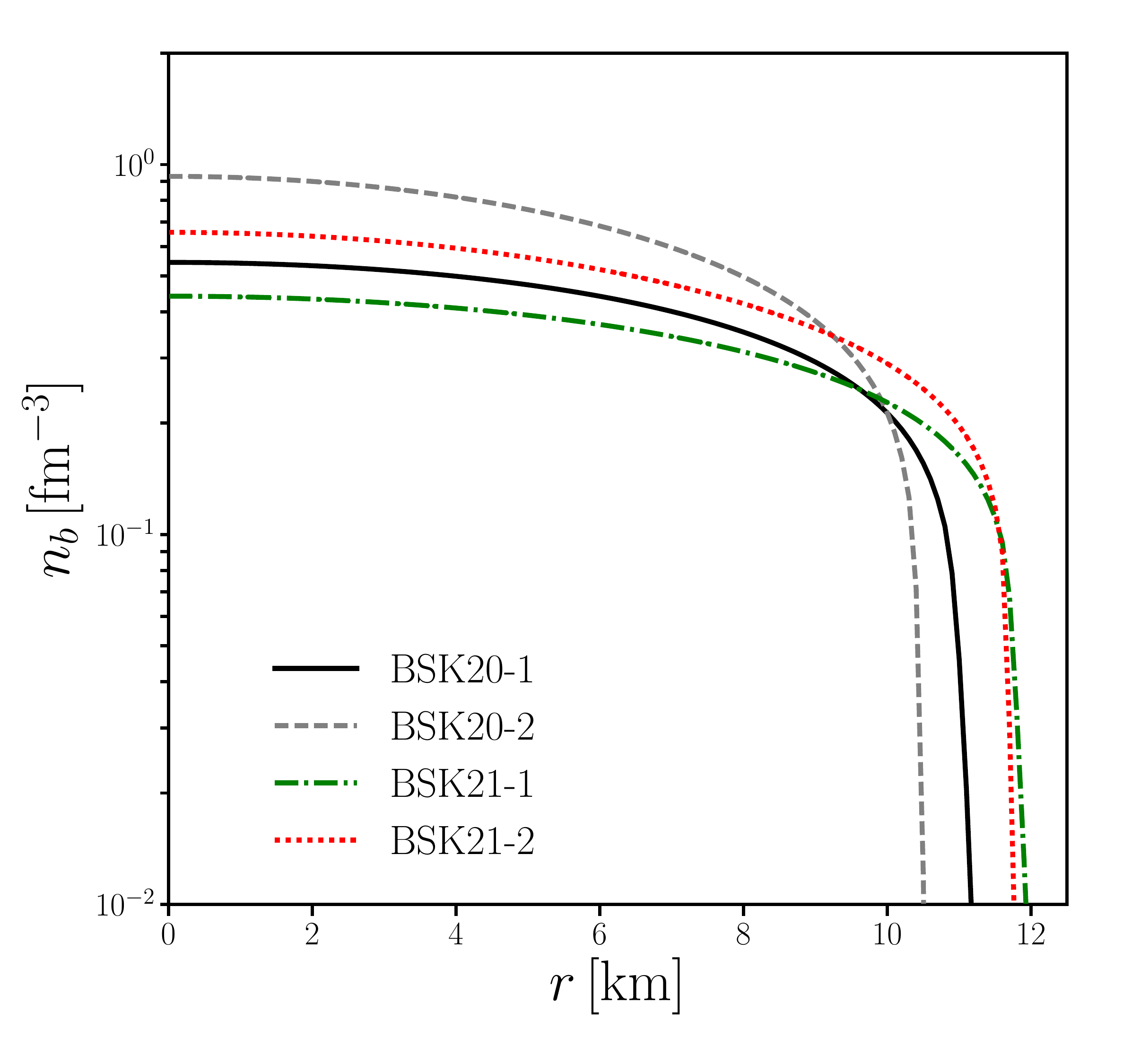}\\
		\includegraphics[width=0.45\linewidth]{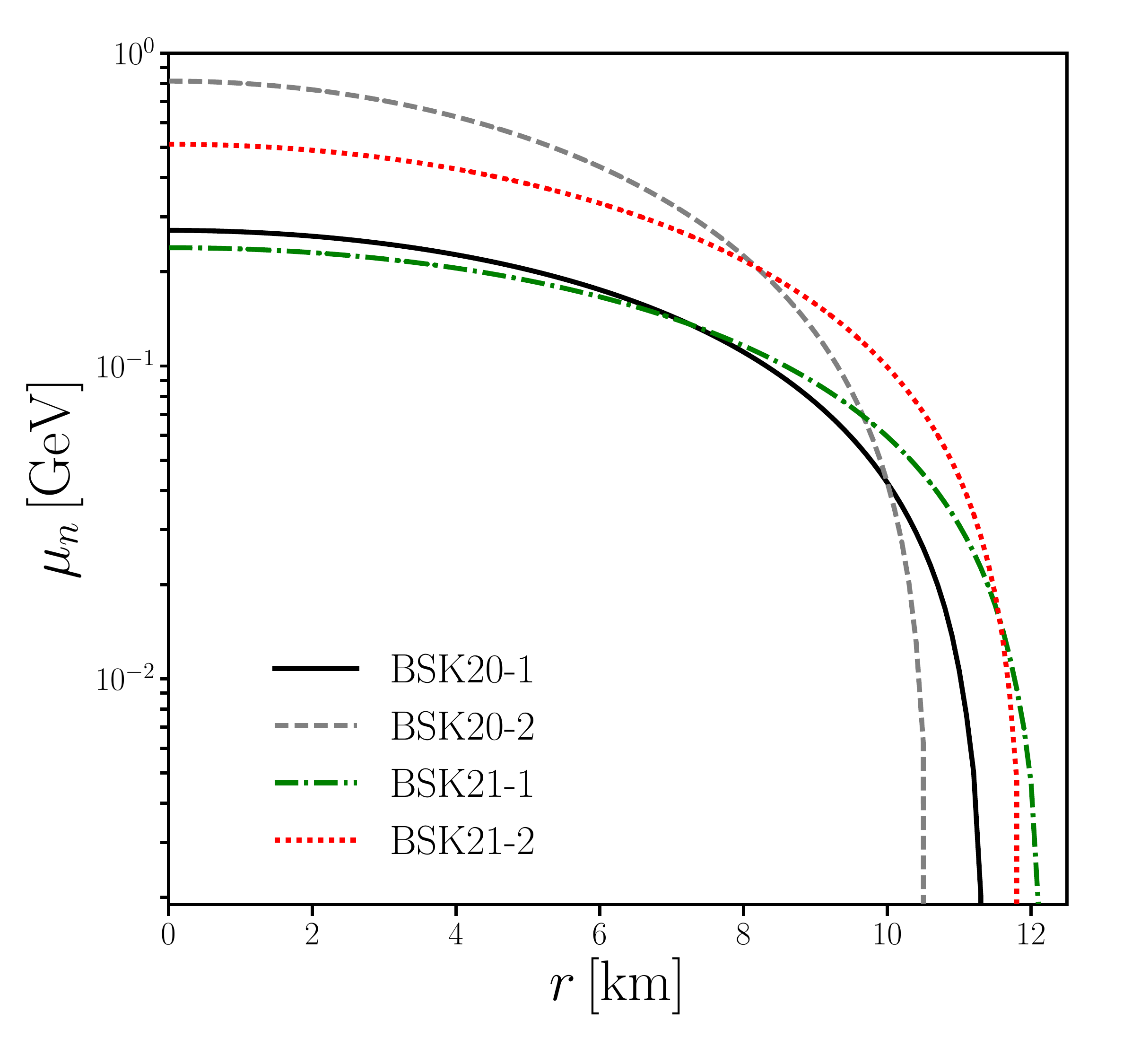}
		\includegraphics[width=0.45\linewidth]{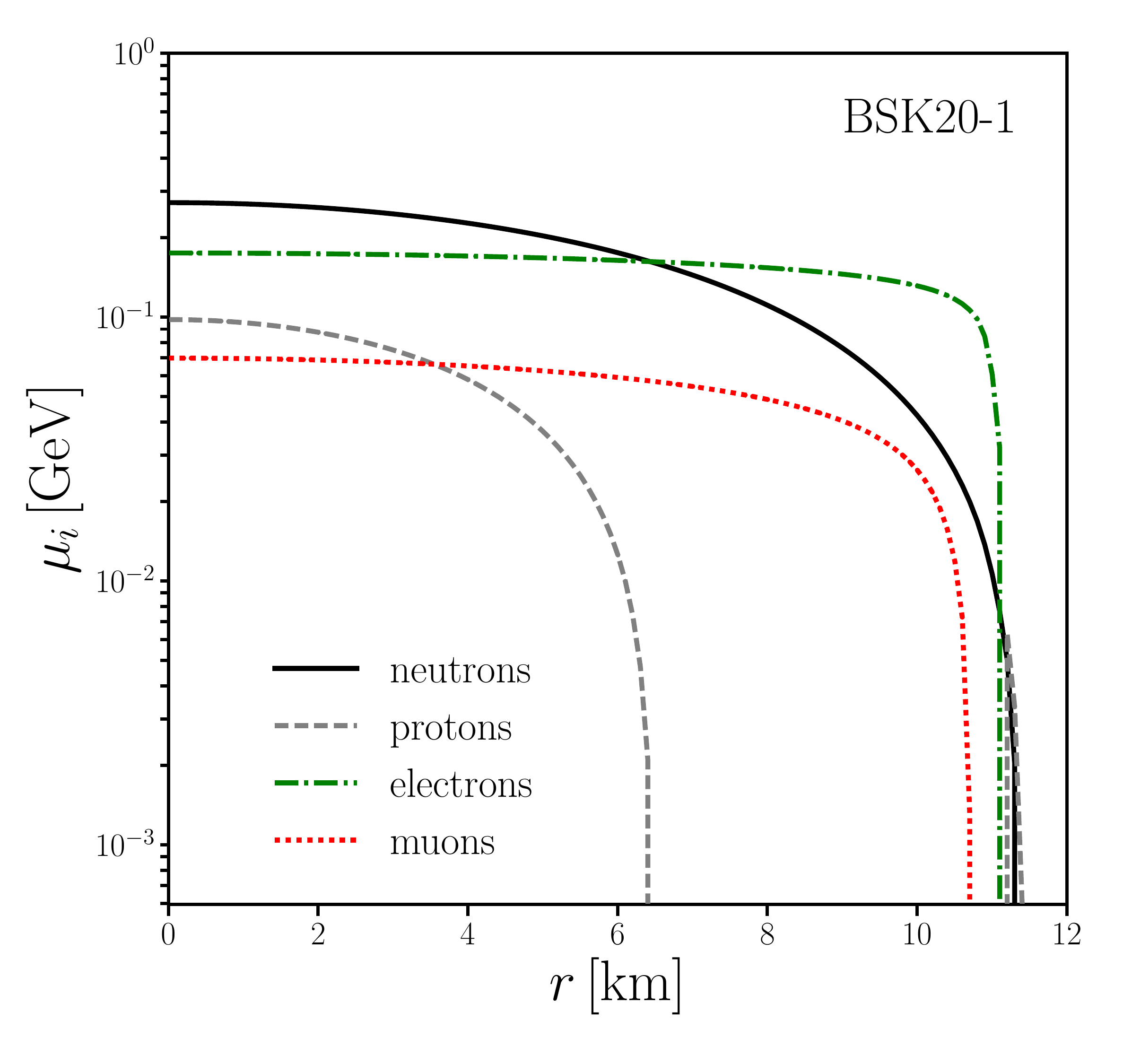}
        \includegraphics[width=0.45\linewidth]{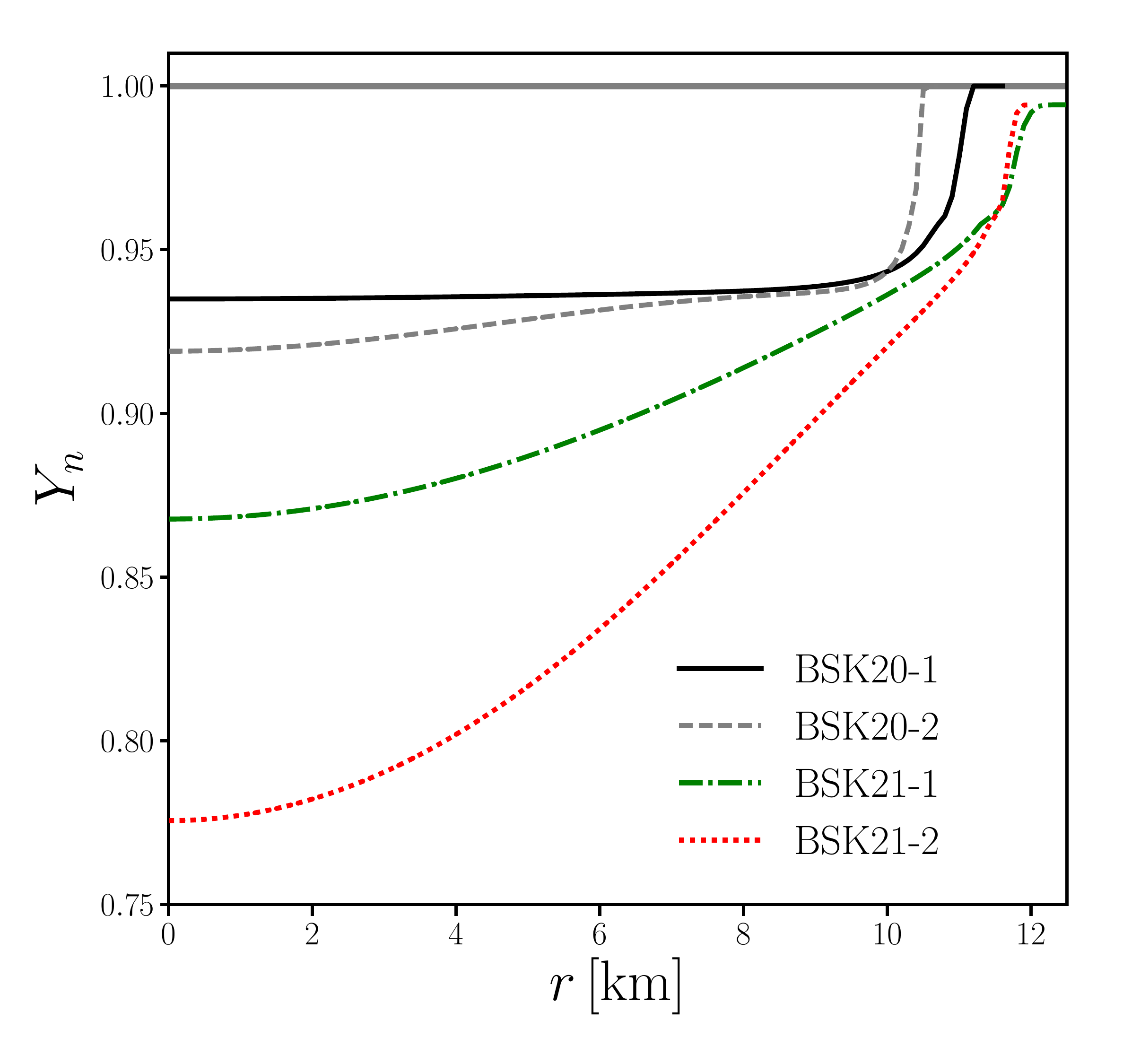}
		\includegraphics[width=0.45\linewidth]{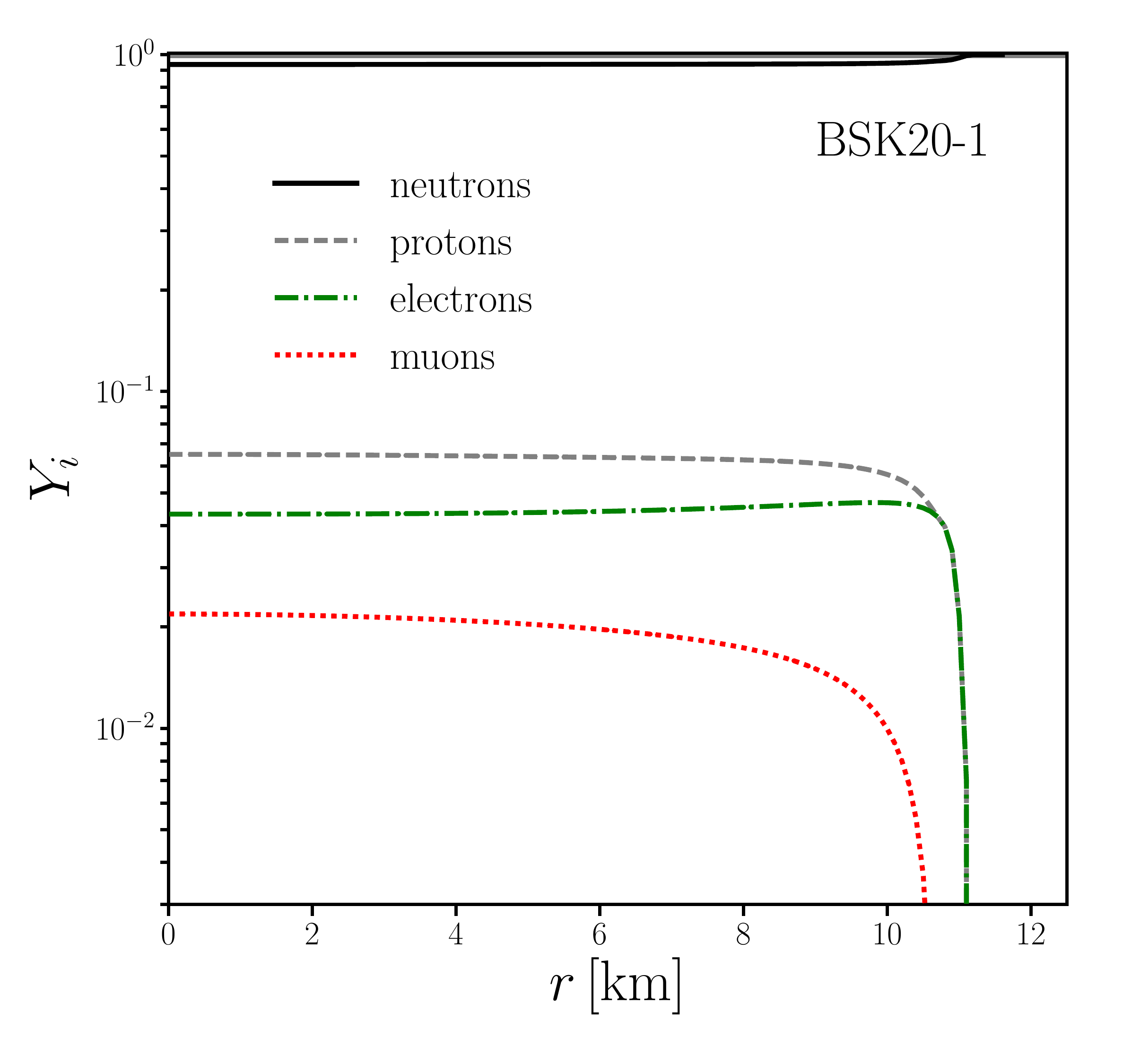}
	\end{center}
	\caption{\textit{Upper panel:} Number density of baryons in the different models. \textit{Middle panels:} Evolution of the chemical potential as a function of radius, for neutrons only in the different models (\textit{left}), and for all the species in the BSK20-1 model (\textit{right}). \textit{Lower panel:} Evolution of the fraction of the different species as a function of radius, for neutrons only in the different models (\textit{left}), and for all the species in the BSK20-1 model (\textit{right}). Note that the fractions $Y_i$ are computed with respect to the baryonic fraction. Hence $Y_n$ + $Y_p$ =1, and $Y_p$ =  $Y_{e}$ +$Y_{\mu}$ for charge conservation. }  
	\label{fig:chem-pot}
\end{figure}

In the case of DM captured by scattering off neutrons, we show in Fig.~\ref{fig:cap_two_mod} (upper panel) the capture rate with respect to DM mass for model BSK20-1 and also for model  BSK21-2 which gives the largest deviation with respect to BSK20-1. For $m_\chi \geqslant 1$GeV, the capture rate {for} BSK21-2 can be completely recovered by multiplying {with a rescaling factor} $\beta=N_n \times M_{BSk21-2}/M_{BSk20-1}$ (the relative ratio of total number of neutrons, and the NS mass ratios, see Tab.~\ref{tab:numbers}) to the capture rate obtained for BSK20-1. Below this mass, for $m_\chi \leqslant 1$GeV, Pauli blocking kicks in and suppresses the capture rate. The variation of the capture rate with the chemical potential is not trivial since it depends on the radial profile of the chemical potential {(see below).}
Since the constraints presented in Figs.~\ref{fig:cap-exclusion-neutrons} to \ref{fig:cap-exclusion-fermions} are directly proportional to the capture rate, the uncertainties on the NS model imply uncertainties on DM constraints. Computing the rescaling factor $\beta$ for the four benchmark models, one observes that the capture rate, and so the DM constraints, vary by a factor of 1.8, 8.4 and 5.3 for neutron, proton and muon, respectively (for $m_\chi \geqslant 1\,{\rm GeV}$). Note that the reference BSK20-1 (Model A) was chosen because it corresponds to the case which gives the most conservative bounds for all the species except protons. \\

\textit{Neutron star profile effects on DM capture}:\\
 
Now we focus on DM captured by neutrons for the model BSK20-1. From the lower left panel of Fig.~\ref{fig:cap_two_mod} we notice that, for $m_\chi=1$ GeV, the capture rate per unit volume follows the neutron density (see top panel of Fig.~\ref{fig:chem-pot}). For this DM mass the capture rate is quite insensitive to Pauli blocking effects. Instead, for $m_\chi=1$ MeV, the capture rate is boosted in the outer regions where the chemical potential is low (see middle panels of Fig.~\ref{fig:chem-pot}), and suppressed in the inner regions {due to} a higher value of the chemical potential. In the lower right panel of Fig.~\ref{fig:cap_two_mod} we show the distribution of DM captured i.e. the fraction of DM accreted at a given radius, which is thus different for different DM masses.

From this observation we come back to the comment on the capture rate in Fig.~\ref{fig:cap_two_mod} for $m_\chi\leqslant 1$ GeV. We note that, in this mass regime, the effect of Pauli blocking is more important in the model BSK20-1 than in BSK21-2, even {after} rescaling by the factor $\alpha$ (see above). This {seems to be in apparent contradiction as the chemical potential in model BSK21-2 is larger by factor 2 in the core. However, this can be understood from the fact that the outer layers are actually responsible for a larger capture fraction for those light DM particles due to smaller chemical potential in those regions (see lower right panel of Fig.~\ref{fig:cap_two_mod}).} Also note that model BSK21-2 has a larger radius than BSK20-1. Thus, the effective volume of neutron responsible for capture is larger in this model, and compensate for the suppression from {regions with} higher chemical potential. Hence, a precise determination of the capture rate in this regime is intractable without taking into account the radial profile of NS chemical potential.

Interestingly, note that for the case of protons in the model BSK20-1 (grey dashed lines in the left panels of Fig.~\ref{fig:chem-pot}), the chemical potential drops to zero for radii above $\approx$~6.5 km, unlike its contribution to the NS material which remains approximately constant. This observation highlights a phase transition, corresponding to the fact that above $\approx$~6.5 km, protons are bound in clusters constituting a solid crust. In computing DM capture, we only take into accounts collisions with the ``free'' fraction of protons, i.e.~for $r\gtrsim 6.5$~km, since scattering amplitudes might change for DM scattering off protons in clusters. We do not correct the neutron number for the clustered ones since the proton fraction is always much smaller than the neutron fraction. For other elements, electrons and muons, no such phase transition is {predicted} in any model. This remark on protons also applies to the other three models BSK20-2, BSK21-1 and BSK21-2.

	\begin{table}[!b]
	\begin{center}
	\begin{tabular}{l c c c c}
	\hline\hline
	Model & A & B & C & D  \\
	 & BSK-20-1 & BSK-20-2 & BSK 21-1 & BSK 21-2  \\
	\hline 
	Radius $R_\star$ [km] &  11.6 & 10.7 & 12.5 & 12.0\\
	Mass $M_\star$ [$\rm M_{\odot}$] & 1.52 & 2.12 & 1.54 & 2.11\\
	Number of {\bf free} particles & & & &  \\
	normalized to BSK-20-1 & & & &  \\
   \multicolumn{1}{c}{$N_n$} & 1.0 & 1.30  & 1.00    & 1.26 \\ 
   \multicolumn{1}{c}{$N_p$} & 1.0 & 3.42  & 0.83    & 6.07 \\
   \multicolumn{1}{c}{$N_\mu$}&1.0 & 1.80  & 1.52    & 3.82 \\

	Core chemical potential [GeV] & & & & \\
	\multicolumn{1}{c}{$\mu_n$} & 0.27    & 0.81  & 0.24  & 0.51 \\
	\multicolumn{1}{c}{$\mu_p$} & 0.098   & 0.60  & 0.38  & 0.25  \\
	\multicolumn{1}{c}{$\mu_\mu$} & 0.065 & 0.11 & 0.095 & 0.16 \\
	\hline
	
	\end{tabular}
	\caption{Relevant parameters for the benchmark NS models considered. }
        	\label{tab:numbers}
	\end{center}
	\end{table}

%%%%%%%%%%%%%%%%%%%%%%%%%%%%%%%%%%%%%%%%%%%%%%%%%%%%%%%%%%%%%

\section{Number Density Rescaling}\label{app:numd-resc}

The dispersion relation for neutrons in an ideal and non relativistic Fermi gas is $E=\frac{p^2}{2 m_n}$. In fact, at extreme nuclear densities, because of pair interactions, this relation is modified and writes $E=\frac{p^2}{2 m_n^*} + U_n$, where the mass $m_n^*$ is the effective mass of the neutron and $U_n$ is the associated potential energy. In this work, in computing capture and thermalization rates, we assume neutrons are ideal and non relativistic Fermi gas, which is also the approximation made in Ref.~\cite{Bertoni:2013bsa}. Since we are using realistic NS profiles with given number densities $n(r)$, we multiply our predictions by the following rescaling factor, 

\begin{equation}
    \zeta_n(r)=\frac{n(r)}{n^{\rm free}(r)}\label{eq:zeta_n}\;,
\end{equation}

where, in the limit of vanishing temperature, $n^{free}=(2 m_n \mu_n^{3/2}/(3 \pi^2)$. Values of $\zeta$ at the NS core can be found in Tab.~\ref{tab:numbers-zeta}.  

\begin{table}[hb]
	\begin{center}
	\begin{tabular}{l c c c c}
	\hline\hline
	Model & A & B & C & D  \\
	 & BSK-20-1 & BSK-20-2 & BSK 21-1 & BSK 21-2  \\
	\hline 
$\zeta=\frac{n}{n^{free}}$ at NS core & & & & \\
	\multicolumn{1}{c}{$\zeta_n$} & 0.32    & 0.10  & 0.29  & 0.12 \\
	\multicolumn{1}{c}{$\zeta_p$} & 0.10   & 0.014  & 0.68  & 0.11  \\
	\multicolumn{1}{c}{$\zeta_\mu$} & 1.5 & 1.9 & 1.7 & 2.3 \\
	\hline
	\hline	
\end{tabular}
  \caption{Rescaling factor for the number density of neutrons, protons and muons. }\label{tab:numbers-zeta}
\end{center}
\end{table}

%%%%%%%%%%%%%%%%%%%%%%%%%%%%%%%%%%%%%%%%%%%%%%%%%%%%%%%%%%%%%

\section{Differential scattering Rate}\label{app:diff-rate}

In this appendix we provide analytic/semi-analytic expressions for the differential scattering rate.
Starting from Eq.~(\ref{eq:cap-diff}), we find convenient to trade the integral over $d^3 \boldsymbol{u} = u^2  du \;d \cos \theta$, where $\theta$ is the angle between incoming DM and neutron ($|\boldsymbol{w}-\boldsymbol{u}| = \sqrt{\boldsymbol{w}^2+\boldsymbol{u}^2 -2 \boldsymbol{w}\boldsymbol{u} \cos \theta}$), for an integral over 
the velocity in the NS frame of  the center of mass (CM) of the scattering, $v_{s}$, and velocity of DM in this  CM frame, $v_{t}$. These integration variables can be expressed as a function of $\boldsymbol{u}$ and $\boldsymbol{w}$:
\begin{equation}
v_s= \frac{|\boldsymbol{u}+\mu \boldsymbol{w}|}{ 1+\mu}\,,\quad v_t= \frac{|\boldsymbol{w}-\boldsymbol{u}|}{1+\mu}\;,
\end{equation}
where $\mu=m_{\chi}/m_n$.
Separating the $R$ integral in 2 pieces, $R=R^+ + R^-$ depending on whether in the collision DM lose energy ($R^-$: $v<w$: accretion) or gains energy ($R^+$: $v>w$: evaporation) this the integrals
\begin{eqnarray}
\label{eq:rate-cm1}
R^\pm(w \rightarrow v) &=&  16\, \mu_+^4 n(r) \zeta(r) \frac{v}{w} \int_0^\infty \dd v_s \int_0^\infty \dd v_t \,v_t \,f_p(u)(1 -f_{p'}(u^\prime)) \nonumber \\
& &\frac{\dd \sigma}{\dd \cos \theta_{cm}}\, H_\pm(v_s,v_t,w,v),  
\end{eqnarray}
where $\theta_{cm}$ is the angle between incoming DM and outgoing DM in CM frame. The functions $H_\pm$ are combinations of $\Theta$ functions coming from the fact that we traded the cyclic $\theta$ variable for a continuous variable
\begin{equation}
\label{eq:H}
H_\pm(v_s,v_t,w,v) \equiv \Theta(w - |v_s - v_t|) \, \Theta(v_s + v_t - w) \, \Theta(v - |v_s - v_t|) \, \Theta(v_s + v_t - v)\,~,
\end{equation}
with
\begin{eqnarray}
\label{eq:mudef}
u^2 &\equiv& 2 \, \mu \, \mu_{+} \, v_t^2 + 2 \, \mu_{+} \, v_s^2 - \mu \, w^2 ~,    \nonumber \\[1ex]
u^{\prime 2} &\equiv& 2 \, \mu \, \mu_{+} \, v_t^2 + 2 \, \mu_{+} \, v_s^2 - \mu \, v^2 \,,\nonumber\\
\mu_{\pm} &\equiv& \frac{\mu \pm 1}{2} ~, \hspace{1cm} 
\end{eqnarray}
Thus for accretion ($H_-$), the limits of integration if we perform $v_t$-integral first are 
\begin{eqnarray}
\label{eq:intlim}
 \frac{w-v}{2} \le v_s \le \frac{v + w}{2}  ~, \hspace{1cm} w - v_s \le v_t \le v + v_s ~, \nonumber \\[1ex]
 \frac{v + w}{2} \le v_s \le \infty  ~, \hspace{1.7cm} v_s - v \le v_t \le v + v_s ~; 
 \end{eqnarray}
 while for evaporation ($H_+$) they are
\begin{eqnarray}\
 \frac{v - w}{2} \le v_s \le \frac{v + w}{2}  ~, \hspace{1cm} v - v_s \le v_t \le w + v_s ~, \nonumber\\[1ex]
 \frac{v + w}{2} \le v_s \le \infty  ~, \hspace{1.7cm} v_s - w \le v_t \le w + v_s ~. \nonumber 
\end{eqnarray}

As discussed in the introduction of Sec.~\ref{sec:accretion}, we consider the case of a cross section which has no momentum dependence and is isotropic.
Thus the differential cross section in Eq.~(\ref{eq:rate-cm1}) reads
\beq
\label{eq:diff-sig}
\frac{\dd \sigma}{\dd \cos\theta_{cm}} = \frac{\sigma_0}{2}\;,
\eeq
and the differential accretion rate can be written as
\begin{eqnarray}
\label{eq:rate-cm2}
R^-(w \rightarrow v) =    8 \mu_+^4 \sigma_0  n(r)  \zeta(r)  \frac{v}{w} \int_0^\infty \dd v_s \int_0^\infty \dd v_t \,v_t f_{p}(v_s,v_t)(1 - f_{p'}(v_s,v_t))  H_-(v_s,v_t,w,v). 
\end{eqnarray}
For this case of a constant cross section, the $v_t$ integral can be calculated analytically. 
To this end, consider the function:

\begin{eqnarray}
\label{eq:rate-int-I}
I^{\pm} &=&   \int_0^\infty \dd v_s \int_0^\infty \dd v_t \,v_t \nonumber \\
&& f_{p}(v_s,v_t,w)(1 - f_{p'}(v_s,v_t,v))  H^{\pm}(v_s,v_t,w,v). 
\end{eqnarray}

We first integrate in variable $v_t$. In order to make the equations tractable
we can define the following variables

\begin{eqnarray}
\label{eq:t-int-}
I_{t1}^{-} &=& \int_{w-v_s}^{v+v_s} \dd v_t \,v_t \,f_{p}(v_s,v_t,w)(1 - f_{p'}(v_s,v_t,v))\nonumber \\
I_{t2}^{-}  &=& \int_{v_s-v}^{v+v_s} \dd v_t \,v_t \,f_{p}(v_s,v_t,w)(1 - f_{p'}(v_s,v_t,v)),  
\end{eqnarray}

and

\begin{eqnarray}
\label{eq:s-int-}
I_{1}^{-} &=& \int_{\frac{w-v}{2}}^{\frac{v+w}{2}} \dd v_s\, I_{t1}^{-} \nonumber \\
I_{2}^{-}  &=& \int_{\frac{v+w}{2}}^{\infty}   \dd v_s \,I_{t2}^{-} \nonumber \\
I^{-} &=& I_{1}^{-} + I_{2}^{-}. 
\end{eqnarray}

Similarly for evaporation, we have

\begin{eqnarray}
\label{eq:t-int+}
I_{t1}^{+} &=& \int_{v-v_s}^{w+v_s} \dd v_t\, v_t\, f_{p}(v_s,v_t,w)(1 - f_{p'}(v_s,v_t,v))\nonumber \\
I_{t2}^{+}  &=& \int_{v_s-w}^{w+v_s} \dd v_t\, v_t\, f_{p}(v_s,v_t,w)(1 - f_{p'}(v_s,v_t,v)), 
\end{eqnarray}

\begin{eqnarray}
\label{eq:s-int+}
I_{1}^{+} &=& \int_{\frac{v-w}{2}}^{\frac{v+w}{2}} \dd v_s \,I_{t1}^{+} \nonumber \\
I_{2}^{+}  &=& \int_{\frac{v+w}{2}}^{\infty}   \dd v_s\, I_{t2}^{+} \nonumber \\
I^{+} &=& I_{1}^{+} + I_{2}^{+}. 
\end{eqnarray}

Analytical results for the $v_t$-integral are then

\begin{eqnarray}
\label{eq:t-int-2}
I_{t1}^{-} &=&  \frac{1}{4 k^2 \mu \mu_+ \left(1- e^{-\frac{q_0}{ T}}\right)} \left(  \log \left[\frac{1 + x_1}{1+ x_2 } \right] - \log\left[\frac{1 + x_3}{1+ x_4 } \right] \right) \nonumber \\
I_{t2}^{-}  &=& \frac{1}{4 k^2 \mu \mu_+ \left(1- e^{-\frac{q_0}{ T}}\right)} \left(  \log \left[\frac{1 + x_5}{1+ x_6 } \right] - \log\left[\frac{1 + x_3}{1+ x_4 } \right] \right),  \nonumber \\
\end{eqnarray}
and
\begin{eqnarray}
\label{eq:t-int+2}
I_{t1}^{+} &=&  \frac{1}{4 k^2 \mu \mu_+ \left(1- e^{-\frac{q_0}{ T}}\right)} \left(  \log \left[\frac{1 + y_1}{1+ y_2 } \right] - \log\left[\frac{1 + y_3}{1+ y_4 } \right] \right) \nonumber \\
I_{t2}^{+}  &=& \frac{1}{4 k^2 \mu \mu_+ \left(1- e^{-\frac{q_0}{ T}}\right)} \left(  \log \left[\frac{1 + y_5}{1+ y_6 } \right] - \log\left[\frac{1 + y_3}{1+ y_4 } \right] \right),  \nonumber \\
\end{eqnarray}
with 

\begin{eqnarray}
\label{eq:t-pars}
k^2 &=& \frac{m_n}{2 T},  \\
q_0 &= &\frac{1}{2 } m_\chi \left(w^2 -v^2 \right), \\
v_f^2 &=& \frac{2 \,\mu_F}{m_n}, \\ 
x_1 &=& \exp\left[k^2 \left(v_f^2 + \mu w^2 -2 \mu_+ \left(v_s^2 + \mu (w-v_s)^2\right) \right)\right], \nonumber \\
& & \\
x_2 &=& \exp\left[k^2 \left(v_f^2 + \mu v^2 -2 \mu_+ \left(v_s^2 + \mu (w-v_s)^2\right) \right)\right], \nonumber \\
& & \\
x_3 &=& \exp\left[k^2 \left(v_f^2 + \mu w^2 -2 \mu_+ \left(v_s^2 + \mu (v+v_s)^2\right) \right)\right], \nonumber \\
& & \\
x_4 &=& \exp\left[k^2 \left(v_f^2 + \mu v^2 -2 \mu_+ \left(v_s^2 + \mu (v+v_s)^2\right) \right)\right], \nonumber \\
& & \\
x_5 &=& \exp\left[k^2 \left(v_f^2 + \mu w^2 -2 \mu_+ \left(v_s^2 + \mu (v-v_s)^2\right) \right)\right], \nonumber \\
& & \\
x_6 &=& \exp\left[k^2 \left(v_f^2 + \mu v^2 -2 \mu_+ \left(v_s^2 + \mu (v-v_s)^2\right) \right)\right], \nonumber \\
& & \\
y_1 & =&  x_5,\\
y_2 & =&  x_6,\\
y_3 &=&  \exp\left[k^2 \left(v_f^2 + \mu w^2 -2 \mu_+ \left(v_s^2 + \mu (w+v_s)^2\right) \right)\right], \nonumber \\
& & \\ 
y_4 &=&  \exp\left[k^2 \left(v_f^2 + \mu v^2 -2 \mu_+ \left(v_s^2 + \mu (w+v_s)^2\right) \right)\right], \nonumber \\
& & \\ 
y_5 & =&  x_1,\\
y_6 & =&  x_2.
\end{eqnarray}

Next, the $v_s$-integral must be performed. For this integral, to get an exact analytical expression for arbitrary degeneracy turns out to be not possible. However we can make approximations to Eq.~(\ref{eq:t-int-2}) and ~(\ref{eq:t-int+2}) in the extremely degenerate and non-relativistic limits ($T \ll \mu_F, m_n$). For instance, a good approximation one can make is to replace,
\bea
\label{eq:s-int-approx}
\lim_{T \ll \mu_F} \frac{1}{k^2 (1-e^{-q_0/T})} \log[1+x_1] \sim \left(v_f^2 + \mu w^2 -2 \mu_+ \left(v_s^2 + \mu (w-v_s)^2\right) \right) \Theta\left(v_f^2 + \mu w^2 -2 \mu_+ \left(v_s^2 + \mu (w-v_s)^2\right) \right),
\ena
with similar limits for all other terms involved in Eq.~(\ref{eq:t-int-2}) and ~(\ref{eq:t-int+2}). With this simplification we can proceed to perform the $v_s$-integral by parts. As all the integrands are of the form
\begin{equation}
\int_{s_1}^{s_2} \dd s \left(a s^2 + b s +c\right)\Theta\left(a s^2 + b s +c\right) 
\end{equation}
they can be rewritten as
\bea
\int_{s_1}^{s_2} \dd s \left(a s^2 + b s +c\right)\Theta\left(a s^2 + b s +c\right) &=& \left(\frac{a}{3} s^3 + \frac{b}{2} s^2  + c s \right)\Theta \left(a s^2 + b s +c\right) \big|_{s_1}^{s_2} \nonumber \\
 & & - \int_{s_1}^{s_2} \dd s \left(\frac{a}{3} s^3 + \frac{b}{2} s^2  + c s \right) \delta \left(a s^2 + b s +c\right)\left(b + 2 a s\right).
\ena
Using the $\delta$ function identity
\beq
\delta\left(f(s) \right) = \mathlarger{\mathlarger{\Sigma}}_i \frac{\delta\left(s-s_i\right)}{|f^\prime(s_i)|},
\eeq
we get
\bea
\int_{s_1}^{s_2} \dd s \left(a s^2 + b s +c\right)\Theta\left(a s^2 + b s +c\right) &=& \left(\frac{a}{3} s^3 + \frac{b}{2} s^2  + c s \right)\Theta \left(a s^2 + b s +c\right) \big|_{s_1}^{s_2} \nonumber \\
&& - \mathlarger{\mathlarger{\Sigma}}_i  \left(\frac{a}{3} s_i^3 + \frac{b}{2} s_i^2  + c s_i \right) \Theta(s_2 -s_i)\Theta(s_i-s_1).
\ena
This gives the final result
\bea
\label{eq:r-factor}
6 \mu_+ I^- &=& -(2 \vf+v \mu ) \left(\Theta \left(\frac{\rho _--\vf}{2 \mu _+}\right)+\Theta \left(\frac{\vf-\rho _+}{2 \mu _+}\right)\right) (\vf-v \mu )^2 \nonumber \\
 & & - 2 \Theta \left(\frac{\vf-\phi _-}{2 \mu _+}\right) \Theta \left(\frac{\rho _+-\vf}{2 \mu _+}\right)
\left(\mu  (2 \mu +3) v^2-6 \mu  \mu _+ v^2-\vf \mu  v+2 \vf^2\right) (\vf-v \mu ) \nonumber \\
& & -3 \mu  \left((-v-w) \left(v^2-w^2\right)
  \Theta \left(\vf^2-\alpha _-^2\right) \right . \nonumber \\ 
&& \left . \qquad \qquad
 -(v-w) \left(\left(v^2-w^2\right) \left(\Theta \left(\vf^2-\alpha _+^2\right)-\Theta
   \left(\vf^2-\beta _+^2\right)\right)-(v+w)^2 \Theta \left(\vf^2-\beta _-^2\right)\right)\right) \mu _+^2 \nonumber \\
   & & +(\vf-w \mu )^2 (2
   \vf+w \mu ) \Theta \left(\frac{\vf+\alpha _-}{2 \mu _+}\right) \Theta \left(\frac{\alpha _+-\vf}{2 \mu _+}\right)\nonumber \\
   &&-(2
   \vf-w \mu ) (\vf+w \mu )^2 \Theta \left(\frac{\alpha _--\vf}{2 \mu _+}\right) \Theta \left(\frac{\vf+\alpha _+}{2 \mu
     _+}\right) \nonumber \\
   &&+(\vf+v \mu ) \left( \Theta \left(\frac{\vf+\rho _-}{2 \mu _+}\right) + \Theta \left(-\frac{\vf+\rho _+}{2 \mu _+}\right) \right) \left(\mu  (2 \mu +3) v^2-6 \mu  \mu _+ v^2+\vf
   \mu  v+2 \vf^2\right) \nonumber \\
   && - \left(\Theta \left(\frac{\rho _--\Lambda _+}{2 \mu _+}\right)   +\Theta \left(\frac{\Lambda _+-\rho _+}{2 \mu
   _+}\right)\right)\left(v \mu -\Lambda _+\right) \left(2
   \mu ^2 v^2-6 \mu  \mu _+ v^2-\mu  \Lambda _+ v+3 \vf^2-\Lambda _+^2+3 w^2 \mu \right) \nonumber \\
   && - \left( \Theta \left(\frac{\rho _-+\Lambda _+}{2 \mu _+}\right) + \Theta \left(-\frac{\Lambda _++\rho _+}{2 \mu _+}\right) \right)\left(v \mu +\Lambda _+\right) \left(2 \mu ^2 v^2-6 \mu  \mu _+ v^2+\mu 
   \Lambda _+ v+3 \vf^2-\Lambda _+^2+3 w^2 \mu \right) \nonumber \\
   && -\Theta \left(\frac{\Lambda
   _+-\phi _-}{2 \mu _+}\right) \Theta \left(\frac{\rho _+-\Lambda _+}{2 \mu _+}\right) \left(v \mu -\Lambda _+\right) \left(-6 \mu  \mu _+
   v^2+2 \vf^2+\mu  \left((2 \mu +1) v^2-\Lambda _+ v+2 w^2\right)\right) \nonumber \\
   && -\Theta \left(-\frac{\phi _-+\Lambda _+}{2 \mu _+}\right) \Theta
   \left(\frac{\Lambda _++\rho _+}{2 \mu _+}\right) \left(v \mu +\Lambda _+\right) \left(-6 \mu  \mu _+ v^2+2 \vf^2+\mu  \left((2 \mu +1)
   v^2+\Lambda _+ v+2 w^2\right)\right) \nonumber \\
   && +\Theta \left(\frac{\alpha _-+\Lambda _-}{2 \mu _+}\right) \Theta \left(\frac{\alpha _+-\Lambda _-}{2 \mu
   _+}\right) \left(w \mu -\Lambda _-\right) \left(2 \vf^2+\mu  \left(2 v^2+w \left(2 \mu  w+w-\Lambda _-\right)\right)-6 w^2 \mu  \mu
   _+\right) \nonumber \\
   && +\Theta \left(\frac{\alpha _--\Lambda _-}{2 \mu _+}\right) \Theta \left(\frac{\Lambda _-+\alpha _+}{2 \mu _+}\right) \left(w \mu
   +\Lambda _-\right) \left(2 \vf^2+\mu  \left(2 v^2+w \left(2 \mu  w+w+\Lambda _-\right)\right)-6 w^2 \mu  \mu _+\right).
\ena
with
\bea
\label{eq:variables-2}
\Lambda_{\pm} &\equiv& \left(v_f^2 \pm \mu (w^2 -v^2) \right)^{1/2}, \nonumber\\
\alpha_\pm &\equiv& \mu_+ v \pm \mu_- w, \nonumber\\
\beta_\pm &\equiv& \mu_- v \pm \mu_+ w, \nonumber\\
\rho_\pm &\equiv&  \mu v \pm \mu_+ \left(v+w\right), \nonumber\\
\phi_\pm &\equiv&  \mu v \pm \mu_+ \left(v -w \right). 
\ena
This result, which hold for $T=0$, has 2 types of terms, the ones which involve a single theta function (with linear and quadratic arguments) and the ones which involve the product of two theta functions (again with linear and quadratic arguments).

\section{Functions for Eq.~(\ref{eq:el_averaged1})}\label{app:fct_eq}
For orbits smaller than the NS size, namely $r_0<R_\star$, the orbits are isochrones with period:
\bea 
 T_{in} = 2\pi \displaystyle\frac{R^{3/2}}{\sqrt{GM}}\;.
\ena
For orbits larger than the NS size, namely $r_0>R_\star$, the orbits are no longer isochrones, and the periods depends on $r_0$,
\bea 
T_{out} &=& 4 \displaystyle\frac{R^{3/2}}{\sqrt{GM}} \left( \arcsin(B) + \frac{1}{2} \left(\frac{r_0}{R}\right)^{\frac{3}{2}} \right . \nonumber \\
 && \left. \qquad \qquad \left[  \frac{R}{r_0}\sqrt{\frac{r_0}{R}-1}+\frac{1}{2}\arctan{ \sqrt{\frac{r_0}{R}-1} - \frac{1}{2}\frac{r_0}{R}(\frac{r_0}{R}-1)^{-1/2} } + \frac{\pi}{4} \right]\right) \,,\\
 \ena
with $B = \left(3-2\frac{R}{r_0}\right)^{-1/2}$. In Eq.~(\ref{eq:el_averaged1}), $\eta(E)$ is the ratio between these two periods, $\eta(E)\equiv\frac{T_{in}}{T_{out}}$. Note that Eq.~(\ref{eq:el_averaged1}) also involves the function $I(E)$, whose form we give here:
\bea
 I(E) &=& \int_0^B(1-x^2)^{5/2} dx \\&=& \frac{1}{48}(B\sqrt{1-B^2}(8B^4-26B^2+33)+15\arcsin(B))\\
\ena

\section{Numerical Solution of Eq.~(\ref{eq:redistrib})}\label{app:num_sol}

To solve Eq.~(\ref{eq:redistrib}) we simply discretize with respect to the energy. The grid indices are chosen so that $E_0$ corresponds to the largest energy (basically $E_{surf}$), and $E_N$ to the smallest one (chosen to be $E_{th}$). 
The fact that the source term has a dependence in energy which doesn't depend on time simplifies the problem.
We present the solutions for two typical time evolution: (i) a \textit{steady source term} in time $q(E,t)=Q(E)\times \Theta(t)$, and (ii) a \textit{burst like source term} $q(E,t)=Q(E)\times \delta(t)$. We introduce the discretized quantities:
\bea
f_i(t)&=&f_\chi(E_i,t)\\
\Gamma_i &=& \Gamma(E_i)\\
\Gamma_{i\rightarrow j}&=&\frac{\dd\Gamma}{\dd E'}(E_i \rightarrow E_j)\\
Q_i&=&Q(E_i)\;.
\ena
Using the trapezoidal rule\footnote{We could have used more advanced scheme such as Simpson's rule, although this method appears to be accurate enough.} to discretize the integral of Eq.~(\ref{eq:redistrib}), we obtain the following differential equation in time for each component $f_i$
\begin{eqnarray}
	\frac{\dd f_i}{\dd t}\;= \;\sum_{j=0}^i \tilde\Gamma_{j\rightarrow i}\,f_j \;- \;\Gamma_i\,f_i \;+\; Q_i,\label{eq:redistrib_disc}\,
\end{eqnarray}
with 
\bea
\tilde\Gamma_{i\rightarrow j}=\frac{1}{2}\,(E_{i-1}-E_{i+1})\,\Gamma_{i\rightarrow j}\,f_i\\
\tilde\Gamma_{0\rightarrow j}=\frac{1}{2}\,(E_{0}-E_{1})\,\Gamma_{0\rightarrow j}\,f_0\;.
\ena
Solving Eq.~(\ref{eq:redistrib_disc}) is handle-able recursively, by going from the highest to the lowest energies. The solution can be cast into the form,
\beq
f_i(t)=\sum_{j=0}^i \alpha_j^i \exp{(-\Gamma_j\,t)}\,, \label{eq:fE_discrete}
\eeq
with, for the case (i)
\bea
j=0:\quad &\alpha_0^i = \displaystyle\frac{1}{\Gamma_i}\,\left(\sum_{k=0}^{i-1}\tilde\Gamma_{k\rightarrow i}\,\alpha_0^k+Q_i \right)\\
0<j<N: \quad &\alpha_j^i =\displaystyle \frac{1}{\Gamma_i-\Gamma_j}\, \sum_{k=0}^{i-1} \tilde\Gamma_{k\rightarrow i}\alpha_j^k\\
j=N: \quad &\alpha_N^i = \displaystyle-\sum_{k=0}^{i-1}\,\alpha_k^i \;.
\ena

For case (ii), only the expression for $\alpha_0^i$ changes, and we get
\bea
j=0:\quad &\alpha_0^i = \displaystyle\frac{ Q_i}{\Gamma_i} \;.
\ena

\section{Combined Results and Results for $T=10^6$~K}\label{app:combinedresults}

\begin{figure}[H]
	\begin{center}
	  \includegraphics[width=0.45\linewidth]{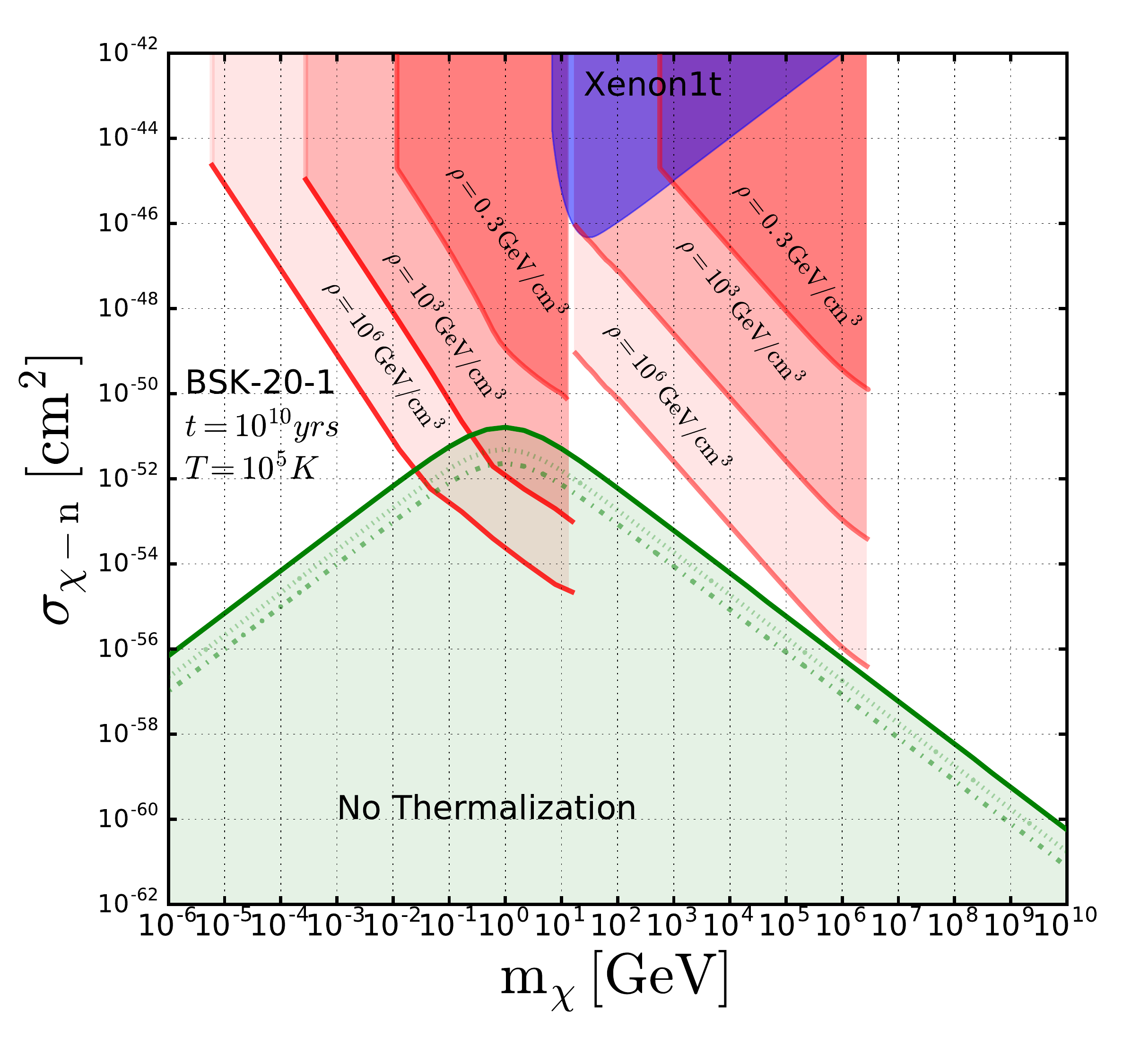}   \includegraphics[width=0.45\linewidth]{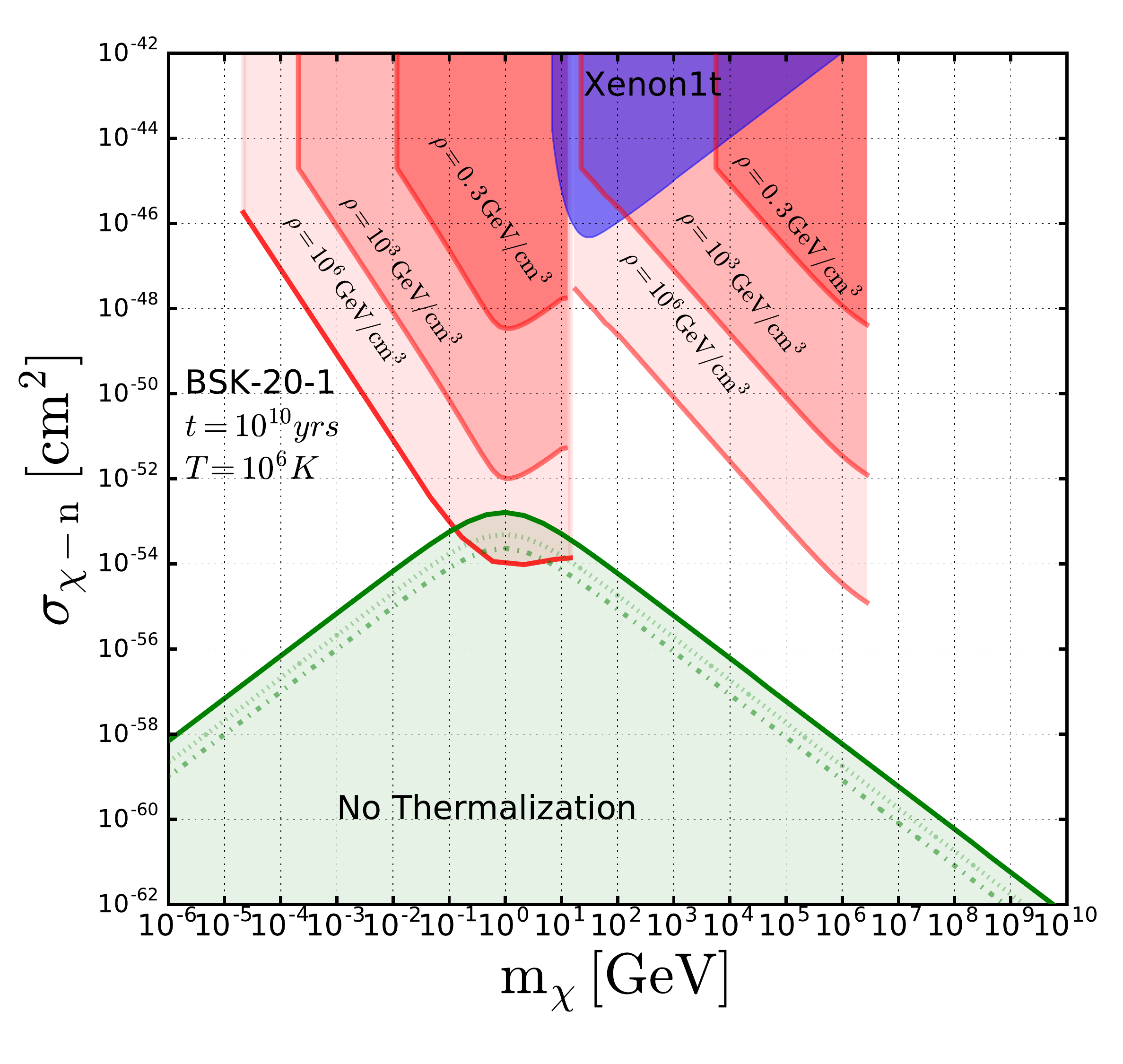}\\
	\end{center}
	\caption{\textbf{\textit{Combined exclusion plot for bosonic DM scattering off neutrons for $T=10^5$~K and $T=10^6$~K:} } Same as Fig.~\ref{fig:cap-exclusion-neutrons}. 
 }
	\label{fig:cap-exclusion-n-comb}
\end{figure}

In the left panel of Fig.~\ref{fig:cap-exclusion-n-comb} we merge both panels of Fig.~\ref{fig:cap-exclusion-neutrons} to summarize the best constraints which hold for bosonic DM scattering off neutrons. In the right panel we present the combined exclusion curve for T$=10^6$ K. The discontinuities in this plot are due to black hole evaporation thresholds.  

\begin{figure}[H]
	\begin{center}
	  \includegraphics[width=0.45\linewidth]{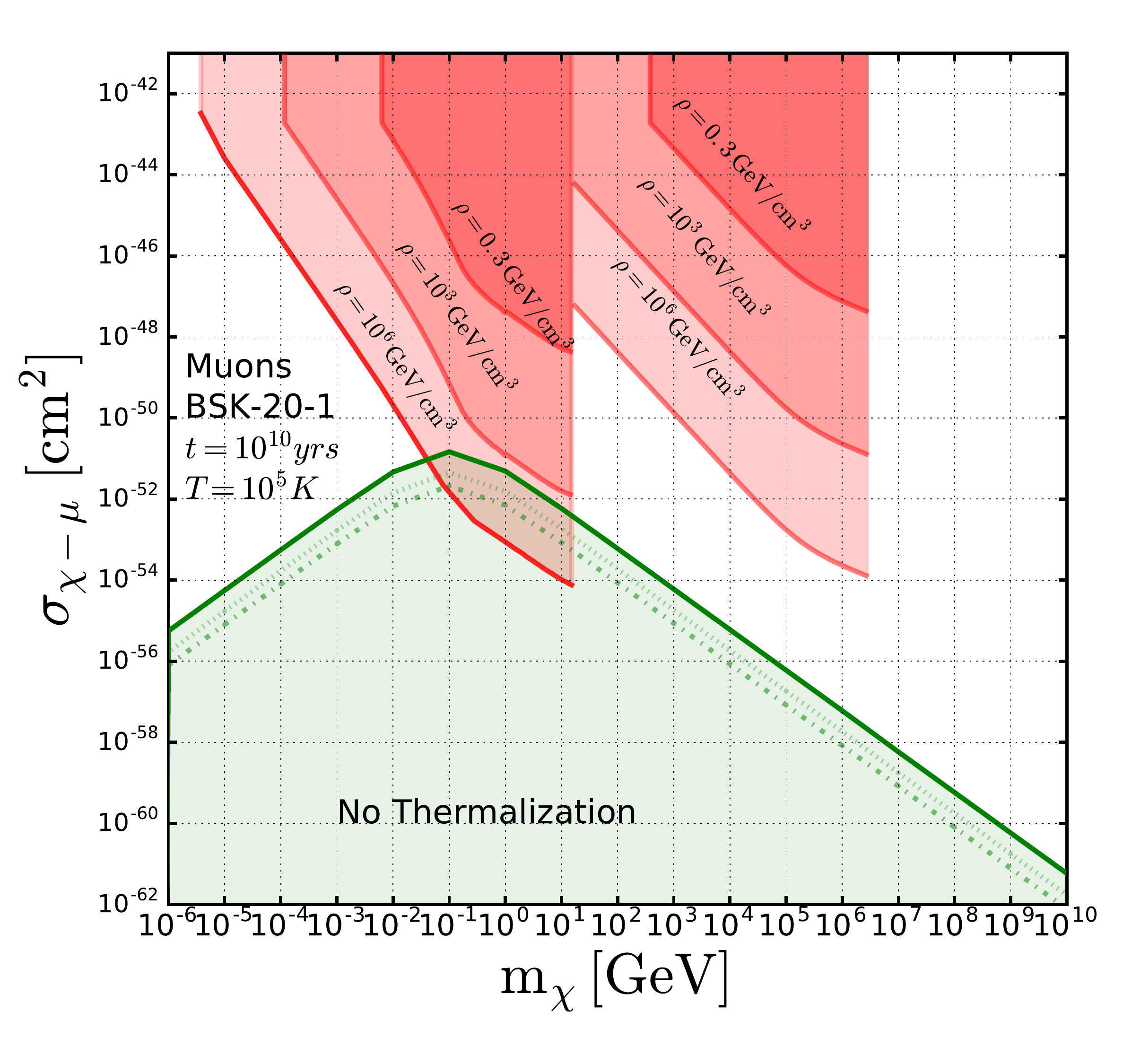}   \includegraphics[width=0.45\linewidth]{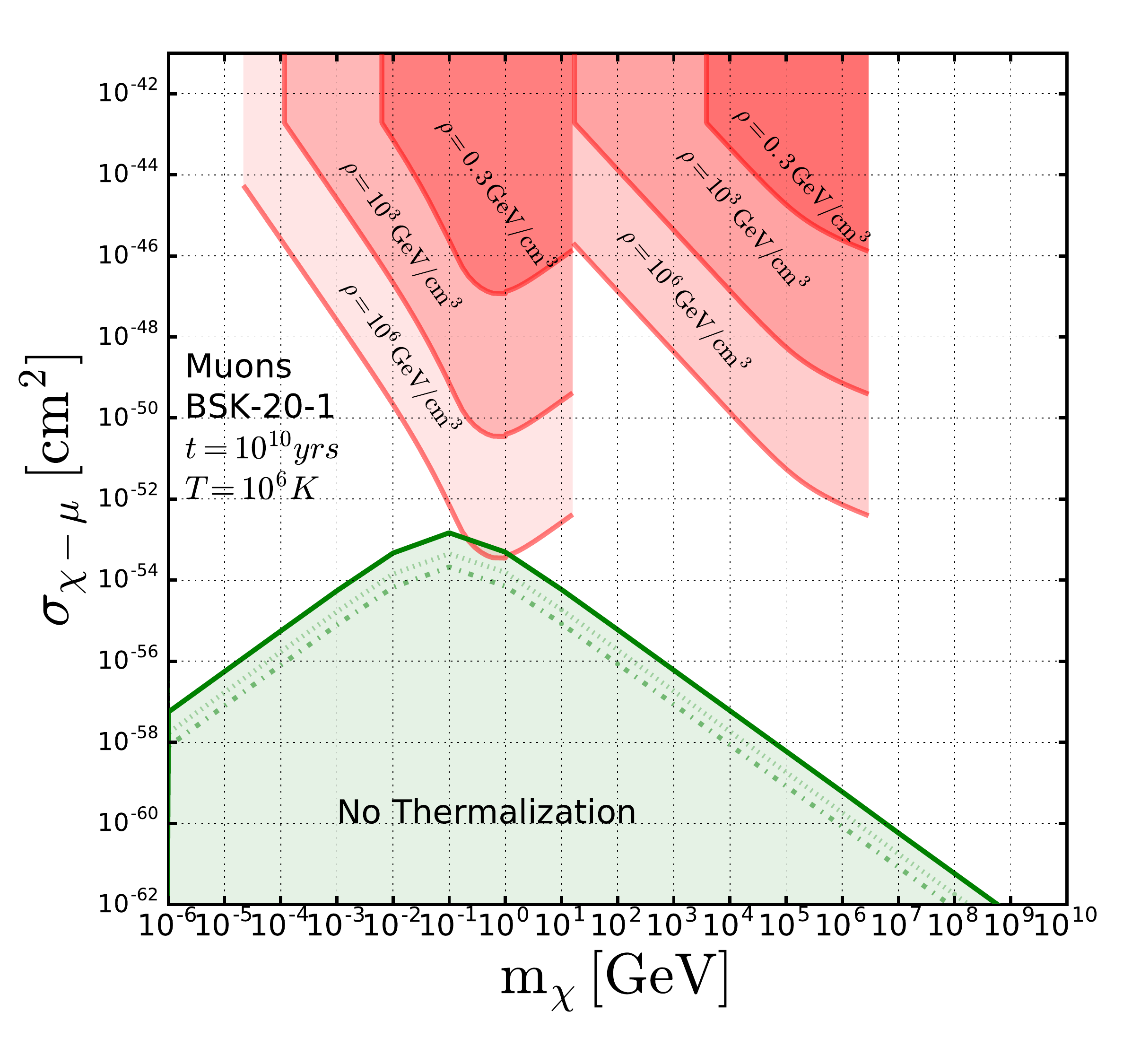}\\
	\end{center}
	\caption{\textbf{\textit{Combined exclusion plot for bosonic DM scattering off muons for $T=10^5$~K and $T=10^6$~K:} } Same as Fig.~\ref{fig:cap-exclusion-neutrons} for DM scattering off muons.
 }
	\label{fig:cap-exclusion-muons-combined}
\end{figure}

Similar to above, in Fig.\ref{fig:cap-exclusion-muons-combined} we merge both panels of Fig.~\ref{fig:cap-exclusion-muons} to summarize the best constraints which hold for DM scattering off muons. In the right panel we present the combined exclusion curve for T$=10^6$ K.

\section{Remark Regarding the Effect of DM on Gravitational Wave Emission in Neutron Star Mergers}\label{app-gw}

We finish the appendix with a comment which is a bit off the main subject of this work but which we find interesting to mention. It concerns the possibility to observe a DM induced modification to the spectrum of gravity waves in binary NS mergers. Of course to leave an observable imprint in the gravitational wave spectrum one needs many more DM particles in the NS than what can be accreted from scattering DM off ordinary matter. If all the DM particles crossing the NS are all trapped (geometric case above) one could hardly get a DM fraction larger than $\sim 10^{-10}$ with respect to ordinary matter\footnote{Note that, as shown in \cite{Brayeur:2011yw}, NS in binary systems can accrete about four times more DM than an isolated NS.} (see Eq.~(\ref{eq:est_tot_M}) for $\rho_\chi=10^4$ GeV/cm$^{3}$). However one could eventually get more DM in other ways\cite{Nelson:2018xtr,McKeen:2018xwc,Cline:2018ami,Grinstein:2018ptl}.
In this case one could hope, see for instance  \cite{Ellis:2017jgp}, that during the final stage of the spiral motion, because of tidal forces, the two DM cores come out of the neutron stars and continue to rotate within the newly-formed matter disk. Thus, the difference of rotation speed of the cores and the NS could leave a characteristic imprint in the gravitational wave signal. We argue in the following that this scenario is not realistic, i.e. the cores cannot escape from the gravitational well. To convince the reader, we simply compute the velocity needed for the core at the center to reach the edge at rest, and compare it with the one of the final collision.

We take the example of GW170817~\cite{GBM:2017lvd}. The collision speed is obtained by multiplying the length of \textit{the last orbit} for which radius is assumed to be the NS radius, by half of the highest frequency measured by LIGO (to account for the quadrupole emission). We also take into account the gravitational redshift of the gravitational wave signal. Thus we get

\begin{eqnarray}
v_{\rm collision}= \frac{600 \rm Hz}{2}\times \frac{1}{\sqrt{1-\frac{4GM_\star}{c^2R_\star}}}\times 2\pi R_\star \approx 0.073 c\,. 
\end{eqnarray}
The escape velocity needed for the core to reach the edge of the NS can be estimated classically for a homogeneous sphere:
\begin{eqnarray}
v_{\rm esc}^N = \sqrt{\frac{GM_\star}{R_\star}} = 0.44\, c\,.
\end{eqnarray}
The exact GR calculation leads to:
\begin{eqnarray}
v_{\rm esc}^{GR}=c\,\sqrt{1-e^{2(\Phi(0)-\Phi(R))}}=0.63 \,c\,.
\end{eqnarray}
In any case we see that the velocity reached during the collision is much lower than the one required to extract the dark matter cores from the gravitational well of the NS. Thus, the fate of the DM cores is completely related to what happens to the baryons. The addition of an extra long range interactions between the DM particles could modify this picture (see e.g. \cite{Kopp:2018jom}). 

\end{document}